\documentclass[journal]{IEEEtran}
\IEEEoverridecommandlockouts
\pdfoutput=1
\def\linspread{0.917}
\def\linspreadalgr{0.8}
\usepackage{cite}
\usepackage{amsfonts}
\usepackage{amssymb}
\usepackage{amsmath}
\usepackage{amsthm}
\usepackage{bm}
\usepackage{algorithm}
\usepackage{algorithmic}
\usepackage{array}
\usepackage{booktabs} 
\usepackage{float}

\usepackage{multirow}
\usepackage{balance}  
\usepackage{subfigure}
\usepackage{textcomp}
\usepackage{stfloats}  % Correctly control big float in double column 
\usepackage{url}
\usepackage{verbatim}  % Use "\begin{verbatim}  \end{verbatim}" or \verb|text| to show origin text 
\usepackage{graphicx}
\usepackage{epstopdf}
\usepackage{balance}  

\usepackage[colorlinks=false,
linkcolor=blue,
anchorcolor=blue,
citecolor=blue]{hyperref}

\graphicspath{ {figs/system-model/} {figs/simulation/} }
\DeclareMathOperator*{\argmax}{arg\,max}
\DeclareMathOperator*{\argmin}{arg\,min}

\theoremstyle{definition}

\theoremstyle{remark}

\allowdisplaybreaks[4]
\usepackage{afterpage}

\def\BibTeX{{\rm B\kern-.05em{\sc i\kern-.025em b}\kern-.08em
    T\kern-.1667em\lower.7ex\hbox{E}\kern-.125emX}}

\makeatletter

\newcommand{\Rmnum}[1]{\expandafter\@slowromancap\romannumeral #1@}
\makeatother

\newfloat{figtab}{thb}{fgtb}
\makeatletter
\newcommand\figcaption{\def\@captype{figure}\caption}
\newcommand\tabcaption{\def\@captype{table}\caption}
\makeatother

\begin{document}

%\title{IRS-Aided Joint Spatial Division and Multiplexing for mmWave Multiuser MISO Systems 
%}
\title{Joint Location Sensing and Channel Estimation for IRS-Aided mmWave ISAC Systems
}

\author{Zijian Chen,  Ming-Min Zhao, Min Li, Fan Xu, Qingqing Wu,  and Min-Jian Zhao
\thanks{Z. Chen,  M. M. Zhao, M. Li and M. J. Zhao  are  with the College of Information
	Science and Electronic Engineering, Zhejiang University, Hangzhou 310027, China
%	and also with the Zhejiang Provincial Key Laboratory of Information Processing, Communication and Networking (IPCAN), Hangzhou 310027, China 
	(e-mail: chenzijian@zju.edu.cn; zmmblack@zju.edu.cn;  min.li@zju.edu.cn; mjzhao@zju.edu.cn).  F. Xu is with Peng Cheng Laboratory, Shenzhen (e-mail: xuf02@pcl.ac.cn). Q. Wu is with the Department of Electronic Engineering, Shanghai Jiao Tong University, 200240, China (e-mail: qingqingwu@sjtu.edu.cn).
		}}

\maketitle

\begin{abstract}

In this paper, we investigate a self-sensing intelligent reflecting surface (IRS) aided millimeter wave (mmWave) integrated sensing and communication (ISAC) system. Unlike the conventional purely passive IRS, the self-sensing IRS can effectively reduce the path loss of sensing-related links, thus rendering it advantageous in ISAC systems. Aiming to jointly sense the target/scatterer/user positions as well as estimate the sensing and communication (SAC) channels  in the considered system, 
%Aiming to enhance the location sensing and channel estimation (CE)  performance in the considered system, 
we propose a two-phase transmission scheme, where the coarse and refined sensing/channel estimation (CE) results are  respectively obtained in the first phase (using scanning-based IRS reflection coefficients) and second phase (using optimized IRS reflection coefficients). For each phase, an angle-based sensing turbo variational Bayesian inference (AS-TVBI) algorithm, which combines the VBI, messaging passing and expectation-maximization (EM) methods, is developed to solve the considered joint location sensing and CE problem. The proposed algorithm effectively exploits the partial overlapping structured (POS) sparsity and  2-dimensional (2D) block sparsity inherent in the SAC channels to enhance the overall performance. Based on the estimation results from the first phase, we formulate a  Cram\'{e}r-Rao bound (CRB) minimization problem for optimizing IRS reflection coefficients, and through proper reformulations, a low-complexity manifold-based optimization algorithm is proposed to solve this problem.  Simulation results are provided to verify the superiority of the proposed transmission scheme and associated algorithms.

\end{abstract}


\vspace{-0.2cm}
\section{Introduction}
\label{sec:Introduction}

\par Intelligent reflecting surface (IRS), due to its low hardware cost, high energy efficiency, and smart wireless environment configuration capability, is regarded as a promising technology  in communication systems to enhance network coverage and spectral efficiency, as well as in radar sensing to improve detection accuracy \cite{wuqq2021irstutorial,Lu2021radarirs,zhao2021two,meng2022multitarget}. Meanwhile, in response to the ever-increasing demands for enhanced spectrum efficiency, resource utilization and intelligent sensing capabilities in the next-generation wireless networks,  integrated sensing and communication (ISAC) technique that shares the same infrastructure and/or spectrum for achieving both  sensing and communication tasks simultaneously, has emerged as a key enabler \cite{liu2022surveyisac}. Recently, there has been a growing body of researches on IRS-aided ISAC systems, aiming at enhancing both the communication and sensing performance, as demonstrated in \cite{Huxl2022isacIRS,wang2023starIRS,song2022irsisac}. Particularly, to enable simultaneous data transmission and user positioning, a novel IRS-based ISAC framework was proposed in \cite{Huxl2022isacIRS}. In \cite{wang2023starIRS}, the concept of simultaneously transmitting and reflecting intelligent surface (STARS) was exploited to partition the space into distinct communication and sensing subspaces, and an efficient phase-shift optimization algorithm was devised to achieve high sensing accuracy under communication constraints. Besides, the work \cite{song2022irsisac} utilized the IRS to establish virtual line-of-sight (LoS) links, facilitating target sensing in scenarios with obstructed LoS links, and also to enhance the communication between the base station (BS) and user.

\par However, the aforementioned studies mainly focused on employing a purely passive IRS in the sensing process, which may cause severe path loss due to multiple signal reflections (e.g., BS$\to$IRS$\to$target$\to$IRS$\to$BS), thereby degrading the sensing performance. To address this issue, a novel self-sensing IRS was introduced in \cite{shaoxd2022selfsenIRS}, comprising an IRS controller, reflecting elements, and dedicated sensors. The IRS controller is deployed near the reflecting elements and transmits omnidirectional probing signals to illuminate the potential targets. Subsequently, the IRS sensors  receive two types of echo signals, i.e.,  the direct echo signal (IRS-controller$\to$target$\to$sensors) and the IRS-reflected echo signal (IRS controller$\to$reflecting elements$\to$target$\to$sensors). This innovative self-sensing IRS architecture can significantly reduce the path loss of the IRS-reflected link due to the short distance between the IRS controller and reflecting elements. Despite a thorough discussion on the self-sensing IRS-based direction-of-arrival (DoA) estimation for a single target in  \cite{shaoxd2022selfsenIRS}, the potential of the self-sensing IRS to sense the locations of multiple targets in the ISAC systems remains unexplored in the existing literature.

\par On the other hand, the prevailing works usually ignore the extensive structured sparsity inherent in  the sensing and communication (SAC) channels within IRS-aided ISAC systems.  To enhance both the SAC performance, in this paper,  we focus on two distinct forms of structured sparsity, namely the partial overlapping structured (POS) sparsity and the 2-dimensional (2D) block sparsity. The former sparsity follows from the observation that certain sensing targets may also  function  as  communication scatterers in practice \cite{Gaudio2020overlap,huang2022jpotdce,liufan2020joint}, which results in the partial overlapping of specific channel parameters within the SAC channels, e.g., the shared angle-of-arrivals (AoAs) of the common targets and scatterers. The latter sparsity characterizes the feature when employing 2D location grids for constructing the sparse representation of the SAC channels within IRS-aided millimeter wave (mmWave) ISAC systems,  which resembles the 1-dimentional (1D) cluster  sparsity of AoAs in mmWave massive multiple-input-multiple-output (MIMO) channels \cite{liuguanying2020tracking}.  For instance, when the targets or scatterers are large (e.g., substantial buildings or trucks) in the SAC channels, they may span across several adjacent grids within the predefined 2D location grids.

\par Motivated by the aforementioned considerations, in this paper, we investigate an IRS-aided mmWave ISAC system, where a self-sensing IRS is employed to reduce the path loss of sensing-related links and  the signals received at the BS and IRS sensors are processed cooperatively to achieve AoA based location sensing and SAC channel estimation (CE). The main contributions of this paper are summarized as follows.

\begin{itemize}
	\item A novel self-sensing IRS based joint sensing and CE (SI-JSCE) scheme is proposed to jointly sense the target/scatterer/user positions  and estimate both the SAC channels. Specifically, the proposed scheme comprises two phases. In phase \Rmnum{1}, the IRS controller and the user transmit $T_1$ sensing pilots and $T_2$ channel estimation (CE) pilots, respectively, during which the scanning-based IRS reflection coefficients are employed.  By processing the observations from the BS and IRS sensors cooperatively, the coarse target/scatterer/user positions and imprecise SAC channels can be jointly estimated. Subsequently, in phase \Rmnum{2}, the IRS reflection coefficients are optimized based on the estimation results from phase \Rmnum{1}, and then $T_3$ sensing pilots and $T_4$ CE pilots are further sent for JSCE.  In this phase, the performance of location sensing and CE can be improved using the enhanced observations yielded by the optimized IRS reflection coefficients.  
	\item  We construct a sparse representation of the SAC channels based on the 2D location grids within the space of interest (SOI), where the position offsets with respect to (w.r.t.) the 2D location grids  are introduced to achieve super-resolution location sensing.  Moreover, we  propose a hierarchical prior model to characterize the abovementioned POS sparsity and 2D block sparsity inherent in the SAC channels, as well as to facilitate the design of the joint location sensing and CE algorithm.
	\item An angle-based sensing turbo variational Bayesian inference (AS-TVBI) algorithm is proposed  by combining the VBI, messaging passing and expectation-maximization (EM) methods
	to obtain the marginal posteriors of the sparse SAC channels and the maximum likelihood (ML) estimates of the considered position offsets. Furthermore, a double direction gradient (DDG)-based method is proposed to replace the commonly used gradient ascent approach employed in updating the position offsets, which exhibits superior performance. 
	\item We derive the Cram\'{e}r-Rao bound (CRB) matrix for the position offsets of the estimated target/scatterer/user and formulate a CRB minimization problem to optimize the IRS reflection coefficients employed in the second phase. To lower the computational complexity, we propose to approximate the  CRB matrix and a manifold-based optimization algorithm is proposed  to efficiently solve the resulting  problem.
\end{itemize}

\par The rest of this paper is organized as follows. In Section \ref{sec:System Model}, we present the system model and the proposed SI-JSCE scheme. In Section \ref{sec: Hierarchical Prior Model}, we construct the sparse representation of the SAC channels and introduce the hierarchical prior model. Section \ref{sec:AS-TVBI Algorithm} proposes an AS-TVBI algorithm to jointly sense the target/scatterer/user positions and estimate the SAC channels. Section \ref{sec: CRLB optimization} optimizes the IRS reflection coefficients using the proposed manifold-based optimization algorithm.  Numerical results are provided in Section \ref{sec:simulation} and finally we conclude the paper in Section \ref{sec:conclusion}.

\par \textit{Notations:} Scalars, vectors and matrices are respectively denoted by lower/upper case, boldface lowercase and boldface uppercase letters. The  transpose, conjugate and conjugate transpose of a general vector $\mathbf{z}$ are denoted as $\mathbf{z}^T$, $\mathbf{z}^*$ and $\mathbf{z}^H$, respectively. $\text{diag}(\mathbf{z})$ yields a diagonal matrix with $\mathbf{z}$ as its diagonal elements.   For a general matrix $\mathbf{Z}$, $\text{vec} (\mathbf{Z})$ denotes its vectorization, $\text{tr} (\mathbf{Z})$ denotes its trace, and  $\text{diag} (\mathbf{Z})$ denotes a vector that contains the diagonal elements of $\mathbf{Z}$. For matrices $\mathbf{Z}_1 \!\! \in \! \mathbb{C}^{M_1 \!\times\! N}$ and $\mathbf{Z}_2 \!\!\in\! \mathbb{C}^{M_2 \!\times\! N}$, $[\mathbf{Z}_1;\!\mathbf{Z}_2] \!\in\! \mathbb{C}^{(M_1\!+\!M_2) \!\times\!  N}$ denotes the row-wise concatenation of $\mathbf{Z}_1$ and $\mathbf{Z}_2$.  $\text{blkdiag}(\mathbf{Z}_1,\mathbf{Z}_2)$  corresponds to the block diagonalization  with $\mathbf{Z}_1$ and $\mathbf{Z}_2$ as the diagonal blocks.  $\circ$ and $\otimes$  denote the Hadamard product and Kronecker product, respectively.

\section{System Model and Proposed Scheme}
\label{sec:System Model}
\subsection{System Model}
\par
\begin{figure}[tbp]
	\setlength{\abovecaptionskip}{-0.1cm}
	\centering
	\includegraphics[width=0.46\textwidth]{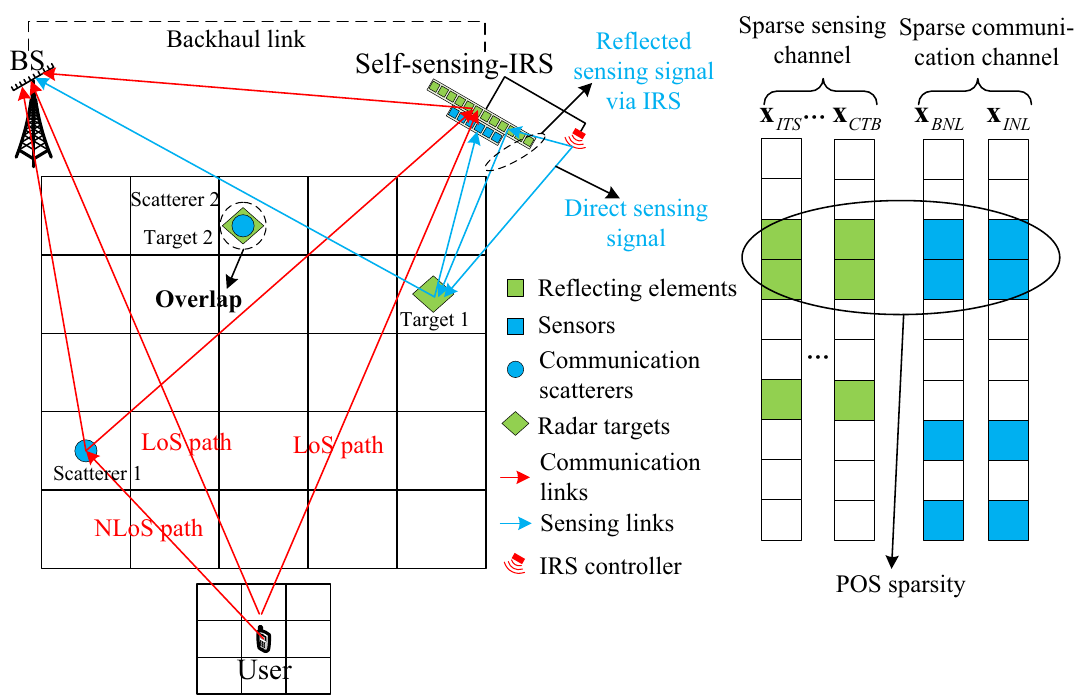}
	\caption{System model of the considered IRS-aided mmWave ISAC system. For clarity, only the communication links of scatterer 1 and sensing links of target 1 are plotted in the figure.}
	\label{pic:system model}	
	\vspace{-0.3cm}
\end{figure}

 As shown in Fig. \ref{pic:system model}, we consider a self-sensing IRS-aided mmWave ISAC system, where a self-sensing IRS composed of $N_p$ passive reflecting elements is deployed to sense the targets and in the meantime enhance the uplink communication from  a single-antenna user to a BS equipped with $M$ antennas.
The BS antennas and IRS reflecting elements are arranged as a uniform linear array (ULA) with the antenna/element interval being half of the carrier wavelength $\lambda$. A single-antenna IRS controller is deployed near the reflecting elements which is capable of transmitting omnidirectional probing signals for target sensing. Moreover, $N_s$ low-cost sensors are deployed in parallel with the reflecting elements on the self-sensing IRS to receive two kinds of target echo signals, i.e., the IRS-reflected echo signal via the IRS controller$
\rightarrow$reflecting elements$\rightarrow$target$\rightarrow$sensors link, and the direct echo signal via the IRS controller$
\rightarrow$target$\rightarrow$sensors link. Note that the BS can also receive these two kinds of echo signals via similar links. Suppose there are $K$ sensing targets and $L$ communication scatterers in a two-dimensional (2D) SOI $\mathcal{R}$. Similar to \cite{huang2022jpotdce}, we assume in this paper that some sensing targets also serve as communication scatterers (e.g., target 2 and scatterer 2 in Fig. \ref{pic:system model} represent the same physical object).  Besides, we further assume that the user
is situated within a smaller SOI denoted as $\mathcal{R}_u$ beneath $\mathcal{R}$ and the communication scatterers are shared between the BS-user and IRS-user channels.
The BS, the far-field reference point of reflecting elements and the IRS controller are located at known positions, denoted by  $\mathbf{p}_B = [p_B^x, p_B^y]^T$, $\mathbf{p}_I = [p_I^x, p_I^y]^T$ and $\mathbf{p}_c = [p_c^x, p_c^y]^T$, respectively. While the  positions of the user, the $k$-th ($\forall k \in \mathcal{K} \triangleq \{1, \cdots, K\}$) sensing target and the $l$-th ($\forall l \in \mathcal{L} \triangleq \{1, \cdots, L\}$) communication scatterer are unknown and denoted by  $\mathbf{p}_u = [p_u^x, p_u^y]^T$, $\mathbf{p}_{T,k} = [p_{T,k}^x, p_{T,k}^y]^T$ and  $\mathbf{p}_{S,l} = [p_{S,l}^x, p_{S,l}^y]^T$, respectively.

\vspace{-0.2cm}
\subsection{Transmission Protocol}
\par
\begin{figure}[htbp]
	\setlength{\abovecaptionskip}{-0.0cm}
	\centering
	\includegraphics[width=0.46\textwidth]{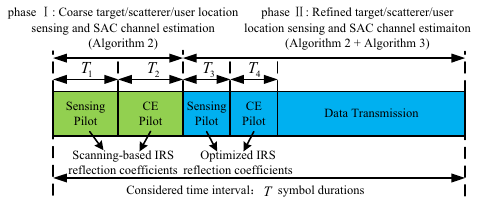}
	\caption{Transmission protocol of the proposed SI-JSCE scheme.}
	\label{pic:transmission protocol}	
	\vspace{-0.5cm}
\end{figure}

 In this paper, we propose to sense the target/scatterer/user positions as well as estimate the SAC channels jointly by exploiting the inherent cooperativity between the BS and the self-sensing IRS. To this end, our proposed scheme is based on a specific transmission protocol, as depicted in Fig. \ref{pic:transmission protocol}. Specifically, there are $T$ symbol durations in the considered time interval, within which the SAC channels are assumed to be unchanged. Moreover, the considered time interval is divided into two phases. In phase \Rmnum{1}, $T_1$ sensing pilots and $T_2$ CE pilots are sequentially transmitted by the IRS controller and the user, respectively. Based on the observations received at the BS and the IRS sensors, the coarse positions of target/scatterer/user and the imprecise SAC channels can be jointly estimated (the details will be shown in Section \ref{sec:AS-TVBI Algorithm}) at the end of phase \Rmnum{1} in the considered transmission protocol. Since the scattering environment is unknown during phase \Rmnum{1}, the IRS reflection coefficients are designed according to the unit-modulus hierarchical codebook \cite{xiao2016codebook} during phase \Rmnum{1} of this protocol such that the SOI $\mathcal{R}$ and  $\mathcal{R}_u$ can be fully scanned. At the beginning of phase \Rmnum{2}, the IRS reflection coefficients are optimized (as will be elaborated in Section \ref{sec: CRLB optimization}) based on the estimated target/scatterer/user positions and SAC channels from phase \Rmnum{1}. Then, the IRS controller and the user transmit $T_3$ sensing pilots and $T_4$ CE pilots, respectively. In this phase,  by leveraging the optimized IRS reflection coefficients, more accurate observations can be obtained, thereby enhancing the location sensing and CE performance. Finally, in the remaining duration of phase \Rmnum{2}, uplink data transmission is conducted between the BS and the user. In this paper, we assume that the time durations of the pilot transmission stages, i.e., $\{T_1, T_2, T_3, T_4\}$, are fixed. 
\vspace{-0.3cm}

\subsection{Sensing Signal Model}
\par To enable target sensing within the SOI $\mathcal{R}$, in the $t$-th symbol duration of phase $R$ ($R = \text{\Rmnum{1},\Rmnum{2}}$), the IRS controller transmits a sensing pilot $s^R(t)$,  and both the IRS sensors and the BS can receive the target echo signals through two distinct links: the direct link and the IRS-reflected link. First, we focus on  the received target echo signal at the IRS sensors and the associated sensing channels. Let $\mathbf{h}_{CI} \!=\! [h_{CI,1}, \cdots, h_{CI,N_p}]^T \!\in\! \mathbb{C}^{N_p \times 1}$ denote the channel vector from the IRS controller to the reflecting elements. Due to the short distance between the IRS controller and the reflecting elements, $\mathbf{h}_{CI}$ is characterized by the near-field channel model \cite{Davide2022nearfild}, 
i.e.,
\begin{equation}
	h_{C\!I\!,n} \!=\! \sqrt{\!\frac{\lambda^2}{16\pi^2 \|\mathbf{p}_{c} \!\!-\! \mathbf{p}_{I,n}\|^2}} e^{-j \frac{2\pi}{\lambda}\|\mathbf{p}_{c} - \mathbf{p}_{I,n}\!\|}, \quad \forall n \!\in\! \mathcal{N}_p,
\end{equation}
where $\mathbf{p}_{I,n}$ denotes the position of the $n$-th reflecting element and $\mathcal{N}_p  \!\triangleq\!\! \{1, \!\cdots\!,\! N_p\}$. Under the far-field assumption, the angle-of-departure (AoD) from the reflecting elements to the $k$-th target is identical to the AoA from the $k$-th target to the IRS sensors. Therefore, the $k$-th target echo channel associated with the IRS-reflected link (i.e., reflecting elements$\rightarrow$target $k$$\rightarrow$sensors), denoted by $\mathbf{H}_{IT_{k}S}$, can be modeled as
\begin{equation}
	\mathbf{H}_{IT_{k}S} = \alpha_{IT_kS} \mathbf{a}_s(\theta_{IT_k}) \mathbf{a}_I^H(\theta_{IT_k}),
\end{equation}
where $\alpha_{IT_kS} \!\!=\!\! \beta_{IT_kS}G_{IT_kS}$ denotes the complex path gain with $\beta_{IT_kS} \sim \mathcal{CN}(0,1)$ being the small-scale fading coefficient, $G_{IT_kS} \!=\!\! \sqrt{\frac{\lambda^2 \kappa_k}{64\pi^3 \|\mathbf{p}_I - \mathbf{p}_{t,k}\|^4}}$ being the large-scale path loss and  $\kappa_k$ being the the radar cross section (RCS)  of the $k$-th target. $\theta_{IT_k} \!\!=\! \arctan \Big( \frac{p_{T,k}^y \!-\! p_{I}^y}{p_{T,k}^x \!-\! p_{I}^x} \Big) \!-\! \pi \!-\! \vartheta_{I}$ represents the AoD from the reflecting elements to the $k$-th target with $\vartheta_{I}$ being the angle between the reflecting elements and the positive $x$-axis while the vectors  $\mathbf{a}_I(\theta) \!\!=\!\! [1, e^{-j\pi \cos(\theta)}, \!\cdots\!, e^{-j\pi(N_p-1) \cos(\theta)}]^T$ and $\mathbf{a}_s(\theta) \!\!=\!\! [1, e^{-j\pi \cos(\theta)}, \!\cdots\!, e^{-j\pi (N_s-1) \cos(\theta)}]^T$ denote the array response vectors of the reflecting elements and IRS sensors with an arbitrary AoD or AoA $\theta$, respectively. Similarly, the  $k$-th target echo channel associated with the direct link (i.e., IRS controller$\rightarrow$target $k$$\rightarrow$sensors), denoted by $\mathbf{h}_{CT_{k}S}$, is modeled as
\vspace{-0.1cm}
\begin{equation}
	\mathbf{h}_{CT_{k}S} = \alpha_{CT_kS} \mathbf{a}_s(\theta_{IT_k}),
\vspace{-0.1cm}
\end{equation}
where $\alpha_{CT_kS} = \beta_{CT_kS}G_{CT_kS}$ denotes the corresponding complex path gain with $\beta_{CT_kS} \sim \mathcal{CN}(0,1)$ and $G_{CT_kS} = \sqrt{\frac{\lambda^2 \kappa_k}{64\pi^3 \|\mathbf{p}_c - \mathbf{p}_{T,k}\|^2\|\mathbf{p}_I - \mathbf{p}_{T,k}\|^2}}$. By defining the IRS reflection vector as  $\boldsymbol{\varphi}_r^R(t) \!\triangleq\! [e^{j{\phi_1}}, \cdots, e^{j\phi_{N_p}}]^T \in \mathbb{C}^{N_p \times 1}$  with $\phi_n$, $\forall n \in \mathcal{N}_p$, being the phase shift of the $n$-th reflecting element and the IRS reflection matrix as $\boldsymbol{\Theta}_r^R(t) \triangleq \text{diag}(\boldsymbol{\varphi}_r^R(t))$, the received target echo signal at the IRS sensors (denoted by $\mathbf{y}_{s,r}^R(t)$) can be written as
\vspace{-0.1cm}
\begin{equation}
	\label{received echo signal: y_s1}
	\mathbf{y}_{s,r}^R(t) \!=\! \sum_{k=1}^K( \mathbf{H}_{IT_{k}S} \boldsymbol{\Theta}_r^R(t) \mathbf{h}_{CI}\! +\! \mathbf{h}_{CT_{k}S}) s^R(t) \!+\! \mathbf{n}_{s,r}^R(t),
\vspace{-0.1cm}
\end{equation}
where $\mathbf{n}_{s,r}^R(t) \!\sim\! \mathcal{CN}(0, \sigma^2\mathbf{I})$ is the additive white Gaussian noise (AWGN) with variance $\sigma^2$. 

\par Next, we consider the received target echo signal at the BS and the corresponding sensing channels. Similar to the abovementioned channel modeling, the $k$-th target echo channels associated with the direct (i.e., IRS controller$\rightarrow$target $k$$\rightarrow$BS) and IRS-reflected (i.e., reflecting elements$\rightarrow$target $k$$\rightarrow$BS) links are respectively given by
\vspace{-0.1cm}
\begin{align}
	\!\!\!\mathbf{h}_{C\!T_{k}\!B} \!\!=\!\!  \alpha_{C\!T_k\!B} \mathbf{a}_{B}(\theta_{BT_k}\!),  \,
	\mathbf{H}_{\!IT_{k}\!B} \!\!=\!\!  \alpha_{\!IT_k\!B} \mathbf{a}_{B}(\theta_{\!B\!T_k}\!)  \mathbf{a}_{I}^H\!(\theta_{\!IT_k}\!), \!
	\vspace{-0.1cm}
\end{align}
 where $\alpha_{CT_kB} \!=\!  \beta_{CT_kB}G_{CT_kB}$ and  $\alpha_{IT_kB} \!= \! \beta_{IT_kB}G_{IT_kB}$ denote the corresponding complex path gains with $\beta_{CT_kB} \!\sim\! \mathcal{CN}(0,1)$, $G_{CT_kB} \!=\! \sqrt{\frac{\lambda^2 \kappa_k}{64\pi^3 \|\mathbf{p}_c \!-\! \mathbf{p}_{T,k}\|^2\|\mathbf{p}_B - \mathbf{p}_{T,k}\|^2}}$,   $\beta_{IT_kB} \!\sim\! \mathcal{CN}(0,1)$  and $G_{IT_kB} = \sqrt{\frac{\lambda^2 \kappa_k}{64\pi^3 \|\mathbf{p}_I - \mathbf{p}_{T,k}\|^2\|\mathbf{p}_B - \mathbf{p}_{T,k}\|^2}}$. $\mathbf{a}_B(\theta) = [1, e^{-j\pi \cos(\theta)}, \cdots, e^{-j\pi(M-1) \cos(\theta)}]^T$ denotes the array response vector of the BS with an arbitrary AoA $\theta$, and  $\theta_{BT_k} = \arctan \Big( \frac{p_{T,k}^y - p_{B}^y}{p_{T,k}^x - p_{B}^x} \Big) + \vartheta_{B}$ represents the AoA from the $k$-th target to the BS with $\vartheta_{B}$ being the angle between the BS antennas and the positive $x$-axis. Moreover, by letting $\mathbf{H}_{IB} \in \mathbb{C}^{M \times N_p}$ and $\mathbf{h}_{CB} \in \mathbb{C}^{M \times 1}$ denote the line-of-sight (LoS) BS-IRS and controller-BS channels, respectively, the received target echo signal at the BS (denoted by $\mathbf{y}_{B,r}^R(t)$) is given by 
\begin{equation}
\begin{aligned}
	\label{received echo signal: y_B1}
	\mathbf{y}_{B,r}^R(t) = & \underbrace{\sum_{k=1}^K \left( \mathbf{H}_{IT_{k}B} \boldsymbol{\Theta}_r^R(t) \mathbf{h}_{CI} + \mathbf{h}_{CT_{k}B}\right) s^R(t)}_{\text{Desired signal}} \\
	 &+  \underbrace{\big( \mathbf{H}_{IB}\boldsymbol{\Theta}_r^R(t) \mathbf{h}_{CI} + \mathbf{h}_{CB} \big) s^R(t)}_{\text{ Interference}}  + \mathbf{n}_{B,r}^R(t),
\end{aligned}
\end{equation}
where $\mathbf{n}_{B,r}^R(t) \sim \mathcal{CN}(0, \sigma^2\mathbf{I})$ denotes the AWGN. Since the BS aims to sense the positions of $K$ targets, we refer to the first term and the second term in \eqref{received echo signal: y_B1} as its desired signal and interference, respectively. In this paper, we assume that  $\mathbf{H}_{IB}$, $\mathbf{h}_{CB}$ and $\mathbf{h}_{CI}$  remain unchanged over a much larger timescale than the considered time interval and are perfectly known \cite{shaoxd2022selfsenIRS}. Therefore, the interference term in \eqref{received echo signal: y_B1} can be canceled easily and we only focus on the desired signal in the sequel of this paper.

\subsection{Communication Signal Model}
\par  Let $\mathbf{h}_{IU} \in \mathbb{C}^{N_p \times 1}$ and $\mathbf{h}_{BU} \in \mathbb{C}^{M \times 1}$ denote the IRS-user and BS-user mmWave communication channels, respectively. Then, they can be modeled according to the Saleh-Valenzuela (SV) channel model \cite{akdeniz2014millimeter}, i.e.,
\begin{align}
	\mathbf{h}_{IU} \!= & \sqrt{\!\!\frac{1}{L\!+\!1}} \Big(\! \sum_{l=1}^{L}\! \alpha_{IU,l} \mathbf{a}_{I}(\theta_{IU,l}) \!+\! \alpha_{IU,0}\mathbf{a}_{I}(\theta_{IU,0}) \Big), \\
	\!\!\mathbf{h}_{BU} \!= & \sqrt{\!\!\frac{1}{L\!+\!1}} \Big(\! \sum_{l=1}^{L}\! \alpha_{BU,l} \mathbf{a}_{B}(\theta_{BU,l}) \!+\! \alpha_{BU,0}\mathbf{a}_{B}(\theta_{BU,0}) \Big),
\end{align}
where $\alpha_{IU,l} \sim \mathcal{CN}(0, \zeta_{IU,l}^2)$ and $\alpha_{BU,l}  \sim \mathcal{CN}(0, \xi_{BU,l}^2)$ denote the complex gains  associated with the $l$-th ($\forall l \in \mathcal{L} \triangleq \{0, \cdots, L\}$) path of $\mathbf{h}_{IU}$ and $\mathbf{h}_{BU}$, respectively, with $\zeta_{IU,l}^2$ and $\xi_{BU,l}^2$ being the large-scale path losses; $\theta_{IU,l}$ and $\theta_{BU,l}$ represent the AoAs associated with $l$-th ($\forall l \in \mathcal{L}$) path of $\mathbf{h}_{IU}$ and $\mathbf{h}_{BU}$, respectively. Since the far-field model is considered, the sensors-user channel $\mathbf{h}_{SU}$ can be modeled similar to $\mathbf{h}_{IU}$, i.e., $\mathbf{h}_{SU} = \sqrt{\frac{1}{L+1}} \Big( \sum_{l=1}^{L} \alpha_{IU,l} \mathbf{a}_{S}(\theta_{IU,l}) + \alpha_{IU,0}\mathbf{a}_{S}(\theta_{IU,0}) \Big)$. Based on the above channel modeling, in the $t$-th symbol duration of phase $R$, the received uplink signals at the IRS sensors and the BS (denoted by $\mathbf{y}_{s,c}^R(t)$ and $\mathbf{y}_{B,c}^R(t)$) are respectively given by
\begin{align}
	&\mathbf{y}_{s,c}^R(t) = \mathbf{h}_{SU} u^R(t) + n_{s,c}^R(t), 	\label{received uplink signal: y_s2} \\
	&\mathbf{y}_{B,c}^R(t) = \left( \mathbf{h}_{BU} + \mathbf{H}_{IB}\boldsymbol{\Theta}_c^R(t)\mathbf{h}_{IU}  \right) u^R(t) + n_{B,c}^R(t) \label{received uplink signal: y_B2},
\end{align}
where $u^R(t)$ denotes the CE pilot transmitted by the user; $\boldsymbol{\Theta}_c^R(t) = \text{diag}(\boldsymbol{\varphi}_c^R(t))$ denotes the IRS reflection matrix with $\boldsymbol{\varphi}_c^R(t)$ being the corresponding IRS reflection vector; $\mathbf{n}_{s,c}^R(t) \sim \mathcal{CN}(0, \sigma^2\mathbf{I})$ and $\mathbf{n}_{B,c}^R(t) \sim \mathcal{CN}(0, \sigma^2\mathbf{I})$ denote the AWGN at the IRS sensors and the BS, respectively.
\vspace{-0.2cm}

\section{Hierarchical Prior Model for Structured Sparsity}
\label{sec: Hierarchical Prior Model}
\par In this section, we begin by constructing the sparse representation of the SAC channels based on the 2D location grids of $\mathcal{R}$ and $\mathcal{R}_u$. Next, we propose a hierarchical prior model to characterize the POS sparsity between the sparse SAC channels, which facilitates the design of the proposed joint location  sensing and CE algorithm. Furthermore, to effectively capture the block-sparse structure of the sparse SAC channels, we introduce  a Markov random field (MRF) prior, which is particularly beneficial for enhancing the location sensing and CE performance  in low signal-to-noise ratio (SNR) regimes.
\vspace{-0.2cm}
\subsection{Sparse Representation of the SAC Channels}
\par To obtain a location-grid-based sparse channel representation, we uniformly pick   $Q \gg K + L$ and $P \gg 1$ grid positions from the 2D SOI $\mathcal{R}$ and $\mathcal{R}_u$, respectively, which are given by $\mathbf{r}=[\mathbf{r}_1; \cdots; \mathbf{r}_Q] \in \mathbb{R}^{Q \times 2}$ and $\mathbf{z}=[\mathbf{z}_1; \cdots; \mathbf{z}_P] \in \mathbb{R}^{P \times 2}$. 
%As there is often a mismatch  between the true target/scatterer/user  positions and the pre-defined grid positions in practice, we introduce two position offset vectors $\Delta\mathbf{r} = [\Delta \mathbf{r}_1; \cdots; \Delta \mathbf{r}_Q]$  and  $\Delta\mathbf{z} = [\Delta \mathbf{z}_1; \cdots; \Delta \mathbf{z}_P]$  for   $\mathbf{r}$ and  $\mathbf{z}$, respectively, to achieve super-resolution location sensing and improve the CE performance. 
Let $q_{T,k} = \argmin_{q} \| \mathbf{p}_{T,k} - \mathbf{r}_q\|$ and $q_{S,l} = \argmin_{q} \| \mathbf{p}_{S,l} - \mathbf{r}_q\|$ denote the indices of the pre-defined grid position $\mathbf{r}_q$ nearest to the $k$-th target and the $l$-th scatterer, respectively. Similarly, let $p_{u} = \argmin_{p} \| \mathbf{p}_u - \mathbf{z}_{p}\|$ denote the index of the pre-defined grid position $\mathbf{z}_p$ nrearest to the user. We assume that $Q$ and $P$ are sufficiently large so that the nearest grid position to each target/scatterer/user is distinct. Thus, for each grid position $\mathbf{r}_q$, $\forall q \in \mathcal{Q} \triangleq \{1, \cdots, Q\}$, we define its position offset to be the difference between its position and the position of the target/scatterer within its  corresponding grid, i.e., 
\begin{equation}
	\Delta \mathbf{r}_q \!=\! \left\{
	\begin{aligned}
	 & \mathbf{p}_{T,k} \!-\! \mathbf{r}_{q_{T,k}}, \quad \text{if} \; q = q_{T,k},  \forall k \in \mathcal{K}, \\ 
	 & \mathbf{p}_{S,l} \!-\! \mathbf{r}_{q_{S,l}}, \quad \text{if} \; q = q_{S,l},   \forall l \in \mathcal{L}, \\ 
	 & \mathbf{0}, \quad \text{if} \; q \!\neq\! q_{T,k} \; \text{and} \; q \neq q_{S,l},   \forall k \in \mathcal{K}, \forall l \in \mathcal{L}.
	\end{aligned}
	\right.
\end{equation}
Similarly, for each grid position $\mathbf{z}_p$, $\forall p \in \mathcal{P} \triangleq \{1, \cdots, P\}$, we define its position offset $\Delta \mathbf{z}_p$ to be the difference between its position and the position of the user  within  its corresponding grid, i.e., 
\begin{equation}
	\Delta \mathbf{z}_p = \left\{
	\begin{aligned}
		& \mathbf{p}_{u} - \mathbf{z}_{p_{u}}, \quad \text{if} \; p = p_{u},  \\ 
		& \mathbf{0}, \quad \text{if} \; p \neq p_{u}.
	\end{aligned}
	\right.
\end{equation}
We further define $\Delta\mathbf{r} = [\Delta \mathbf{r}_1; \cdots; \Delta \mathbf{r}_Q]$  and  $\Delta\mathbf{z} = [\Delta \mathbf{z}_1; \cdots; \Delta \mathbf{z}_P]$  to be the position offset vectors for grid positions $\mathbf{r}$ and $\mathbf{z}$, respectively. These position offset vectors will help us achieve super resolution location sensing and improve the CE performance in the rest of this paper.

\par Based on the above definitions, we can define the sensor-related sparse dictionaries w.r.t. the grid positions $\mathbf{r}$ and $\mathbf{z}$ as  $\mathbf{A}_{s}(\Delta\mathbf{r}) \!=\! [\mathbf{a}_s(\theta_I(\mathbf{r}_1 \!+\! \Delta \mathbf{r}_1)), \cdots,  \mathbf{a}_s(\theta_I(\mathbf{r}_Q \!+\! \Delta \mathbf{r}_Q))]$ and  $\mathbf{A}_{s}(\Delta\mathbf{z}) \!=\! [\mathbf{a}_s(\theta_I(\mathbf{z}_1 \!+\! \Delta \mathbf{z}_1)), \!\cdots\!,  \mathbf{a}_s(\theta_I(\mathbf{z}_P \!+\! \Delta \mathbf{z}_P))]$, respectively. Furthermore, by defining the reflecting elements-related sparse dictionary w.r.t. the grid positions $\mathbf{r}$  as  $\mathbf{A}_{I}(\Delta\mathbf{r}) \!=\! [\mathbf{a}_I(\theta_I(\mathbf{r}_1 \!+\! \Delta \mathbf{r}_1)), \!\cdots\!,  \mathbf{a}_I(\theta_I(\mathbf{r}_Q \!+\! \Delta \mathbf{r}_Q))]$, the sparse representation of the SAC channels in \eqref{received echo signal: y_s1} and \eqref{received uplink signal: y_s2} can be rewritten as 
%\begin{align}
%	\label{sparse representation of channels, IRS sensors}
%	\mathbf{H}_{ITS} = &\sum\nolimits_{k=1}^K\mathbf{H}_{IT_kS} = \mathbf{A}_{s}(\Delta \mathbf{r}) \text{diag}(\mathbf{x}_{ITS}) \mathbf{A}_I^H(\Delta \mathbf{r}), \\
%	&\mathbf{h}_{CTS} = \sum\nolimits_{k=1}^K\mathbf{h}_{IT_kS} = \mathbf{A}_{s}(\Delta \mathbf{r}) \mathbf{x}_{CTS}, \\
%	&\mathbf{h}_{SU} = \mathbf{A}_{s}(\Delta \mathbf{z})\mathbf{x}_{IL} +  \mathbf{A}_{s}(\Delta \mathbf{r}) \mathbf{x}_{INL},
%\end{align}
$\mathbf{H}_{\!IT\!S} \!=\! \sum\nolimits_{k=1}^K\mathbf{H}_{\!IT_k\!S}\!\! =\! \mathbf{A}_{s}(\Delta \mathbf{r}) \text{diag}(\mathbf{x}_{\!IT\!S}) \mathbf{A}_I^H(\Delta \mathbf{r})$, 
$\mathbf{h}_{C\!T\!S} \!=\! \sum\nolimits_{k=1}^K\mathbf{h}_{IT_kS}\! =\! \mathbf{A}_{s}(\Delta \mathbf{r}) \mathbf{x}_{\!C\!T\!S}$ and  
$\mathbf{h}_{SU} \!=\! \mathbf{A}_{s}(\Delta \mathbf{z})\mathbf{x}_{IL} \!+\!  \mathbf{A}_{s}(\Delta \mathbf{r}) \mathbf{x}_{INL}$, respectively,
where $\mathbf{x}_{\!IT\!S} \!\in\! \mathbb{C}^{Q \!\times\! 1}$, $\mathbf{x}_{\!C\!T\!S} \!\in\!\mathbb{C}^{Q \!\times\! 1}$, $\mathbf{x}_{\!I\!L} \!\in\! \mathbb{C}^{P \!\times\! 1}$ and $ \mathbf{x}_{\!I\!N\!L} \!\in\! \mathbb{C}^{Q \!\times\! 1}$ denote the corresponding sparse SAC channels. Similarly, we can obtain the sparse representation of the SAC channels in \eqref{received echo signal: y_B1} and \eqref{received uplink signal: y_B2} as follows:
%\begin{align}
%	\label{sparse representation of channels, BS}
%	\mathbf{H}_{ITB} = &\sum\nolimits_{k=1}^K\mathbf{H}_{IT_kB} = \mathbf{A}_{B}(\Delta \mathbf{r}) \text{diag}(\mathbf{x}_{ITB}) \mathbf{A}_I^H(\Delta \mathbf{r}), \\
%	&\mathbf{h}_{CTB} = \sum\nolimits_{k=1}^K\mathbf{h}_{IT_kB} = \mathbf{A}_{B}(\Delta \mathbf{r}) \mathbf{x}_{CTB}, \\
%	&\mathbf{h}_{BU} = \mathbf{A}_{B}(\Delta \mathbf{z})\mathbf{x}_{BL} +  \mathbf{A}_{B}(\Delta \mathbf{r}) \mathbf{x}_{BNL}, \\
%	&\mathbf{h}_{IU} = \mathbf{A}_{I}(\Delta \mathbf{z})\mathbf{x}_{IL} +  \mathbf{A}_{I}(\Delta \mathbf{r}) \mathbf{x}_{INL},
%\end{align}
$\mathbf{H}_{\!IT\!B} \!=\! \sum\nolimits_{k\!=\!1}^K\mathbf{H}_{\!IT_k\!B} \!=\! \mathbf{A}_{B}(\Delta \mathbf{r}) \text{diag}(\mathbf{x}_{\!IT\!B}) \mathbf{A}_I^H(\Delta \mathbf{r})$, 
$\mathbf{h}_{\!C\!T\!B} \!=\! \sum\nolimits_{k\!=\!1}^K\mathbf{h}_{\!IT_k\!B} \!=\! \mathbf{A}_{B}(\Delta \mathbf{r}) \mathbf{x}_{\!C\!T\!B}$, 
$\mathbf{h}_{\!B\!U} \!=\! \mathbf{A}_{B}(\Delta \mathbf{z})\mathbf{x}_{\!B\!L} \!+\!  \mathbf{A}_{B}(\Delta \mathbf{r}) \mathbf{x}_{B\!N\!L}$ and 
$\mathbf{h}_{\!I\!U} = \mathbf{A}_{\!I}(\Delta \mathbf{z})\mathbf{x}_{IL} \!+\!  \mathbf{A}_{\!I}(\Delta \mathbf{r}) \mathbf{x}_{INL}$,
where $\mathbf{A}_{\!B}(\Delta \mathbf{r})$, $\mathbf{A}_{\!B}(\Delta \mathbf{z})$ and $\mathbf{A}_{\!I}(\Delta \mathbf{z})$ are the associated sparse dictionaries; $\mathbf{x}_{\!IT\!B} \!\in\! \mathbb{C}^{Q \!\times\! 1}$, $\mathbf{x}_{\!C\!T\!B} \!\in\! \mathbb{C}^{Q \!\times\! 1}$, $\mathbf{x}_{\!B\!L} \!\in\!\mathbb{C}^{P \!\times\! 1}$ and $\mathbf{x}_{\!B\!N\!L} \!\in\! \mathbb{C}^{Q \!\times\! 1}$ denote the corresponding sparse SAC channels.

By leveraging the constructed sparse channel representation and  stacking the $T_1$ received echo signals  in phase \Rmnum{1}, the overall observation vectors at the IRS sensors and the BS can be respectively expressed as
\vspace{-0.2cm}
\begin{align}
	\mathbf{y}_{\!s,r}^{\text{\Rmnum{1}}} \!\!=\! \mathbf{F}_{\!s,r}^{\text{\Rmnum{1}}}(\Delta \mathbf{r}) \mathbf{x}_{TS} \!+\! \mathbf{n}_{\!s,r}^{\text{\Rmnum{1}}}, \;
	\mathbf{y}_{\!B,r}^{\text{\Rmnum{1}}} \!\!=\! \mathbf{F}_{\!B,r}^{\text{\Rmnum{1}}}(\Delta \mathbf{r}) \mathbf{x}_{TB} \!+\! \mathbf{n}_{\!B,r}^{\text{\Rmnum{1}}},
\end{align}
where $\mathbf{F}_{s,r}^{\text{\Rmnum{1}}}(\Delta \mathbf{r}) = [ (\tilde{\boldsymbol{\Phi}}_r^{\text{\Rmnum{1}}}) ^T \big(\mathbf{A}_I^*(\Delta \mathbf{r})  \odot \mathbf{A}_{s}(\Delta \mathbf{r}) \big), \mathbf{1}_{T_1} \otimes \mathbf{A}_{s}(\Delta \mathbf{r})]$ and $\mathbf{F}_{B,r}^{\text{\Rmnum{1}}}(\Delta \mathbf{r}) = [(\tilde{\boldsymbol{\Phi}}_r^{\text{\Rmnum{1}}}) ^T \big(\mathbf{A}_I^*(\Delta \mathbf{r})  \odot \mathbf{A}_{B\!}(\Delta \!\mathbf{r}) \big), \!\mathbf{1}_{\!T_1} \otimes \mathbf{A}_{\!B}(\!\Delta \mathbf{r})]$ denote the sensing measurement matrices with $\tilde{\boldsymbol{\Phi}}_r^{\text{\Rmnum{1}}} = [\tilde{\boldsymbol{\varphi}}_r^{\text{\Rmnum{1}}}(1),\!\cdots\!,\tilde{\boldsymbol{\varphi}}_r^{\text{\Rmnum{1}}}(T_1)]  = \text{diag}(\mathbf{h}_{CI}) \boldsymbol{\Phi}_r^{\text{\Rmnum{1}}}$ and $\boldsymbol{\Phi}_r^{\text{\Rmnum{1}}} = [\boldsymbol{\varphi}_r^{\text{\Rmnum{1}}}(1), \cdots, \boldsymbol{\varphi}_r^{\text{\Rmnum{1}}}(T_1)]$ respectively; $\mathbf{x}_{TS} = [\mathbf{x}_{ITS};\mathbf{x}_{CTS}]$ and $\mathbf{x}_{TB} = [\mathbf{x}_{ITB};\mathbf{x}_{CTB}]$ denote the aggregated sparse sensing channels associated with the IRS sensors and the BS, respectively; $\mathbf{n}_{s,r}^{\text{\Rmnum{1}}}$ and $\mathbf{n}_{B,r}^{\text{\Rmnum{1}}}$ are the  aggregated noise vectors. Similarly, by stacking the $T_2$ received uplink signals at the IRS sensors and the BS, we can obtain
\begin{equation}
	\mathbf{y}_{c}^{\text{\Rmnum{1}}}  = \mathbf{F}_{c}^{\text{\Rmnum{1}}}(\Delta \mathbf{r},\Delta \mathbf{z}) \mathbf{x}_{c} + \mathbf{n}_{c}^{\text{\Rmnum{1}}},
\end{equation}
where $\mathbf{F}_{c}^{\text{\Rmnum{1}}}(\Delta \mathbf{r},\Delta \mathbf{z}) = [\mathbf{F}_{s,c}^{\text{\Rmnum{1}}}(\Delta \mathbf{r},\Delta \mathbf{z});\mathbf{F}_{B,c}^{\text{\Rmnum{1}}}(\Delta \mathbf{r},\Delta \mathbf{z})]$ denotes the communication measurement matrix with $\mathbf{F}_{s,c}^{\text{\Rmnum{1}}}(\Delta \mathbf{r},\Delta \mathbf{z}) = [\mathbf{0}, \mathbf{1}_{T_2} \otimes  \mathbf{A}_{s}(\Delta \mathbf{r}), \mathbf{0}, \mathbf{1}_{T_2} \otimes  \mathbf{A}_{s}(\Delta \mathbf{z})]$, $\mathbf{F}_{B,c}^{\text{\Rmnum{1}}}(\Delta \mathbf{r},\Delta \mathbf{z}) = [\mathbf{1}_{T_2} \otimes  \mathbf{A}_{B}(\Delta \mathbf{r}),  \big((\boldsymbol{\Phi}_c^{\text{\Rmnum{1}}})^T \odot \mathbf{H}_{IB}\big)  \mathbf{A}_I(\Delta \mathbf{r}),	\mathbf{1}_{T_2} \otimes  \mathbf{A}_{B}(\Delta \mathbf{z}), \big( (\boldsymbol{\Phi}_c^{\text{\Rmnum{1}}})^T \odot \mathbf{H}_{IB} \big)  \mathbf{A}_I(\Delta \mathbf{z})]$ 
and $\boldsymbol{\Phi}_c^{\text{\Rmnum{1}}} = [\boldsymbol{\varphi}^{\text{\Rmnum{1}}}_c(
 1), \cdots,   \boldsymbol{\varphi}^{\text{\Rmnum{1}}}_c(T_2)]$; $\mathbf{x}_{c} = [ \mathbf{x}_{BNL};\mathbf{x}_{INL};\mathbf{x}_{BL};\mathbf{x}_{IL}]$ denotes the aggregated sparse communication channel  and $\mathbf{n}_{c}^{\text{\Rmnum{1}}} $ is the  aggregated noise vector. As such, the overall observation vector (denoted by $\mathbf{y}^{\text{\Rmnum{1}}}$) in phase \Rmnum{1} can be expressed as
\begin{equation}
	\label{overall oberv vec}
	\mathbf{y}^{\text{\Rmnum{1}}} = \mathbf{F}^{\text{\Rmnum{1}}}(\Delta \mathbf{r},\Delta \mathbf{z}) \mathbf{x} + \mathbf{n}^{\text{\Rmnum{1}}},
\end{equation}
where $\mathbf{y}^{\text{\Rmnum{1}}} \!=\! [\mathbf{y}_{s,r}^{\text{\Rmnum{1}}};\mathbf{y}_{B,r}^{\text{\Rmnum{1}}};\mathbf{y}_{c}^{\text{\Rmnum{1}}}]$, $\mathbf{F}^{\text{\Rmnum{1}}}(\Delta \mathbf{r},\Delta \mathbf{z}) = \text{blkdiag}\big(\mathbf{F}_{s,r}^{\text{\Rmnum{1}}}(\Delta \mathbf{r}), \\ \mathbf{F}_{B,r}^{\text{\Rmnum{1}}}(\Delta \mathbf{r}),\mathbf{F}_{c}^{\text{\Rmnum{1}}}(\Delta \mathbf{r},\Delta \mathbf{z})\big)$, $\mathbf{x} = [\mathbf{x}_{TS};  \mathbf{x}_{TB};\mathbf{x}_{c}]$ and $\mathbf{n}_{c}^{\text{\Rmnum{1}}} = [\mathbf{n}_{s,r}^{\text{\Rmnum{1}}};\mathbf{n}_{B,r}^{\text{\Rmnum{1}}};\mathbf{n}_{c}^{\text{\Rmnum{1}}}]$. In this work, for simplicity, the sensing pilot $s^R(t)$ and CE pilot $u^R(t)$ are simply set to $s^R(t) = 1$ and $u^R(t) = 1$, $\forall R$, $\forall t$. Following a similar derivation, we can readily  obtain the overall observation vector $\mathbf{y}^{\text{\Rmnum{2}}}$ and measurement matrix $\mathbf{F}^{\text{\Rmnum{2}}}(\Delta \mathbf{r},\Delta \mathbf{z})$ in phase \Rmnum{2} by replacing the number of  pilots $T_1$, $T_2$ and the stacked IRS reflection vectors $\boldsymbol{\Phi}^{\text{\Rmnum{1}}}_r$, $\boldsymbol{\Phi}^{\text{\Rmnum{1}}}_c$ by $T_3$, $T_4$, $\boldsymbol{\Phi}^{\text{\Rmnum{2}}}_r = [\boldsymbol{\varphi}^{\text{\Rmnum{2}}}_r(
1), \cdots, \boldsymbol{\varphi}^{\text{\Rmnum{2}}}_r(T_3)]$ and $\boldsymbol{\Phi}^{\text{\Rmnum{2}}}_c =  [\boldsymbol{\varphi}^{\text{\Rmnum{2}}}_c(
1), \cdots, \boldsymbol{\varphi}^{\text{\Rmnum{2}}}_c(T_4)]$, respectively.
The details are omitted here for brevity. Note that  the number of  available observations in phase \Rmnum{2} (including $\mathbf{y}^{\text{\Rmnum{1}}}$ and $\mathbf{y}^{\text{\Rmnum{2}}}$ ) is larger than that in phase \Rmnum{1} (i.e., $\mathbf{y}^{\text{\Rmnum{1}}}$) when performing joint location sensing and CE.  Besides, since the same joint location sensing and CE algorithm is applied in both phases,   we drop the superscripts \Rmnum{1} and \Rmnum{2}, and simply use  $\mathbf{y}$ and $\mathbf{F}(\Delta \mathbf{r},\Delta \mathbf{z})$ in the following. 
%Specifically, in phase \Rmnum{1}, we have $\mathbf{y} = \mathbf{y}^{\text{\Rmnum{1}}}$ and $\mathbf{F}(\Delta \mathbf{r}, \Delta\mathbf{z}) = \mathbf{F}^{\text{\Rmnum{1}}}(\Delta \mathbf{r}, \Delta\mathbf{z})$. While in phase \Rmnum{2}, we have $\mathbf{y} = [\mathbf{y}^{\text{\Rmnum{1}}};\mathbf{y}^{\text{\Rmnum{2}}}]$ and $\mathbf{F}(\Delta \mathbf{r}, \Delta\mathbf{z}) = [\mathbf{F}^{\text{\Rmnum{1}}}(\Delta \mathbf{r}, \Delta\mathbf{z});\mathbf{F}^{\text{\Rmnum{2}}}(\Delta \mathbf{r}, \Delta\mathbf{z})]$.
\vspace{-0.1cm}

\subsection{Probability Model for Structured Sparsity }
\label{3-B}
To characterize the POS sparsity between the SAC channels (as illustrated in Fig. \ref{pic:system model}), we propose a hierarchical prior model for sparse Bayesian inference. In particular, let $\mathbf{s}_T \triangleq [s_{T,1}, \cdots,  s_{T,Q}]^T$, $\mathbf{s}_{NL} \triangleq [s_{NL,1}, \cdots, s_{NL,Q}]^T$ and $\mathbf{s}_{L} \triangleq [s_{L,1}, \cdots, s_{L,P}]^T$ denote the support vectors of the sparse SAC channels. Therein,  $s_{T,q} = 1$ ($s_{NL,q} = 1$) indicates the presence of a target (scatterer) in the $q$-th grid of $\mathbf{r}$; otherwise, $s_{T,q} = -1$ ($s_{NL,q} = -1$). Similarly,  $s_{L,p} = 1$ signifies the user's location at the $p$-th grid of $\mathbf{z}$; otherwise, $s_{L,p} = -1$. Moreover, let $\boldsymbol{\rho} \triangleq [\boldsymbol{\rho}_{\text{ITS}};\boldsymbol{\rho}_{\text{CTS}};\boldsymbol{\rho}_{\text{ITB}};\boldsymbol{\rho}_{\text{CTB}};\boldsymbol{\rho}_{\text{BNL}};\boldsymbol{\rho}_{\text{INL}};\boldsymbol{\rho}_{\text{BL}};\boldsymbol{\rho}_{\text{IL}}] \in \mathbb{R}^{ (6Q+2P) \times1}$ denote the precision vector of $\mathbf{x}$, then we model  $\mathbf{x}$  as a Gaussian prior distribution, i.e., 
\vspace{-0.1cm}
\begin{equation}
	\label{prior model, x}
	p(\mathbf{x}|\boldsymbol{\rho}) = \prod_{j \in \mathcal{J}_1 \cup \mathcal{J}_2 \cup \mathcal{J}_3} \mathcal{CN}( \mathbf{x}_j; \mathbf{0}, \text{diag}^{-1}(\boldsymbol{\rho}_j)),
	\vspace{-0.1cm}
\end{equation}
where $\mathcal{J}_1 \triangleq \{\text{ITS, CTS, ITB, CTB}\}$, $\mathcal{J}_2 \triangleq \{\text{BNL, INL}\}$ and $\mathcal{J}_3 \triangleq \{\text{BL, IL}\}$. Conditioned on the aggregated  support vector $\mathbf{s} = [\mathbf{s}_T;\mathbf{s}_{NL};\mathbf{s}_L]$, the prior model for $\boldsymbol{\rho}$ is given by 
\begin{equation}
	\label{prior model, rho}
	p(\boldsymbol{\rho}|\mathbf{s}) \!=\!\!\! \prod_{j_1 \!\in\! \mathcal{J}_1}  \prod_{j_2 \!\in\! \mathcal{J}_2}  \prod_{j_3 \!\in\! \mathcal{J}_3}\! p(\boldsymbol{\rho}_{j_1}|\mathbf{s}_T)p(\boldsymbol{\rho}_{j_2}|\mathbf{s}_{NL})p(\boldsymbol{\rho}_{j_3}|\mathbf{s}_L) ,
	\vspace{-0.1cm}
\end{equation}
where $p(\boldsymbol{\rho}_{j_1}|\mathbf{s}_T) \!=\! \prod_{q=1}^Q\! \big( \Gamma(\rho_{j_1,q};a_{j_1,q},b_{j_1,q})\big)^{\frac{1+s_{T,q}}{2}} \big(\Gamma(\rho_{j_1,q}; \\ \bar{a}_{j_1,q},  \bar{b}_{j_1,q})\big)^{ \frac{1-s_{T,q}}{2}}$ with  $\Gamma(\rho;a,b)$ denoting a Gamma distribution with shape parameter $a$ and rate parameter $b$. $p(\boldsymbol{\rho}_{j_2}|\mathbf{s}_{NL})$ and $p(\boldsymbol{\rho}_{j_3}|\mathbf{s}_L)$ are defined similarly as $p(\boldsymbol{\rho}_{j_1}|\mathbf{s}_T)$ but with different  hyperparameters. It is worth noting that in order to control the amplitude of $x_{j_1,q}$ based on $s_{T,q}$, the shape and rate parameters $a_{j_1,q}$ and $b_{j_1,q}$ should be chosen such that $a_{j_1,q}/b_{j_1,q} = \mathbb{E}[\rho_{j_1,q}] \approx \mathcal{O}(1/|\alpha_{j_1,q}|^2)$ with $\alpha_{j_1,q}$ denoting the corresponding path gain, when $s_{T,q} = 1$. Conversely, $\bar{a}_{j_1,q}$ and $\bar{b}_{j_1,q}$ are chosen to satisfy $\bar{a}_{j_1,q}/\bar{b}_{j_1,q} = \mathbb{E}[\rho_{j_1,q}] \gg 1$ for the case of   $s_{T,q} = -1$.
 
Let $\mathbf{s}_U \triangleq [s_{U,1}, \cdots, s_{U,Q}]^T$ denote a union support vector, where $s_{U,q} = 1$ if either $s_{T,q} = 1$ or $s_{NL,q} = 1$; otherwise, $s_{U,q} = -1$, $\forall q \in \mathcal{Q}$. Then, the joint distribution of all the support vectors  can be obtained as
$p(\mathbf{s}_T,\mathbf{s}_{NL},\mathbf{s}_{L},\mathbf{s}_{U}) = p(\mathbf{s}_L)p(\mathbf{s}_U)p(\mathbf{s}_T | \mathbf{s}_U)p(\mathbf{s}_{NL} | \mathbf{s}_U)$. In order to characterize the POS sparsity between the sparse SAC channels, the conditional priors 
$p(\mathbf{s}_T | \mathbf{s}_U)$ and $p(\mathbf{s}_{NL} | \mathbf{s}_U)$ are
given by 
\vspace{-0.2cm}
\begin{align}
	\label{prior model, pT_U, pNL_U}
	&p(\!\mathbf{s}_{\!T} | \mathbf{s}_{U}\!)  \!\!=\!\!\! \prod_{q=1}^Q \!\! \Big[\! \frac{1 \!\!- \!\!s_{U\!\!,q}}{2}\delta(\!s_{T\!,q} \!\!+\!\! 1\!) \!\!+\!\! \frac{1\!\!+\!\!  s_{U\!\!,q}}{2} p_{\!T}^{\!\!\frac{1\!+\!s_{T\!,q}}{2}}\!(\!1\!\!-\!\!p_T\!)^{\!\!\!\frac{1\!-\!s_{T\!,q}}{2}}\!\Big],  \\
	&\!\!\!p(\!\mathbf{s}_{\!N\!\!L} \!| \mathbf{s}_{U}\!) \!\! =\!\!\! \prod_{q\!=\!1}^Q \!\!\Big[\! \frac{1\!\!\!-\!\!s_{U\!\!,q}}{2}\!\delta(\!s_{\!N\!\!L\!,q} \!\!+\!\! 1\!) \!\!+\!\! \frac{1\!\!\!+\!\!s_{U\!\!,q}}{2}  p_{\!N\!\!L}^{\frac{1\!+\!s_{\!N\!\!L\!,q}}{2}}\!\!(\!1\!\!-\!\!p_{\!N\!\!L}\!)^{\!\frac{1\!-\!s_{\!N\!\!L\!,q}}{2}}\!\Big]\!,\!
	\vspace{-0.2cm}
\end{align}
where $p_T \triangleq \frac{K}{K+L-O}$ denotes the probability of $s_{T,q}=1$ conditioned on  $s_{U,q}=1$ and  $p_{NL} \triangleq \frac{L}{K+L-O}$ denotes the probability of $s_{NL,q}=1$ conditioned on $s_{U,q}=1$, $\forall q \in \mathcal{Q}$, with $O$ being the number of  overlapping targets and scatterers. Note that larger values of $p_T$ and $p_{NL} $ indicate a higher degree of overlap between the targets and scatterers.

\par
\begin{figure}[tbp]
	\setlength{\abovecaptionskip}{-0.0cm}
	\centering
	\includegraphics[width=0.45\textwidth]{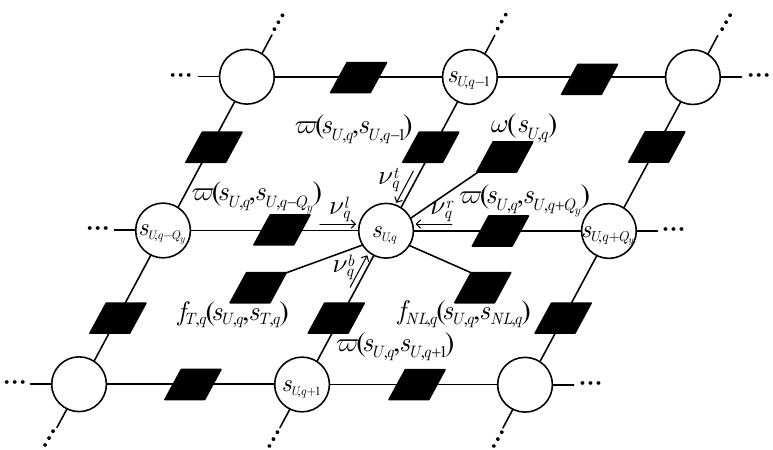}
	\caption{Factor graph of the 4-connected MRF. For clarity, only the factor nodes associated with the central variable node $s_{U,q}$ are plotted in the figure.}
	\label{pic:facgph_MRF}	
	\vspace{-0.5cm}
\end{figure}

 On the other hand, to effectively capture the  2D block sparsity of the sparse SAC channels $\mathbf{x}$, we employ the MRF prior to characterize the union support vector $\mathbf{s}_U$, which can be modeled by the Ising model\cite{som2011approximate} as
\vspace{-0.1cm}
\begin{equation}
	\label{Ising model}
	p(\mathbf{s}_U) = \frac{1}{C} \prod_{q=1}^Q \Big[ \omega(s_{U,q}) \prod_{q^\prime \in \mathcal{D}(q)}  \varpi^{\frac{1}{2}} (s_{U,q},s_{U,q^\prime}) \Big],
\end{equation}
where $C$ is a normalization constant, $\mathcal{D}(q) \subset \mathcal{Q}$ represents the set of all neighboring indices of $q$, $\omega(s_{U,q}) = e^{-\alpha s_{U,q}}$ and $\varpi (s_{U,q},s_{U,q^\prime}) = e^{\beta s_{U,q}, s_{U,q^\prime}}$ with $\alpha$ and $\beta$ being the bias and interaction parameters of the MRF. To illustrate the internal structure of the MRF  intuitively, a factor graph of the 4-connected MRF\footnote{In other sensing/CE tasks, the sparse signal $\mathbf{x}$ may exhibit 1D or 3-dimensional (3D) block sparsity \cite{kuai2019struct,liuguanying2020tracking}, which can be characterized using 2-connected or 8-connected MRF, respectively.} is depicted in Fig. \ref{pic:facgph_MRF}, which is able to characterize  the 2D block sparsity of the location-grid-based sparse SAC channels. By adjusting $\alpha$ and  $\beta$, the degree of sparsity and the block size of $\mathbf{x}$ can be effectively controlled. To be specific,  a larger value of $\alpha$ promotes a sparser $\mathbf{x}$, while a larger value of $\beta$ indicates a larger size of each non-zero block. Besides, assuming that there is no  prior information about the user location,  $p(\mathbf{s}_L)$ can be  modeled by an independent and identically distributed (i.i.d.)  Bernoulli distribution as 
\vspace{-0.1cm}
\begin{equation}
	\label{prior model, p_sL}
	p(\mathbf{s}_{L}) =  \prod_{p=1}^P p_{L}^{ \frac{1+s_{L,p}}{2}}(1-p_{L})^{ \frac{1-s_{L,p}}{2}}, 
\end{equation}
where  $p_{L} = \frac{1}{P}$ denotes the sparsity level of $\mathbf{x}_{IL}$ and $\mathbf{x}_{BL}$. 
\par Based on the above hierarchical prior model, the joint distribution of all the considered random variables can be written as
\vspace{-0.2cm}
\begin{align}
	\label{joint distribution of all}
	&\!\!\! p(\mathbf{y}\!,\!\mathbf{x},\! \boldsymbol{\rho},\!\mathbf{s}_{\!T}\!,\mathbf{s}_{\!N\!\! L}\!,\!\mathbf{s}_{\!L}\!,\!\mathbf{s}_U\!; \!\Delta\! \mathbf{r}\! , \!\Delta\!\mathbf{z}) \!\!=\!\! p( \mathbf{y} \!| \mathbf{x};  \!\Delta \!\mathbf{r}\!, \!\Delta\!\mathbf{z})\!\! \! 
	\prod_{j_{\!1} \!\in\! \mathcal{J}_{\!1}}\!\! \!  p(\mathbf{x}_{j_{\!1}} \! |  \boldsymbol{\rho}_{j_{\! 1}}\! )  p(\boldsymbol{\rho}_{j_{\! 1}}\! |\mathbf{s}_{\! T}\! ) \nonumber\\ 
	& \!\!\!  \times\!\! \!\!    \prod_{j_{\!2} \!\in\! \mathcal{J}_{\!2}}\!\! p(\!\mathbf{x}_{\!j_{\!2}}\!|\boldsymbol{\rho}_{\!j_{\!2}}\!)  p(\!\boldsymbol{\rho}_{\!j_{\!2}}|\mathbf{s}_{\!N\!\!L}\!)\!\!\!
	\prod_{j_{\!3} \!\in\! \mathcal{J}_{\!3}}\!\!
	p(\!\mathbf{x}_{j_{\!3}}\!|\boldsymbol{\rho}_{j_{\!3}}\!)   p(\!\boldsymbol{\rho}_{j_{\!3}}\!|\mathbf{s}_{\!L}\!) p(\!\mathbf{s}_{\!T}\!,\!\mathbf{s}_{\!N\!\!L}\!,\!\mathbf{s}_{\!L}\!,\!\mathbf{s}_{\!U}\!),
\end{align}
where $p(\mathbf{y} | \mathbf{x};  \Delta \mathbf{r}, \Delta \mathbf{z}) = \mathcal{CN}\big(\mathbf{F}(\Delta \mathbf{r}, \Delta\mathbf{z})\mathbf{x}, \sigma^2\mathbf{I}\big)$.  With the given observation vector $\mathbf{y}$ and position offset $\{\Delta \mathbf{r}, \Delta \mathbf{z}\}$,  our main objective is to calculate the accurate marginal posteriors $p(\mathbf{x}|\mathbf{y};\Delta \mathbf{r}, \Delta \mathbf{z})$,  $p(\boldsymbol{\rho}|\mathbf{y};\Delta \mathbf{r}, \Delta \mathbf{z})$ and  $p(\mathbf{s}|\mathbf{y};\Delta \mathbf{r}, \Delta \mathbf{z})$ by performing Bayesian inference for $\mathbf{x}$, $\boldsymbol{\rho}$ and $\mathbf{s}$, respectively. Furthermore, the optimal position offset $\{\Delta \mathbf{r}^*, \Delta\mathbf{z}^*\}$ is determined by solving the following ML problem:
\vspace{-0.1cm} 
\begin{equation}
	\label{ML problem in M step}
	 \{\Delta \mathbf{r}^*, \Delta\mathbf{z}^*\} = \argmax_{\Delta \mathbf{r}, \Delta \mathbf{z}} \ln p(\mathbf{y}|\Delta \mathbf{r}, \Delta\mathbf{z}).
	\vspace{-0.1cm} 
\end{equation}
However, achieving these two objectives poses significant challenges due to the following two reasons. First, obtaining the accurate marginal posteriors w.r.t. $\mathbf{x}$, $\boldsymbol{\rho}$ and $\mathbf{s}$  is difficult due to the
presence of loops in the factor graph of the joint distribution \eqref{joint distribution of all}. Second, deriving the likelihood function $\ln p(\mathbf{y}|\Delta \mathbf{r}, \Delta\mathbf{z})$ in closed form is difficult  as it requires to calculate  an intractable multidimensional integration over all the hidden variables $\mathbf{x}$, $\boldsymbol{\rho}$, $\mathbf{s}$, $\mathbf{s}_U$.
To tackle these challenges, we propose in the subsequent section  the AS-TVBI algorithm,  which combines the VBI, message passing and EM methods to provide high-performance approximate solutions for the desired marginal posteriors and ML estimates.

\vspace{-0.0cm}

\section{AS-TVBI Algorithm}
\label{sec:AS-TVBI Algorithm}

\par
\begin{figure}[thbp]
	\setlength{\abovecaptionskip}{-0.0cm}
	\centering
	\includegraphics[width=0.48\textwidth]{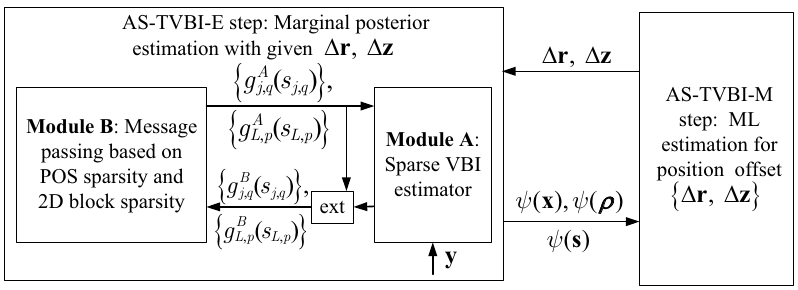}
	\caption{Top-level diagram of the proposed  AS-TVBI algorithm.}
	\label{pic:AS-TVBI Algorithm}	
	\vspace{-0.4cm}
\end{figure}

 The proposed AS-TVBI algorithm is primarily based on the EM framework \cite{Liu2003EM} and  consists of two major steps: the E step and the M step, as illustrated in Fig. \ref{pic:AS-TVBI Algorithm}. In the E step, for given  observation vector $\mathbf{y}$ and position offset $\{\Delta \mathbf{r},\Delta \mathbf{z}\}$, the accurate marginal posteriors $p(\mathbf{x}|\mathbf{y};\Delta \mathbf{r},\Delta \mathbf{z})$, $p(\boldsymbol{\rho}|\mathbf{y};\Delta \mathbf{r},\Delta \mathbf{z})$ and $p(\mathbf{s}|\mathbf{y};\Delta \mathbf{r},\Delta \mathbf{z})$ are approximated using tractable variational distributions $\psi(\mathbf{x})$, $\psi(\boldsymbol{\rho})$ and $\psi(\mathbf{s})$ which can be obtained by combing the VBI and message passing methods within the turbo framework. While in the M step,    we construct a surrogate function for $\ln p(\mathbf{y}|\Delta \mathbf{r},\Delta \mathbf{z})$ by exploiting the approximate marginal posteriors derived from the E step,  and present a DDG-based method to find an approximate solution of problem \eqref{ML problem in M step}. These two steps are iteratively executed until convergence is achieved.  In the following, we illustrate the E step and the M step for the AS-TVBI algorithm in detail. 
\vspace{-0.2cm}

\subsection{AS-TVBI-E Step}
\label{sec:AS-TVBI-E step}
\par Due to the numerous loops in the factor graph of the joint distribution \eqref{joint distribution of all}, computing the accurate marginal posteriors w.r.t. $\mathbf{x}$, $\boldsymbol{\rho}$ and $\mathbf{s}$ is an NP-hard problem \cite{dagum1993approximating}. To address this issue, we propose to compute the marginal posteriors approximately by partitioning the original factor graph into two sub-graphs corresponding to Module A and Module B as shown in Fig. \ref{pic:AS-TVBI Algorithm}. In particular,
based on the observation vector $\mathbf{y}$ and the output messages from Module B, denoted by $\{g^A_{j,q}(s_{j,q})\}$ and $\{g^A_{L,p}(s_{L,p})\}$, $\forall j \in \mathcal{J}_4 \triangleq \{\text{T, NL}\}$, $\forall q \in \mathcal{Q}$, $\forall p \in \mathcal{P}$, Module A employs the VBI method to compute the approximate marginal posteriors $\psi(\mathbf{x})$, $\psi(\boldsymbol{\rho})$ and $\psi(\mathbf{s})$ under the following prior distribution:
%\vspace{-0.1cm}
\begin{equation}
\begin{aligned}
	\label{prior distribution in Module A}
	\hat{p}(\mathbf{x},\! \boldsymbol{\rho},\! \mathbf{s}) \!\!=&\!\!\!\! \prod_{j_1 \in \mathcal{J}_1} \!\!\!\! p(\mathbf{x}_{j_1} \!|  \boldsymbol{\rho}_{j_1}\!)  p(\boldsymbol{\rho}_{j_1}\!|\mathbf{s}_T\!) \!\!\!
	\prod_{j_2 \in \mathcal{J}_2}\! \!\! p(\mathbf{x}_{j_2}\!|\boldsymbol{\rho}_{j_2}\!)  p(\boldsymbol{\rho}_{j_2}|\mathbf{s}_{N\!L}\!)\!\!\! \\
	& \times \prod_{j_3 \in  \mathcal{J}_3}\!\!\!
	p(\mathbf{x}_{j_3}\!|\boldsymbol{\rho}_{j_3}\!) p(\boldsymbol{\rho}_{j_3}|\mathbf{s}_L\!)\, \hat{p}(\mathbf{s}_T\!,\!\mathbf{s}_{N\!L},\!\mathbf{s}_{L}\!),
\end{aligned}
\vspace{-0.1cm}
\end{equation}
where $\hat{p}(\!\mathbf{s}_T,\mathbf{s}_{N\!L},\mathbf{s}_{L}\!) \!\!=\!\! \prod_{j\in\! \mathcal{J}_{\!4}}\!\prod_{q\in \mathcal{Q}} (\!\pi_{j,q}\!)^{\!\frac{1\!+\!s_{j\!,q}}{2}}\!(\!1\!-\!\pi_{j\!,q})^{\!\!\frac{1\!-\!s_{j\!,q}}{2}} \!\! \prod_{p\in\! \mathcal{P}}\\(\pi_{L,p})^{\frac{1+s_{L,p}}{2}}(1-\pi_{L,p})^{\frac{1-s_{L,p}}{2}}$ with $\pi_{j,q} = g^A_{j,q}(1) \big/ \big(g^A_{j,q}(0) + g^A_{j,q}(1)\big) $ and $\pi_{L,p} = g^A_{L,p}(1) \big/ \big(g^A_{L,p}(0) + g^A_{L,p}(1)\big)$. Then, the output messages from Module A (also known as the extrinsic information), denoted by  $\{g^B_{j,q}(s_{j,q})\}$ and $\{g^B_{L,p}(s_{L,p})\}$, $\forall j \in \mathcal{J}_4$, $\forall q \in \mathcal{Q}$, $\forall p \in \mathcal{P}$, can be calculated as
\begin{equation}
	\label{the calculation of the extrinsic information}
	g^B_{j,q}(s_{j,q}) = \frac{\psi(s_{j,q})}{g^A_{j,q}(s_{j,q})},   \quad 
	g^B_{L,p}(s_{L,p}) = \frac{\psi(s_{L,p})}{g^A_{L,p}(s_{L,p})}.
\end{equation}
On the other hand, based on the prior information $\{g^B_{j,q}(s_{j,q})\}$ and $\{g^B_{L,p}(s_{L,p})\}$  provided by Module A,  Module B leverages the POS sparsity and 2D block sparsity, as characterized by the support prior distribution $p(\mathbf{s}_T,\mathbf{s}_{NL},\mathbf{s}_{L},\mathbf{s}_{U})$, to enhance the estimation performance. By performing  sum-product message passing over the associated factor graph, the output messages from Module B, i.e., $\{g^A_{j,q}(s_{j,q})\}$, $\{g^A_{L,p}(s_{L,p})\}$, can be obtained. Module A and Module B are iteratively executed  until  convergence. The details within these two modules are presented as follows.

\begin{figtab*}
	\begin{minipage}{0.4\linewidth}
		\centering
		\includegraphics[width=0.99\textwidth]{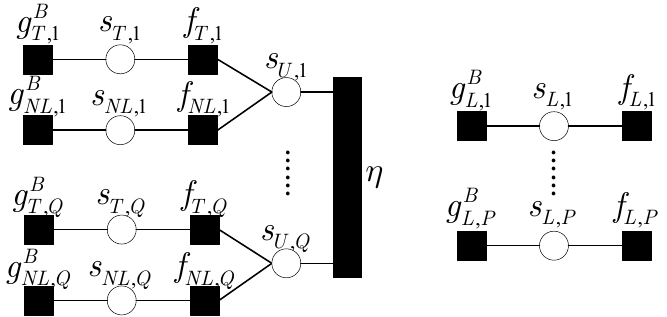}
		\figcaption{Factor graph of the joint distribution of all support vectors $p(\mathbf{s}_T,\mathbf{s}_{NL},\mathbf{s}_{L},\mathbf{s}_{U})$.}
		\label{pic:support factor graph}	
	\end{minipage}\;
	\begin{minipage}{0.59\linewidth}
		\centering
		\tabcaption{Summary of factor nodes and their corresponding expressions in Fig. \ref{pic:support factor graph}.}
		\vspace{-0.2cm}
		\label{tab:1}  
		\small
		\setlength{\tabcolsep}{0.7mm}{\begin{tabular}{ccc}
				\hline\hline\noalign{\smallskip}	
				Factor node & Distribution & Detailed expression   \\
				\noalign{\smallskip}\hline\noalign{\smallskip}
				$g_{\!T\!,q}^B(s_{\!T\!,q})$ & $\psi(s_{\!T\!,q})/g_{\!T\!,q}^A(s_{\!T\!,q})$ & $\pi_{\!T\!,q}^B \delta(s_{\!T\!,q}\!\!-\!\!1) \!\!+\!\! (1\!-\!\pi_{\!T\!,q}^B) \delta(s_{\!T\!,q}\!\!+\!\!1)$ \\
				$g_{\!N\!L\!,q}^B(s_{\!N\!L\!,q})$ & $\psi(\!s_{\!N\!L\!,q}\!)/g_{\!N\!L\!,q}^A(\!s_{N\!L\!,q}\!)$ & $\pi_{\!N\!L,q}^B \delta(s_{\!N\!L,q}\!\!-\!\!1) \!\!+\!\! (1\!\!-\!\!\pi_{\!N\!L,q}^B) \delta(s_{\!N\!L,q}\!\!+\!\!1)$ \\
				$f_{\!T\!,q}\!(\!s_{\!T\!,q},\! s_{\!U\!,q}\!)$ & $p(s_{\!T\!,q}|s_{\!U\!,q})$ & $\frac{1\!-\!s_{\!U\!,q}}{2}\!\delta(\!s_{\!T\!,q} \!\!+\!\! 1\!) \!\!+\!\! \frac{1\!+\!s_{\!U\!,q}}{2} p_T^{\frac{1\!+\!s_{\!T\!,q}}{2}}(1\!\!-\!\!p_{\!T})^{\frac{1\!-\!s_{\!T\!,q}}{2}}$ \\
				$f_{\!N\!L\!,q}\!(\!s_{\!N\!L\!,q}\!,\! s_{\!U\!\!,q}\!)$ & $p(\!s_{N\!L,q}|s_{U,q}\!)$ & $\frac{1\!-\!s_{\!U\!,q}}{2}\!\delta(\!s_{\!N\!L\!,q} \!\!+\!\! 1\!) \!\!+\!\! \frac{1\!+\!s_{\!U\!\!,q}}{2}  p_{\!N\!L}^{\!\frac{1\!+\!s_{N\!L,q}}{2}}\!\!(\!1\!\!-\!\!p_{\!N\textbf{}\!L}\!)^{\!\frac{1\!-\!s_{\!N\!L\!,q}}{2}}$ \\
				$f_{L,p}(s_{L,p})$ & $p(s_{L,p})$ & $ p_{L}^{ \frac{1\!+\!s_{L,p}}{2}}(1\!-\!p_{L})^{ \frac{1\!-\!s_{L,p}}{2}}$ \\
				$\eta(\mathbf{s}_U)$ & $p(\mathbf{s}_U)$ & $\frac{1}{C} \!\prod_{q\!=\!1}^Q \!\! \Big[\! \omega(s_{\!U\!,q}) \!\prod_{\!q^\prime \!\in\! \mathcal{D}(q)} \! \varpi^{\frac{1}{2}} (s_{U,q},s_{U,q^\prime}) \Big]$ \\
				\noalign{\smallskip}\hline
				%		\vspace{-0.0cm}
		\end{tabular}}
	\end{minipage}
\end{figtab*}

\subsubsection{Module A}
\label{sec: Module A}
The goal of Module A is to compute the approximate marginal posteriors $\psi(\mathbf{w}), \forall \mathbf{w} \in \Omega \triangleq \{ \mathbf{x}, \boldsymbol{\rho}, \mathbf{s} \}$ by minimizing the Kullback-Leibler divergence (KLD) between  $\psi(\mathbf{x},\boldsymbol{\rho},\mathbf{s}) =  \prod_{\mathbf{w} \in \Omega}\psi(\mathbf{w})$ and the posterior distribution $\hat{p}(\mathbf{x},\boldsymbol{\rho},\mathbf{s}|\mathbf{y};\Delta \mathbf{r},\Delta \mathbf{z})$ with the prior \eqref{prior distribution in Module A}, i.e.,
\begin{equation}
\label{KLD min}
\min_{\psi(\mathbf{x},\boldsymbol{\rho},\mathbf{s})} \int \psi(\mathbf{x},\boldsymbol{\rho},\mathbf{s}) \ln \frac{\psi(\mathbf{x},\boldsymbol{\rho},\mathbf{s})}{\hat{p}(\mathbf{x},\boldsymbol{\rho},\mathbf{s}|\mathbf{y};\Delta \mathbf{r},\Delta \mathbf{z})} d\mathbf{x} d\boldsymbol{\rho} d\mathbf{s}. 
\end{equation}
By exploiting the alternating optimization (AO) method, we can obtain  the stationary solution  $\psi^*(\mathbf{x},\boldsymbol{\rho},\mathbf{s}) = \prod_{\mathbf{w} \in \Omega}\psi^*(\mathbf{w})$ of  problem \eqref{KLD min}. Specifically,  with given $\psi(\mathbf{v})$, $ \forall \mathbf{v} \in \Omega \setminus{\mathbf{w}}$, the optimal $\psi(\mathbf{w})$ that minimizes the KLD in \eqref{KLD min}  can be expressed as \cite{Tzikas2008vbi}
\begin{equation}
	\label{stationary solution for KLD min}
	\psi(\mathbf{w}) \propto \text{exp}\Big( \big\langle \ln p(\mathbf{x},\boldsymbol{\rho},\mathbf{s},\mathbf{y};\Delta \mathbf{r},\Delta \mathbf{z})\big\rangle_{ \prod\nolimits_{\mathbf{v} \in \Omega \setminus{\mathbf{w}}} \psi(\mathbf{v})}  \Big),
\end{equation}
where $\langle f(x) \rangle_{ \psi(x) } \triangleq \int f(x) \psi(x) dx$. With the help of  the hierarchical prior design discussed in Section \ref{3-B}, we can obtain the closed-form update equations of  $\psi(\mathbf{w})$, $\forall \mathbf{w} \in \Omega$, which is presented in Appendix \ref{Appendix: Derivation of the VBI update expression}.

\par To ensure a favorable starting point for the AO method,  we initialize the approximate marginal posteriors $\psi(\mathbf{w}), \forall \mathbf{w} \in \Omega$ as follows. 
\begin{itemize}
	\item $\psi\!(\mathbf{s})$ is initialized as $\hat{p}(\!\mathbf{s}_{\!T}\!,\!\mathbf{s}_{\!N\!\!L}\!,\!\mathbf{s}_{\!L}\!)$ which is specified in \eqref{prior distribution in Module A}.
	
	\item $\psi(\boldsymbol{\rho})$ is initialized as a gamma distribution, i.e., 
	 $\psi(\boldsymbol{\rho}) \!=\! \prod_{j_1\!\in\mathcal{J}_1 \!\cup\! \mathcal{J}_2} \! \prod_{q \in\! \mathcal{Q}}  \! \Gamma(\rho_{j_1\!,q};\tilde{a}_{j_1\!,q},\tilde{b}_{j_1\!,q})\! 
	 \prod_{j_2\in\mathcal{J}_3}  \!\prod_{p\in\mathcal{P}} \! \Gamma(\rho_{j_2,p}; \\ \tilde{a}_{j_2,p},\tilde{b}_{j_2,p})$, where $\tilde{a}_{j_1,q} = \pi_{T,q}a_{j_1,q} + (1-\pi_{T,q})\bar{a}_{j_1,q}$, $\tilde{b}_{j_1,q} = \pi_{T,q}b_{j_1,q} + (1-\pi_{T,q})\bar{b}_{j_1,q}$, $\forall j_1 \in \mathcal{J}_1$, $\forall q \in \mathcal{Q}$. The remaining parameters of the gamma distribution  are defined similarly.
	
	\item $\psi(\mathbf{x})$ is initialized as a Gaussian distribution, i.e.,  $\psi(\mathbf{x}) = \prod_{j \in \mathcal{J}_1 \cup \mathcal{J}_2 \cup \mathcal{J}_3} \mathcal{CN}( \mathbf{x}_j; \boldsymbol{\mu}_j, \boldsymbol{\Sigma}_j)$, where for the cases of $j = \text{ITS}$ and $j = \text{CTS}$, $\boldsymbol{\Sigma}_{\text{ITS}}$ and $\boldsymbol{\Sigma}_{\text{CTS}}$ are diagonal matrices with their elements being the diagonal of $\boldsymbol{\Sigma}_{\text{T\!S}} \!=\! \big(\! \text{diag}(\mathbf{c}_{\text{T\!S}}) \!+\! \frac{1}{\sigma^2}\mathbf{F}^H_{\!s,r\!}(\!\Delta\mathbf{r}\!)\mathbf{F}_{\!\!s,r\!}(\!\Delta\mathbf{r}\!) \big)^{\!\!-\!1}$\!\!, $\mathbf{c}_{\text{TS}} \!\!=\!\! [\tilde{a}_{\text{IT\!S},\!1}\!/\tilde{b}_{\text{IT\!S},\!1}, \\ \!\cdots\!,\tilde{a}_{\text{IT\!S},Q}/\tilde{b}_{\text{IT\!S},Q},\tilde{a}_{\text{CTS},\!1}\!/\tilde{b}_{\text{C\!T\!S},\!1},\!\cdots\!,\tilde{a}_{\text{C\!T\!S}\!,Q}\!/\tilde{b}_{\text{C\!T\!S}\!,Q}]^{\!T}$ and $\boldsymbol{\mu}_{\text{TS}} \\ \!=\!  [\boldsymbol{\mu}_{\text{ITS}};\boldsymbol{\mu}_{\text{CTS}}] \!=\!  \frac{1}{\sigma^2}  \boldsymbol{\Sigma}_{\text{TS}}\mathbf{F}^H_{s,r}(\Delta\mathbf{r})\mathbf{y}_{s,r}$. The expressions of $\boldsymbol{\mu}_j$ and $\boldsymbol{\Sigma}_j$ in the remaining cases can be calculated in a similar manner.
\end{itemize}
After initialization, the approximate marginal posteriors  $\psi(\mathbf{w}), \forall \mathbf{w} \in \Omega$ are updated iteratively until convergence. During each iteration, we  sequentially update the three marginal posteriors  $\psi(\mathbf{x})$, $\psi(\boldsymbol{\rho})$, $\psi(\mathbf{s})$, which is detailed as follows. First, with  given $\psi(\boldsymbol{\rho})$ and $\psi(\mathbf{s})$, $\psi(\mathbf{x})$ can be updated by
\begin{equation}
	\label{update for x}
	\psi(\mathbf{x}) = \prod_{j \in \mathcal{J}_1 \cup \mathcal{J}_2 \cup \mathcal{J}_3} \mathcal{CN}( \mathbf{x}_j; \boldsymbol{\mu}_j, \boldsymbol{\Sigma}_j),
\end{equation}
where $\boldsymbol{\Sigma}_{\text{ITS}} \!=\! \big( \text{diag}(\mathbf{c}_{\text{ITS}}) \!+\! \frac{1}{\sigma^2}\mathbf{F}^H_{\text{ITS}}(\Delta \mathbf{r})\mathbf{F}_{\text{ITS}}(\Delta \mathbf{r}) \big)^{\!\!-\!1}$ and  $\boldsymbol{\mu}_{\text{ITS}} \\ =\!\!  \frac{1}{\sigma^2}\boldsymbol{\Sigma}_{\text{ITS}}\mathbf{F}^H_{\text{\!ITS}}(\!\Delta\mathbf{r}\!)\big(\mathbf{y}_{\!s,r} \!-\! \mathbf{F}_{\text{\!CTS}}(\!\Delta\mathbf{r}\!)  \boldsymbol{\mu}_{\text{CTS}}\big)$ denote the posterior covariance matrix  and  mean vector  w.r.t. $\mathbf{x}_{\text{ITS}}$, respectively,
with  $\mathbf{F}_{\text{ITS}}(\Delta \mathbf{r}) \!=\! [\mathbf{F}_{s,r}(\Delta \mathbf{r}\!)]_{\!:,1:Q}$, $\mathbf{F}_{\text{\!CTS}}(\!\Delta \mathbf{r}\!) \!\!=\!\! [\mathbf{F}_{\!\!s,r}(\!\Delta \mathbf{r}\!)]_{:,Q+1:2Q}$ and  $\mathbf{c}_{\text{ITS}} \!\!=\!\! [\mathbf{c}_{\text{TS}}]_{\!1\!:Q\!}$; $\boldsymbol{\Sigma}_{\text{B\!NL}}\!\! =\!\! \big( \text{diag}(\mathbf{c}_{\text{BNL}}) 
\!\!+\!\! \frac{1}{\sigma^2}\mathbf{F}^H_{\!\text{BNL}}\!(\!\Delta \mathbf{r}\!)  \mathbf{F}_{\!\text{BNL}}\!(\!\Delta \mathbf{r}\!) \!\big)^{\!\!-\!1}$ 
and  $\boldsymbol{\mu}_{\text{BNL}} \!=\! \frac{1}{\sigma^2}\boldsymbol{\Sigma}_{\text{BNL}}\mathbf{F}^H_{\text{BNL}}(\Delta\mathbf{r})\big(\mathbf{y}_{c} - \mathbf{F}_{\text{INL}}(\Delta\mathbf{r})  \boldsymbol{\mu}_{\text{INL}} - \mathbf{F}_{\text{BL}}(\Delta\mathbf{z})  \boldsymbol{\mu}_{\text{BL}} - \mathbf{F}_{\text{IL}}(\Delta\mathbf{z})  \boldsymbol{\mu}_{\text{IL}}\big)$ denote the posterior covariance matrix  and  mean vector  w.r.t. $\mathbf{x}_{\text{BNL}}$, respectively,  with $\mathbf{F}_{\text{INL}}(\Delta \mathbf{r}) \!\!=\!\! [\mathbf{F}_{c}(\Delta \mathbf{r},\Delta \mathbf{z})]_{:, Q+1:2Q}$, $\mathbf{F}_{\text{IL}}(\Delta \mathbf{z})  \!=\! [\mathbf{F}_{c}(\!\Delta \mathbf{r}\!,  \Delta \mathbf{z}\!)]_{:, 2Q\!+\!P\!+\!1:2Q\!+\!2P}$,  $\mathbf{F}_{\text{BNL}}(\Delta \mathbf{r}) \!=\! [\mathbf{F}_{c}(\Delta \mathbf{r},\Delta \mathbf{z})]_{:,1:Q}$,  $\mathbf{F}_{\!\text{BL}}(\!\Delta \mathbf{z}\!) \!=\! [\mathbf{F}_{c}(\!\Delta \mathbf{r},\Delta \mathbf{z}\!)]_{:, 2Q\!+\!1:2Q\!+\!P}$   and  $\mathbf{c}_{\text{B\!NL}} \!=\! [\tilde{a}_{\text{BNL},1}/\tilde{b}_{\text{BNL},1}, \\ \!\cdots\!,\tilde{a}_{\text{BNL},Q}/\tilde{b}_{\text{BNL},Q}]^T$.
Due to the similarity in calculating the remaining posterior covariance matrices and mean vectors, we omit the detailed expressions for simplicity.

\par Then, with fixed  $\psi(\mathbf{x})$ and $\psi(\mathbf{s})$, $\psi(\boldsymbol{\rho})$ can be obtained by 
\begin{equation}
	\label{update for rho}
	\!\!\!\psi(\boldsymbol{\rho}) \!\!=\!\!\!\!\!\! \prod_{j_{\!1}\!\in\!\mathcal{J}_{\!1}  \!\cup\! \mathcal{J}_{\!2}}   \prod_{q \!\in\! \mathcal{Q}}  \! \Gamma(\rho_{\!j_{\!1}\!,q}\!;\tilde{a}_{\!j_{\!1}\!,q},\tilde{b}_{\!j_{\!1}\!,q})\!\!
	\prod_{\!j_{\!2}\!\in\!\mathcal{J}_{\!3}}  \prod_{p\!\in\!\mathcal{P}}\! \Gamma(\rho_{\!j_{\!2},p};\!\tilde{a}_{\!j_{\!2},p},\!\tilde{b}_{\!j_2,p}\!), 
\end{equation}
where  $\tilde{a}_{j_1,q} \!=\! \tilde{\pi}_{T,q}a_{j_1,q} \!+\! (1\!-\!\tilde{\pi}_{T,q})\bar{a}_{j_1,q} \!+ \!1$, $\tilde{b}_{j_1,q} \!=\! \tilde{\pi}_{T\!,q}  b_{j_1,q} + (1-\tilde{\pi}_{T,q})\bar{b}_{j_1,q} + |\mu_{j_1,q}|^2 + \Sigma_{j_1,q}$, $\forall j_1 \in \mathcal{J}_1$, $\forall q \in \mathcal{Q}$, with $\mu_{j_1,q}$ being the $q$-th element of $\boldsymbol{\mu}_{j_1}$, $\Sigma_{j_1,q}$ being the $q$-th diagonal element of $\boldsymbol{\Sigma}_{j_1}$ and $\tilde{\pi}_{T,q}$ being the posterior probability of $s_{T,q} = 1$. Note that in the first iteration, we have $\tilde{\pi}_{T,q} = \pi_{T,q}$, while in the subsequent iterations, $\tilde{\pi}_{T,q}$ is determined by the update equation for $\psi(s_{T,q})$. The remaining parameters of the gamma distributions can be updated similarly.

\par Finally, with  $\psi(\mathbf{x})$ and $\psi(\boldsymbol{\rho})$ fixed, $\psi(\mathbf{s})$ can be derived as
\begin{equation}
	\label{update for s}
	\psi\!(\mathbf{s}\!)  \!=\!\! \prod_{j\!\in \!\mathcal{J}_{\!4}} \prod_{q\in \mathcal{Q}} (\tilde{\pi}_{\!j,q}\!)^{\!\!\frac{1\!+\!s_{j,q}}{2}}\!(\!1\!-\!\tilde{\pi}_{j,q}\!)^{\!\!\frac{1\!-\!s_{j,q}}{2}} \!\! \prod_{p\in \mathcal{P}}(\tilde{\pi}_{\!L,p})^{\!\!\frac{1\!+\!s_{\!L,p}}{2}}\!(\!1\!-\!\tilde{\pi}_{\!L,p})^{\!\!\frac{1\!-\!s_{\!L,p}}{2}}\!,
\end{equation}
where $\tilde{\pi}_{\!T,q} \!=\! \frac{1}{Z}\! \prod_{j \!\in\! \mathcal{J}_{\!1}} \!\! \frac{\pi_{\!T,q}b_{\!j\!,q}^{\!a_{\!j\!,q}}}{\Gamma(a_{\!j\!,q})} e^{ (a_{\!j\!,q} \!-\! 1)\langle\ln \rho_{\!j,q}\rangle_{\psi(\rho_{\!j,q})} \!-\! b_{\!j,q}\langle \rho_{j,q}\rangle_{\!\psi(\rho_{\!j,q})}  }$, $Z$ is a normalization constant, $\langle \rho_{j,q}\rangle_{\psi(\rho_{j,q})} = \tilde{a}_{j,q}/\tilde{b}_{j,q}$, $\langle\ln \rho_{j,q}\rangle_{\psi(\rho_{j,q})} = \Psi(\tilde{a}_{j,q})-\ln(\tilde{b}_{j,q})$ and $\Psi(x)$ denotes the digamma function. The expressions of $\tilde{\pi}_{NL,q}$, $\forall q \in \mathcal{Q}$ and $\tilde{\pi}_{L,p}$, $\forall p \in \mathcal{P}$ can be obtained in similarly.

\subsubsection{Module B}
\par The factor graph of the joint distribution of all support vectors $p(\mathbf{s}_T,\mathbf{s}_{NL},\mathbf{s}_{L},\mathbf{s}_{U})$ is depicted in Fig. \ref{pic:support factor graph}, and the associated factor nodes along with their functional expressions are summarized in Table \ref{tab:1}. Note that the output messages from Module A, i.e., $\{g^B_{j,q}(s_{j,q})\}$ and $\{g^B_{L,p}(s_{L,p})\}$, are also used to represent the prior factor nodes in Fig. \ref{pic:support factor graph}.  By performing  sum-product message passing over the support factor graph, all the messages can be readily obtained. Please refer to Appendix \ref{Appendix: message passing over the support factor graph} for the details. It is worth pointing out that the loopy belief propagation method \cite{li2009markov} is employed to update the messages associated with the MRF due to the presence of loops. After completing the update of all messages, the output messages from Module B are given by
$\{g^A_{j,q}(s_{j,q}) \!=\! \nu_{f_{j,q} \to s_{j,q}}(s_{j,q})\}$, $\forall j \in \mathcal{J}_4$ and $\{g^A_{L,p}(s_{L,p}) \!=\! \nu_{f_{L,p} \to s_{L,p}}(s_{L,p}) \!=\!  p_{L}^{ \frac{1\!+\!s_{L,p}}{2}}(1\!-\!p_{L})^{ \frac{1\!-\!s_{L,p}}{2}} \}$. Since the specific structure of the i.i.d. prior $p(\mathbf{s}_L)$ is sufficiently revealed by the factor node $f_{L,p}$, $\forall p \!\in \!\mathcal{P}$, the output messages $\{g^A_{L,p}(s_{L,p})\}$ from Module B are fixed in the E step.
\vspace{-0.0cm}

\renewcommand\baselinestretch{\linspreadalgr}\selectfont 
\begin{algorithm}[thb]
	\caption{Proposed DDG-based method}
	\label{Algorithm1, DDG-based method}
	\textbf{Input}: $\Delta \mathbf{r}^{\!n}\!, \Delta \mathbf{z}^{\!n}\!,  \varsigma_{r}^n\!, \varsigma_{z}^n, \psi(\mathbf{s}), \boldsymbol{\mu}_j^n\!, \boldsymbol{\Sigma}_j^n\!, \forall j$. \; \textbf{Initialize}: $\mathbf{d}^r\! =\! [\mathbf{d}^{r\!,x}\!,\mathbf{d}^{r\!,y}] \!=\!  \mathbf{0}_{Q \!\times\! 2}$, $\mathbf{d}^z \!=\! [\mathbf{d}^{z\!,x},\mathbf{d}^{z\!,y}] \!=\! \mathbf{0}_{P \!\times\! 2}$. \\
	\textbf{Output}: $\Delta \mathbf{r}^{\!n + 1}\!, \Delta \mathbf{z}^{\!n + 1}\!$.  
	\begin{algorithmic}[1]
		\STATE {Create a set $\mathcal{Q}_{\mathbf{r}} \!\subseteq\! \{1,\!\cdots, \!Q\}$ based on $\psi(\mathbf{s})$ and  $|\mu_{j,q}^n|^2$, $\forall j$, $\forall q$, which includes the potential grid indices of the targets and scatterers.} 
		\STATE {\textbf{for} $q \in \mathcal{Q}_{\mathbf{r}}$ \textbf{do}}
		\STATE {\; Calculate the gradients $\mathbf{g}_{\text{IRS},q}^r$ and $\mathbf{g}_{\text{BS},q}^r$  according to \\ \; Appendix \ref{Appendix: gradient Mstep}.}
		\STATE {\; \textbf{if} $\mathbf{g}_{\text{BS},q}^{r,x}\mathbf{g}_{\text{IRS},q}^{r,x}\!\!\leq\!\!0$ \textbf{then} set the update direction $\mathbf{d}_{q}^{r,x}$ to 0.\!}
		\STATE {\; \textbf{else} set the update direction as $\mathbf{d}_{q}^{r\!,x} \!\!=\!\! \text{sign}( \mathbf{g}_{\text{BS},q}^{r\!,x})$.} \textbf{end if}\!
		\STATE {\;  Calculate $\mathbf{d}_{q}^{r,y}$  as in steps 4-5.   }
		\STATE {\textbf{end for}}
		\STATE {Calculate the update direction $\mathbf{d}^z$ as in steps 1-7.}
		\STATE {Update $\Delta \mathbf{r}^{n+1} = \Delta \mathbf{r}^{n} + \varsigma_{r}^n \mathbf{d}^r$, $\Delta \mathbf{z}^{n+1} = \Delta \mathbf{z}^{n} + \varsigma_{z}^n \mathbf{d}^z$.}
	\end{algorithmic}
\end{algorithm}
\vspace{-0.0cm}
\renewcommand\baselinestretch{\linspread}\selectfont

\subsection{AS-TVBI-M Step}
\par In order to estimate the position offset $\{ \Delta \mathbf{r}, \Delta\mathbf{z} \}$, we need to solve the ML problem \eqref{ML problem in M step}, which, however, is quite challenging since there is no explicit expression of  $\ln p(\mathbf{y}|\Delta \mathbf{r}, \Delta \mathbf{z})$ as previously discussed. To tackle this challenge, we employ the EM method and propose to  construct a sequence of surrogate functions for  $\ln p(\mathbf{y}|\Delta \mathbf{r}, \Delta \mathbf{z})$. In particular, by exploiting the approximate marginal posteriors $\psi(\mathbf{w}), \forall \mathbf{w} \in \Omega$ obtained in the E step, a tractable surrogate function (in the $n$-th EM iteration) is given by
\begin{align}
	\label{Q_funciton}
	&Q(\!\Delta \mathbf{r}\!,\! \Delta \mathbf{z}; \!\Delta \mathbf{r}^{\!n}\!\!,\! \Delta \mathbf{z}^{\!n})  \!\!=\!\!\! \int\!\! \psi(\mathbf{x},\!  \boldsymbol{\rho},\! \mathbf{s}) \ln\! \frac{p(\mathbf{x},\!\boldsymbol{\rho},\!\mathbf{s},\!\mathbf{y}; \!\Delta \mathbf{r}\!, \Delta \mathbf{z})}{\psi(\mathbf{x}, \boldsymbol{\rho}, \mathbf{s}) } d\mathbf{x}d\boldsymbol{\rho}d\mathbf{s} \nonumber \\
    &\!=\! -\frac{1}{\sigma^2} \Big[ \| \mathbf{y}_{\!s,r}  \!-\! \mathbf{F}_{\!s,r}(\Delta \mathbf{r})  \boldsymbol{\mu}_{TS}\|^2 + \| \mathbf{y}_{B,r} - \mathbf{F}_{B,r}(\Delta \mathbf{r})\boldsymbol{\mu}_{TB}   \|^2 \nonumber \\
    &\!+\!\! \| \mathbf{y}_{\!c} \!\!-\! \mathbf{F}_{\!c}(\Delta \mathbf{r}\!,\!\Delta \mathbf{z})\boldsymbol{\mu}_{c}\|^2 
     \!\!+ \!  \text{tr}\big( \mathbf{F}_{\!\!s,r}(\!\Delta \mathbf{r}\!) \boldsymbol{\Sigma}_{T\!S} \mathbf{F}_{\!\!s,r}^H(\!\Delta \mathbf{r}\!)\big) 
   	 \!\! +\!\text{tr}\big( \mathbf{F}_{\!\!B\!,r}(\!\Delta \mathbf{r}\!) \nonumber \\
   	 & \boldsymbol{\Sigma}_{T\!B} \mathbf{F}_{\!B,r}^H(\!\Delta \mathbf{r}\!)\big)	\!+\! \text{tr}\big( \mathbf{F}_{\!c}(\!\Delta \mathbf{r}\!,\!\Delta \mathbf{z}) \boldsymbol{\Sigma}_{c} \mathbf{F}_{c}^H\!(\!\Delta \mathbf{r}\!,\!\Delta \mathbf{z})	  \big) \! \Big] \!\!+\! C,
\end{align}
where  $C$ is a constant, $\boldsymbol{\mu}_{TS} \triangleq [\boldsymbol{\mu}_{ITS};  \boldsymbol{\mu}_{CTS}]$, $\boldsymbol{\Sigma}_{TS}   \triangleq \text{blkdiag}(\boldsymbol{\Sigma}_{ITS}, \boldsymbol{\Sigma}_{CTS})$, $\boldsymbol{\mu}_{TB}$, $\boldsymbol{\Sigma}_{TB}$, $\boldsymbol{\mu}_{c}$ and $\boldsymbol{\Sigma}_{c}$ are defined similarly. Based on this  surrogate function, a straightforward approach is using  the gradient ascent method \cite{boyd2004convex}  to update  $\{ \Delta \mathbf{r},  \Delta \mathbf{z} \}$. Nevertheless, our simulation shows that the performance achieved by this method is unsatisfactory if  the target/scatterer is in close proximity to the BS (IRS sensors). In this case, the 
received signal power at the BS (IRS sensors) is significantly higher than that at the IRS sensors (BS) due to the considered mmWave band, which makes the position-related information provided by the IRS sensors (BS) useless. To address this issue, we propose to use  the DDG-based method outlined in Algorithm \ref{Algorithm1, DDG-based method} to update  $\{ \Delta \mathbf{r},  \Delta \mathbf{z} \}$, where  the update directions of $\Delta\mathbf{r}$ and $\Delta\mathbf{z}$ are determined by the BS and IRS sensors jointly, even if the target/scatterer is close to either the BS or IRS sensors. Note that in Algorithm \ref{Algorithm1, DDG-based method} $\mathbf{g}_{\text{BS}}^r = [\mathbf{g}_{\text{BS},1}^r; \cdots; \mathbf{g}_{\text{BS},Q}^r] \in \mathbb{R}^{Q\times2}$ and $\mathbf{g}_{\text{IRS}}^r = [\mathbf{g}_{\text{IRS},1}^r; \cdots; \mathbf{g}_{\text{IRS},Q}^r] \in \mathbb{R}^{Q\times2}$ denote the BS-related and IRS-related gradient terms of $\nabla_{\Delta \mathbf{r}} Q(\Delta \mathbf{r}, \Delta \mathbf{z};\Delta \mathbf{r}^n, \Delta \mathbf{z}^n)$, $\mathbf{g}_{\text{BS}}^z = [\mathbf{g}_{\text{BS},1}^z; \cdots; \mathbf{g}_{\text{BS},P}^z] \in \mathbb{R}^{P\times2}$ and $\mathbf{g}_{\text{IRS}}^z = [\mathbf{g}_{\text{IRS},1}^z; \cdots; \mathbf{g}_{\text{IRS},P}^z] \in \mathbb{R}^{P\times2}$ denote the BS-related and IRS-related gradient terms of $\nabla_{\!\Delta \mathbf{z}} Q(\Delta \mathbf{r}, \!\Delta \mathbf{z};\!\Delta \mathbf{r\!}^n,\! \Delta \mathbf{z}^n)$,
$\varsigma_{r}^n$ and  $\varsigma_{z}^n$ are the step sizes for the updates of $\Delta\mathbf{r}$ and $\Delta\mathbf{z}$, respectively. The details of $\mathbf{g}_{\text{BS}}^r$, $\mathbf{g}_{\text{IRS}}^r$, $\mathbf{g}_{\text{BS}}^z$ and $\mathbf{g}_{\text{IRS}}^z$ can be found in Appendix \ref{Appendix: gradient Mstep} and the overall AS-TVBI algorithm is summarized in Algorithm \ref{Algorithm2, AS-TVBI algorithm}.
\vspace{-0.1cm}

\renewcommand\baselinestretch{\linspreadalgr}\selectfont 
\begin{algorithm}[htb]
	\caption{Proposed AS-TVBI algorithm}
	\label{Algorithm2, AS-TVBI algorithm}
	\textbf{Input}: $\mathbf{y}$, $\Delta \mathbf{r}^1 \!\!=\!\! \mathbf{0}$, $\Delta \mathbf{z}^1  \!\!=\!\! \mathbf{0}$, $\mathbf{F}(\mathbf{0}, \mathbf{0})$, $p_T$, $p_{NL}$, $p_L$, hyperparameters in \eqref{prior model, rho}: $a_{j,q}$, $b_{j,q}$, $\bar{a}_{j,q}$, $\bar{b}_{j,q}$, $\forall j$, $\forall q$, hyperparameters of MRF: $\alpha$, $\beta$, maximum iteration number $N_{out}$, threshold $\epsilon_{out} >0 $, $n=1$. \quad \textbf{Output}: $\boldsymbol{\mu}_j$, $\forall j$, $\psi(\mathbf{s})$,   $\Delta \mathbf{r}, \Delta \mathbf{z}$. \!\!\!\!\!\!\!\!\!\!\!\!\!\!\!\!\!\!\!\!\!\!\!\!\!\!\!\!\!\!\!\!
	\begin{algorithmic}[1]
		\STATE {\textbf{Repeat}}
		\STATE {\; \textbf{AS-TVBI-E Step:}}  
		%		\STATE {\; \textbf{\%Module A:}  }
		\STATE {\; Initialize the distribution $\psi(\mathbf{w})$, $\forall \mathbf{w} \in \Omega$ according to \\ \; \ref{sec: Module A}. \quad \textbf{\%Module A}} 
		\STATE {\; \textbf{While} not convergence \textbf{do} }
		\STATE {\;\;\;\; Update  $\psi(\mathbf{x})$, $\psi(\boldsymbol{\rho})$ and $\psi(\mathbf{s})$ using \eqref{update for x}, \eqref{update for rho} and \eqref{update for s}, \\ \;\;\;\; respectively.}
		\STATE {\; \textbf{end While} }
		\STATE {\; Calculate the output messages   $\{g^{\!B}_{\!j\!,q}\!(\!s_{\!j\!,q}\!)\}$ and $\{g^{\!B}_{\!L\!,p}(\!s_{\!L\!,p}\!)\}.$}\!\!\!
		\STATE {\; Perform sum-product message passing over the  support \\  \;  factor graph  and   calculate the output messages $\{\!\nu_{f_{j,q} \!\to s_{\!j,q}}$ \\ \; $(s_{j,q})\}$ according to Appendix \ref{Appendix: message passing over the support factor graph}. \quad \textbf{\%Module B}}
		\STATE {\; \textbf{AS-TVBI-M Step:}} 
		\STATE {\; Update the position offset $\{\Delta \mathbf{r}, \Delta \mathbf{z}\}$ using Algorithm \ref{Algorithm1, DDG-based method}.}\!\! 
		\STATE {\; $n \leftarrow n + 1$.}
		\STATE {\textbf{Until} $\sum_{j} \| \boldsymbol{\mu}_j^{n-1} - \boldsymbol{\mu}_j^{n-2} \| < \epsilon_{out}$ or $n > N_{out}$.}
	\end{algorithmic}
\end{algorithm}
\vspace{-0.1cm}
\renewcommand\baselinestretch{\linspread}\selectfont

\section{IRS Reflection Design Based on Cram\'{e}r-Rao Bound}
\label{sec: CRLB optimization}
In this section, we focus on  the IRS reflection coefficients design in phase \Rmnum{2} based on the estimated target/scatterer/user positions and SAC channels from phase \Rmnum{1}. Specifically, the  CRB matrix of the overall position vector $\boldsymbol{\xi}  \triangleq   [\boldsymbol{\xi}^{ \dot{T}} ;\boldsymbol{\xi}^{ O} ;\boldsymbol{\xi}^{ S}  ;\boldsymbol{\xi}^{ U}]  \in \!  \mathbb{R}^{2(\tilde{N} + 1)}$  is first derived, where $\boldsymbol{\xi}^{\dot{T}}    \in \mathbb{R}^{2\tilde{K}}$, $\boldsymbol{\xi}^{O} \in  \mathbb{R}^{2O}  $,  $\boldsymbol{\xi}^{S}  \in \mathbb{R}^{2\tilde{L}}$ and $\boldsymbol{\xi}^{U}   \in \mathbb{R}^{2}$ denote the position vectors of the pure sensing targets, the overlapping parts between the communication scatterers and sensing targets, the pure communication scatterers and the user, respectively. Then, we propose to optimize the IRS reflection vectors  $\varphi_r^{\text{\Rmnum{2}}}(t)$ and $\varphi_c^{\text{\Rmnum{2}}}(t)$, $\forall t$, by minimizing the trace of the CRB matrix of $\boldsymbol{\xi}$, subject to the unit-modulus constraints on $\varphi_r^{\text{\Rmnum{2}}}(t)$ and $\varphi_c^{\text{\Rmnum{2}}}(t)$. Finally, we develop a manifold-based optimization algorithm to solve the CRB minimization problem with low computational complexity.
\vspace{-0.3cm}

\subsection{ Cram\'{e}r-Rao Bound }
Since the position estimation mean squared error (MSE) of the proposed AS-TVBI algorithm is difficult to obtain, we  instead adopt the CRB of $\boldsymbol{\xi}_n$, $\forall n \in \tilde{\mathcal{N}} \triangleq \{ 1, \cdots, 2(\tilde{N} + 1) \}$, which is  a lower bound of the position estimation MSE. Moreover, for ease of practical implementation, the CRB matrix of $\boldsymbol{\xi}$ is derived based on the estimation results from phase \Rmnum{1}, including the target positions $\hat{\mathbf{p}}_{T,k} = \mathbf{r}_{q_{T,k}} + \Delta \mathbf{r}_{q_{T,k}}$, $\forall k$, the scatterer positions $\hat{\mathbf{p}}_{S,l} = \mathbf{r}_{q_{S,l}} + \Delta \mathbf{r}_{q_{S,l}}$, $\forall l$, the user position $\hat{\mathbf{p}}_{u} = \mathbf{z}_{p_{u}} + \Delta \mathbf{z}_{p_{u}}$ and the SAC channels $\hat{\mathbf{x}}_j = \boldsymbol{\mu}_j$, $\forall j$. According to 
\cite{kay1993fundamentals}, the Fisher information matrix (FIM) for estimating the overall position vector $\boldsymbol{\xi}$ is given by 
\begin{align}
	\label{FIM}
	\mathbf{J}_{\!\boldsymbol{\xi}\!,\boldsymbol{\xi}}\! \!=  \!
	2\Re\Big\{\! \frac{\partial\boldsymbol{\mu}^H}{\partial\boldsymbol{\xi}}\! \boldsymbol{\Sigma}^{\!-\!1}\! \frac{\partial\boldsymbol{\mu}}{\partial\boldsymbol{\xi}^{\!T} } \!\!\Big\}
\!	\!=\!\!\!
	\left[ \setlength{\arraycolsep}{0.5mm}
	\begin{array}{cccc}
		\mathbf{J}_{\!\boldsymbol{\xi}^{\!\dot{T}}\!,\boldsymbol{\xi}^{\!\dot{T}}} & \mathbf{J}_{\!\boldsymbol{\xi}^{\!\dot{T}}\!,\boldsymbol{\xi}^{\!O}} & \mathbf{0} & \mathbf{0}  \\
		\mathbf{J}_{\!\boldsymbol{\xi}^{\!\dot{T}}\!,\boldsymbol{\xi}^{\!O}}^{\!T} & \mathbf{J}_{\!\boldsymbol{\xi}^{\!O}\!,\boldsymbol{\xi}^{\!O}} & \mathbf{J}_{\!\boldsymbol{\xi}^{\!O}\!,\boldsymbol{\xi}^{\!S}} & \mathbf{0}  \\
		\mathbf{0}  & \mathbf{J}_{\!\boldsymbol{\xi}^{\!O}\!,\boldsymbol{\xi}^{\!S}}^{\!T} & \mathbf{J}_{\!\boldsymbol{\xi}^{\!S}\!,\boldsymbol{\xi}^{\!S}}  & \mathbf{0}\\
		\mathbf{0} & \mathbf{0} & \mathbf{0} & \mathbf{J}_{\!\boldsymbol{\xi}^{\!U}\!,\boldsymbol{\xi}^{\!U}}
	\end{array}
	\right]\!\!,
\end{align}
where $\boldsymbol{\mu}$ and $\boldsymbol{\Sigma}$ denote the mean vector and covariance matrix of the observation $\mathbf{y}$, respectively. The details of the submatrices in $\mathbf{J}_{\boldsymbol{\xi},\boldsymbol{\xi}}$ can be found in Appendix \ref{Appendix: submatrices in FIM}.
Based on the above, the CRB matrix of $\boldsymbol{\xi}$ can be  given by  $\mathbf{C}(\boldsymbol{\Phi}^{\text{\Rmnum{2}}}_{r}, \boldsymbol{\Phi}^{\text{\Rmnum{2}}}_{c}) = \mathbf{J}_{\boldsymbol{\xi},\boldsymbol{\xi}}^{-1}$, whose diagonal elements characterize the lower bound of the  estimation MSE of $\boldsymbol{\xi}$.
\vspace{-0.3cm}

\subsection{Problem Formulation}
To ensure the estimation performance of all the position parameters in $\boldsymbol{\xi}$, 
we aim to minimize $\text{tr}\big(\mathbf{C}(\boldsymbol{\Phi}^{\text{\Rmnum{2}}}_{r}, \boldsymbol{\Phi}^{\text{\Rmnum{2}}}_{c})\big)$, which corresponds to reducing the average MSE of $\boldsymbol{\xi}_n$, $\forall n \!\in\! \tilde{\mathcal{N}}$. Mathematically, the optimization problem for  IRS reflection design can be formulated as
\vspace{-0.1cm}
\begin{subequations}
	\label{origin problem formulation}
	\begin{align}
		\min_{\boldsymbol{\Phi}^{\text{\Rmnum{2}}}_{r}, \boldsymbol{\Phi}^{\text{\Rmnum{2}}}_{c}} & \;\; \text{tr}\big(\mathbf{C}(\boldsymbol{\Phi}^{\text{\Rmnum{2}}}_{r}, \boldsymbol{\Phi}^{\text{\Rmnum{2}}}_{c})\big)  \label{origin problem formulation, obj func} \\
		\text{s.t.} & \;\;  | [\boldsymbol{\Phi}^{\text{\Rmnum{2}}}_r]_{n,t} | \!=\! 1, \; \forall n \!\in\! \mathcal{N}_p, \; \forall t \!\in\! \mathcal{T}_3 \!\triangleq\!  \{ 1,  \!\cdots\!, T_3\},  \label{unit-modulus constant, Phi_r} \\
		& \;\;  | [\boldsymbol{\Phi}^{\text{\Rmnum{2}}}_c]_{n,t} | \!=\! 1, \; \forall n \!\in\! \mathcal{N}_p, \; \forall t \!\in\! \mathcal{T}_4 \triangleq  \{ 1,  \!\cdots\!, T_4\}\label{unit-modulus constant, Phi_c}.
		\vspace{-0.1cm}
	\end{align}
\end{subequations}
It is challenging to solve problem \eqref{origin problem formulation} due to: 1) the non-convexity of the objective function w.r.t. $\boldsymbol{\Phi}^{\text{\Rmnum{2}}}_{r}$ and $ \boldsymbol{\Phi}^{\text{\Rmnum{2}}}_{c}$, 2)  lack of an explicit expression for  $\mathbf{C}(\boldsymbol{\Phi}^{\text{\Rmnum{2}}}_{r}, \boldsymbol{\Phi}^{\text{\Rmnum{2}}}_{c})$ due to the inversion operation, i.e.,  $\mathbf{J}_{\boldsymbol{\xi},\boldsymbol{\xi}}^{-1}$, 3) the tricky unit-modulus constraints \eqref{unit-modulus constant, Phi_r} and \eqref{unit-modulus constant, Phi_c}. Inspired by the SDR method \cite{luo2010semidefinite}, we introduce new variables $\bar{\boldsymbol{\Phi}}^{\text{\Rmnum{2}}}_r(t) \triangleq \bar{\boldsymbol{\varphi}}^{\text{\Rmnum{2}}}_r(t)(\bar{\boldsymbol{\varphi}}^{\text{\Rmnum{2}}}_r(t))^H$ with $\bar{\boldsymbol{\varphi}}^{\text{\Rmnum{2}}}_r(t) = [(\boldsymbol{\varphi}^{\text{\Rmnum{2}}}_r(t))^*;1]$, $\forall t \in \mathcal{T}_3$ and $\bar{\boldsymbol{\Phi}}^{\text{\Rmnum{2}}}_c(t) \triangleq \bar{\boldsymbol{\varphi}}^{\text{\Rmnum{2}}}_c(t)(\bar{\boldsymbol{\varphi}}^{\text{\Rmnum{2}}}_c(t))^H$ with $\bar{\boldsymbol{\varphi}}^{\text{\Rmnum{2}}}_c(t) = [(\boldsymbol{\varphi}^{\text{\Rmnum{2}}}_c(t))^*;1]$, $\forall t \in \mathcal{T}_4$, based on which all the submatrices in $\mathbf{J}_{\boldsymbol{\xi},\boldsymbol{\xi}}$ can be expressed as linear functions of  $\{\bar{\boldsymbol{\Phi}}^{\text{\Rmnum{2}}}_r(t)\}$ and $\{\bar{\boldsymbol{\Phi}}^{\text{\Rmnum{2}}}_c(t)\}$. Accordingly, by relaxing the rank-1 constraints of $\{\bar{\boldsymbol{\Phi}}^{\text{\Rmnum{2}}}_r(t)\}$ and $\{\bar{\boldsymbol{\Phi}}^{\text{\Rmnum{2}}}_c(t)\}$, problem \eqref{origin problem formulation} can be transformed into
\vspace{-0.1cm}
\begin{subequations}
	\label{SDR problem formulation}
	\begin{align}
		\min_{\{\!\bar{\boldsymbol{\Phi}}^{\!\text{\Rmnum{2}}}_{\!r}(t)\!\}\!,\{\!\bar{\boldsymbol{\Phi}}^{\!\text{\Rmnum{2}}}_{\!c}(t)\!\}} & \;\; \text{tr}\big(\mathbf{C}(\{\bar{\boldsymbol{\Phi}}^{\text{\Rmnum{2}}}_r(t)\}, \{\bar{\boldsymbol{\Phi}}^{\text{\Rmnum{2}}}_c(t)\})\big)  \label{SDR problem formulation, obj func} \\
		\text{s.t.} & \;\;   \big[\bar{\boldsymbol{\Phi}}^{\text{\Rmnum{2}}}_r(t)\big]_{\!n,n} \! \!=\! 1,   n \!=\! 1, \!\cdots\!, N_p \!+\! 1,  \forall t \!\in\! \mathcal{T}_3,  \label{SDR unit-modulus constant, Phi_r} \\
		& \;\; \big[\bar{\boldsymbol{\Phi}}^{\text{\Rmnum{2}}}_c(t)\big]_{\!n,n}  \!\!=\!\! 1,   n \!=\! 1, \!\cdots\!, N_p \!+\! 1,  \forall t \!\in\! \mathcal{T}_4 \label{SDR unit-modulus constant, Phi_c}, \\
		& \;\; \bar{\boldsymbol{\Phi}}^{\text{\Rmnum{2}}}_r(t) \!\succeq\! \mathbf{0},  \forall t \in \mathcal{T}_3, \; \bar{\boldsymbol{\Phi}}^{\text{\Rmnum{2}}}_c(t) \!\succeq\! \mathbf{0},  \forall t \in \mathcal{T}_4, \label{SDR semidefinite constant}
		\vspace{-0.1cm}
	\end{align}
\end{subequations}
which is a convex semidefinite programming (SDP) problem \cite{Lijian2008waveformopt} and can be solved by existing convex optimization solvers, such as CVX \cite{grant2014cvx}. Then, the Gaussian randomization technique can be employed to obtain a feasible solution for problem \eqref{origin problem formulation}. However, the computational complexity of the SDR method is extremely high, especially when there is  a large number of IRS elements and pilots. To address this issue and inspired by \cite{cheng2021onebitdfrc}, we propose to approximate $\mathbf{C}(\boldsymbol{\Phi}^{\text{\Rmnum{2}}}_{r}, \boldsymbol{\Phi}^{\text{\Rmnum{2}}}_{c})$ by $\hat{\mathbf{C}}(\boldsymbol{\Phi}^{\text{\Rmnum{2}}}_{r}, \boldsymbol{\Phi}^{\text{\Rmnum{2}}}_{c}) = \tilde{\mathbf{J}}_{\boldsymbol{\xi},\boldsymbol{\xi}}^{-1}$, where $\tilde{\mathbf{J}}_{\boldsymbol{\xi},\boldsymbol{\xi}}$ represents a diagonal matrix with elements extracted from the diagonal of $\mathbf{J}_{\boldsymbol{\xi},\boldsymbol{\xi}}$. In this case, the approximate CRB of $\boldsymbol{\xi}_{n}$, i.e., $ [\hat{\mathbf{C}}(\boldsymbol{\Phi}^{\text{\Rmnum{2}}}_{r}, \boldsymbol{\Phi}^{\text{\Rmnum{2}}}_{c})]_{n,n} $, $\forall n \in \tilde{\mathcal{N}}$, can be interpreted as being derived under the condition that the remaining parameters $\boldsymbol{\xi}_{n^\prime}$, $\forall n^\prime \in \tilde{N}\setminus{n}$ are given, or that the estimation of all the parameters is inherently decoupled (i.e., the off-diagonal elements of $\mathbf{J}_{\boldsymbol{\xi},\boldsymbol{\xi}}$ are equal to zeros). 

\par Based on  $\hat{\mathbf{C}}(\boldsymbol{\Phi}^{\text{\Rmnum{2}}}_{r}, \boldsymbol{\Phi}^{\text{\Rmnum{2}}}_{c})$, the original CRB minimization problem, i.e., problem \eqref{origin problem formulation}, can be approximated as 
\begin{align}
	\label{approximate problem formulation}
	\min_{\boldsymbol{\varphi}^{\!\text{\Rmnum{2}}}_{\!r}, \boldsymbol{\!\varphi}^{\text{\Rmnum{2}}}_{\!c}} & \;
	\sum_{n \!=\! 1}^{2\tilde{K}}\! \frac{1}{	\mathbf{J}_{\!\boldsymbol{\xi}^{\!\dot{T}}_n\!,\boldsymbol{\xi}^{\!\dot{T}}_n}} \!+\!
	\sum_{n \!=\! 1}^{2O} \frac{1}{	\mathbf{J}_{\!\boldsymbol{\xi}^{\!O}_n\!,\boldsymbol{\xi}^{\!O}_n}} \!+\!
	\sum_{n \!=\! 1}^{2\tilde{L}} \frac{1}{	\mathbf{J}_{\!\boldsymbol{\xi}^{\!S}_n\!,\boldsymbol{\xi}^{\!S}_n}} \!+\!
	\sum_{n \!=\! 1}^{2} \frac{1}{\mathbf{J}_{\!\boldsymbol{\xi}^{\!U}_n\!,\boldsymbol{\xi}^{\!U}_n}} 
	\\
	\text{s.t.} & \; \eqref{unit-modulus constant, Phi_r}, \eqref{unit-modulus constant, Phi_c}, \nonumber
\end{align}
where $\boldsymbol{\!\varphi}^{\text{\Rmnum{2}}}_{\!r} \!\triangleq\! \text{vec}(\boldsymbol{\Phi}^{\!\text{\Rmnum{2}}}_{\!r})$, $\boldsymbol{\varphi}^{\!\text{\Rmnum{2}}}_{\!c} \!\triangleq\! \text{vec}(\boldsymbol{\Phi}^{\!\text{\Rmnum{2}}}_{\!c})$, $\mathbf{J}_{\!\boldsymbol{\xi}^{\!\dot{T}}_n\!,\boldsymbol{\xi}^{\!\dot{T}}_n} \!\!=\!\! \frac{2}{\sigma^2}\!\big( (\boldsymbol{\varphi}^{\!\text{\Rmnum{2}}}_{r})^{\!T} \! \bar{\mathbf{A}}^{\!n}_{\!\dot{T}} (\boldsymbol{\varphi}^{\!\text{\Rmnum{2}}}_{\!r})^* \\ + (\bar{\mathbf{b}}^n_{\dot{T}})^H(\boldsymbol{\varphi}^{\text{\Rmnum{2}}}_{r})^* + (\boldsymbol{\varphi}^{\text{\Rmnum{2}}}_{r})^T \bar{\mathbf{b}}^n_{\dot{T}}  \big) + C^n_{\dot{T}}$ with $C^n_{\dot{T}}$ being a constant, $\bar{\mathbf{A}}^n_{\dot{T}} = \bar{\mathbf{A}}^n_{CTS,\dot{T}} + \bar{\mathbf{A}}^n_{CTB,\dot{T}}$, $\bar{\mathbf{A}}^n_{CTS,\dot{T}} \!=\! \text{Blkdiag}(\mathbf{A}^n_{CTS,\dot{T}}, \! \cdots\!, \\ \mathbf{A}^n_{CTS,\dot{T}})$, $\mathbf{A}^n_{CTS,\dot{T}} = \sum_{n = 1}^{N_s} \!\frac{\partial [\tilde{\mathbf{H}}_{CTS}\!]_{n,:}^{\!T} \partial [\tilde{\mathbf{H}}_{C\!T\!S}]_{n,:}^*}{\partial \boldsymbol{\xi}^{\dot{T}}_n \partial \boldsymbol{\xi}^{\dot{T}}_n}$, 
$\bar{\mathbf{A}}^n_{CTB,\dot{T}} = \\ \text{Blkdiag}(\!\mathbf{A}^{\!n}_{C\!T\!B\!,\dot{T}},\! \cdots\!, \mathbf{A}^{\!n}_{\!C\!T\!B,\dot{T}}\!)$, $\mathbf{A}^{\!n}_{\!C\!T\!B\!,\dot{T}} \!\!\!=\!\!\!\! \sum_{\!n \!=\! 1}^{\!M}\!\! \frac{\partial [\!\tilde{\mathbf{H}}_{\!C\!T\!B}]_{\!n,:}^{\!T} \!\partial [\!\tilde{\mathbf{H}}_{\!C\!T\!B}\!]_{\!n\!,:}^{\!*}}{\partial \boldsymbol{\xi}^{\dot{T}}_n \partial \boldsymbol{\xi}^{\dot{T}}_n}$,
$\bar{\mathbf{b}}^n_{\dot{T}} = \bar{\mathbf{b}}^n_{CTS,\dot{T}} + \bar{\mathbf{b}}^n_{CTB,\dot{T}}$, $\bar{\mathbf{b}}^n_{CTS,\dot{T}} = [\mathbf{b}^n_{CTS,\dot{T}}; \cdots ; \mathbf{b}^n_{CTS,\dot{T}}]$, $\mathbf{b}^n_{CTS,\dot{T}} = \sum_{n = 1}^{N_s}\!\! \frac{\partial [\tilde{\mathbf{H}}_{CTS}]_{n,:}^T \partial \mathbf{h}_{CTS,n}^*}{\partial \boldsymbol{\xi}^{\dot{T}}_n \partial \boldsymbol{\xi}^{\dot{T}}_n}$, $\bar{\mathbf{b}}^n_{CTB,\dot{T}} = [\mathbf{b}^n_{CTB,\dot{T}};  \!\cdots\! ; \\ \mathbf{b}^n_{CTB,\dot{T}}]$, $\mathbf{b}^n_{CTB,\dot{T}} \!\!=\! \sum_{n = 1}^{M}\!\! \frac{\partial [\tilde{\mathbf{H}}_{CTB}]_{n,:}^T \partial \mathbf{h}_{CTB,n}^*}{\partial \boldsymbol{\xi}^{\dot{T}}_n \partial \boldsymbol{\xi}^{\dot{T}}_n}$, 
$\mathbf{J}_{\boldsymbol{\xi}^{S}_n,\boldsymbol{\xi}^{S}_n} \!=\! \frac{2}{\sigma^2}\big( (\boldsymbol{\varphi}^{\text{\Rmnum{2}}}_{c})^T \bar{\mathbf{A}}^n_{S} (\boldsymbol{\varphi}^{\text{\Rmnum{2}}}_{c})^* + (\bar{\mathbf{b}}^n_{S})^H(\boldsymbol{\varphi}^{\text{\Rmnum{2}}}_{c})^* + (\boldsymbol{\varphi}^{\text{\Rmnum{2}}}_{c})^T \bar{\mathbf{b}}^n_{S}  \big) + C^n_{S}$,
$\mathbf{J}_{\boldsymbol{\xi}^{U}_n,\boldsymbol{\xi}^{U}_n} \!=\! \frac{2}{\sigma^2}\big( (\boldsymbol{\varphi}^{\text{\Rmnum{2}}}_{c})^T \bar{\mathbf{A}}^n_{U} (\boldsymbol{\varphi}^{\text{\Rmnum{2}}}_{c})^* + (\bar{\mathbf{b}}^n_{U})^H(\boldsymbol{\varphi}^{\text{\Rmnum{2}}}_{c})^* + (\boldsymbol{\varphi}^{\text{\Rmnum{2}}}_{c})^T \bar{\mathbf{b}}^n_{U}  \big) + C^n_{U}$, 
$\mathbf{J}_{\boldsymbol{\xi}^{O}_n,\boldsymbol{\xi}^{O}_n} \!=\! \frac{2}{\sigma^2}\big( (\boldsymbol{\varphi}^{\text{\Rmnum{2}}}_{r})^T \bar{\mathbf{A}}^n_{r,O} (\boldsymbol{\varphi}^{\text{\Rmnum{2}}}_{r})^* + (\bar{\mathbf{b}}^n_{r,O})^H(\boldsymbol{\varphi}^{\text{\Rmnum{2}}}_{r})^* + (\boldsymbol{\varphi}^{\text{\Rmnum{2}}}_{r})^T \bar{\mathbf{b}}^n_{r,O} +
(\boldsymbol{\varphi}^{\text{\Rmnum{2}}}_{c})^T \bar{\mathbf{A}}^n_{c,O} (\boldsymbol{\varphi}^{\text{\Rmnum{2}}}_{c})^* + (\bar{\mathbf{b}}^n_{c,O})^H(\boldsymbol{\varphi}^{\text{\Rmnum{2}}}_{c})^* + (\boldsymbol{\varphi}^{\text{\Rmnum{2}}}_{c})^T \bar{\mathbf{b}}^n_{c,O}  \big) + C^n_{c,O}$. The coefficients in 
$\mathbf{J}_{\boldsymbol{\xi}^{S}_n,\boldsymbol{\xi}^{S}_n}$, $\mathbf{J}_{\boldsymbol{\xi}^{U}_n,\boldsymbol{\xi}^{U}_n}$ and $\mathbf{J}_{\boldsymbol{\xi}^{O}_n,\boldsymbol{\xi}^{O}_n}$, such as $\bar{\mathbf{A}}^n_{S}$ and $\bar{\mathbf{A}}^n_{r,O}$, share similar expressions with those in $\mathbf{J}_{\boldsymbol{\xi}^{\dot{T}}_n,\boldsymbol{\xi}^{\dot{T}}_n}$. The details are omitted here due to space limitations.
\vspace{-0.3cm}
\renewcommand\baselinestretch{\linspreadalgr}\selectfont 
\begin{algorithm}[htb]
	\caption{Proposed RCG-based Algorithm for Solving Problem \eqref{approximate problem formulation}}
	\label{Algorithm3, RCG-based Algorithm}
	\textbf{Input}: $\{\bar{\mathbf{A}}^n_{\dot{T}}, \bar{\mathbf{A}}^n_{r,O}, \bar{\mathbf{A}}^n_{c,O}, \bar{\mathbf{A}}^n_{S}, \bar{\mathbf{A}}^n_{U}, \bar{\mathbf{b}}^n_{\dot{T}}, \bar{\mathbf{b}}^n_{r,O}, \bar{\mathbf{b}}^n_{c,O}, \bar{\mathbf{b}}^n_{S}, \bar{\mathbf{b}}^n_{U}\}$. \textbf{Initialize}: Choose a feasible initial point $\boldsymbol{\varphi}^0 \!\in\! \mathcal{M}$. Set $k = 0$ and the threshold $\epsilon_{R} \!>\! 0$. \textbf{Output}: $\boldsymbol{\varphi}^k$.\!\!\!\!\!\!\!\!\!\!\!\!\!\!\!
	\begin{algorithmic}[1]
		\STATE {\textbf{Repeat}}
		\STATE {\; Calculate the Euclidean gradient $\nabla_{\boldsymbol{\varphi}} f(\boldsymbol{\varphi}^k)$ and the \\ \; Riemannian gradient $G_{\mathcal{M}} f(\boldsymbol{\varphi}^k )$.} 
		\STATE {\; Choose the  Fletcher-Reeves parameter $\varrho_R$ and update the \\ \; search direction $\mathbf{d}^{k}$.} 
		\STATE {\; Choose the  Armijo parameter $\varrho_A$ and update $\boldsymbol{\varphi}^k $ accord- \\ \; ing to \eqref{update for phi_k}.} 
		\STATE {\; Apply the retraction operation for $\dot{\boldsymbol{\varphi}}^k $.} 
		\STATE {\; $k \leftarrow k+1$.} 
		\STATE {\textbf{Until} The decrease of the objective value of \eqref{approximate problem formulation} is below $\epsilon_{R}$.}
	\end{algorithmic}
\end{algorithm}
\vspace{-0.3cm}
\renewcommand\baselinestretch{\linspread}\selectfont 

\subsection{Manifold-Based Algorithm for IRS Reflection Design}
Although problem \eqref{approximate problem formulation} is much simplified as compared to problem \eqref{origin problem formulation}, it is still difficult to solve due to the non-convex unit-modulus constraints. Given the fact that  $\boldsymbol{\varphi} \triangleq [\boldsymbol{\varphi}^{\text{\Rmnum{2}}}_{r}; \boldsymbol{\varphi}^{\text{\Rmnum{2}}}_{c}]$ constitutes a complex circle manifold $\mathcal{M} = \{ \boldsymbol{\varphi} \in \mathbb{C}^{N_p(T_3 + T_4)} : |\varphi_n| = 1, n = 1, \cdots, N_p(T_3 + T_4)  \}$, we employ the manifold optimization techniques \cite{absil2009optimization}  and pose a Riemannian conjugate gradient (RCG)-based algorithm to solve problem \eqref{approximate problem formulation}. To be specific, we first define the tangent space of the manifold $\mathcal{M}$ at point $\boldsymbol{\varphi}^k$ as $\mathcal{T}_{\boldsymbol{\varphi}^k}\mathcal{M} = \{ \mathbf{c} \in \mathbb{C}^{N_p(T_3 + T_4)}: \Re\{ \mathbf{c} \circ (\boldsymbol{\varphi}^k)^* \} = \mathbf{0} \}$ with $k$ denoting the iteration index. Then, in order to perform the gradient decent on  $\mathcal{M}$ at point $\boldsymbol{\varphi}^k$, we need to determine the corresponding  Riemannian gradient $G_{\mathcal{M}} f(\boldsymbol{\varphi}^k )$ (i.e., the direction of the steepest ascent within the tangent space $\mathcal{T}_{\boldsymbol{\varphi}^k}\mathcal{M}$), which is given by
\begin{equation}
\label{Riemannian gradient}
G_{\mathcal{M}} f(\boldsymbol{\varphi}^k ) \!=\! \nabla_{\boldsymbol{\varphi}} f(\boldsymbol{\varphi}^k ) \!-\! \Re\{\nabla_{\boldsymbol{\varphi}} f(\boldsymbol{\varphi}^k ) \!\circ\! (\boldsymbol{\varphi}^k)^*\}  \!\circ\! \boldsymbol{\varphi}^k. 
\end{equation}
Therein, $\nabla_{\boldsymbol{\varphi}} f(\boldsymbol{\varphi}^k) = [\nabla_{\boldsymbol{\varphi}^{\text{\Rmnum{2}}}_r} f((\boldsymbol{\varphi}^{\text{\Rmnum{2}}}_r)^k);\nabla_{\boldsymbol{\varphi}^{\text{\Rmnum{2}}}_c} f((\boldsymbol{\varphi}^{\text{\Rmnum{2}}}_c)^k)]$ is the Euclidean gradient with $\nabla_{\boldsymbol{\varphi}^{\text{\Rmnum{2}}}_r} f((\boldsymbol{\varphi}^{\text{\Rmnum{2}}}_r)^k) \!=\! -\frac{2}{\sigma^2} \!\big[\! \sum_{n=1}^{2\tilde{K}}  \!\frac{1}{(\mathbf{J}_{\boldsymbol{\xi}^{\dot{T}}_n,\boldsymbol{\xi}^{\dot{T}}_n})^2} \\ \big(\!(\bar{\mathbf{A}}^n_{\dot{T}})^{\!*} (\boldsymbol{\varphi}^{\text{\Rmnum{2}}}_r)^{\!k} + (\bar{\mathbf{b}}^n_{\dot{T}})^{\!*} \big) \!+\! \sum_{n=1}^{2O} \! \frac{1}{(\mathbf{J}_{\boldsymbol{\xi}^{O}_n,\boldsymbol{\xi}^{O}_n})^2} \big(\!(\bar{\mathbf{A}}^n_{r,O})^{\!*}  (\boldsymbol{\varphi}^{\text{\Rmnum{2}}}_r)^{\!k} \!+\! (\bar{\mathbf{b}}^n_{r,O})^{\!*} \big)  \! \big]  $ and
$\nabla_{\!\boldsymbol{\varphi}^{\!\text{\Rmnum{2}}}_{\!c}} f((\boldsymbol{\varphi}^{\!\text{\Rmnum{2}}}_{\!c})^{\!k}) \! \!=\!\! -\!\frac{2}{\sigma^2}\!\big[\! \sum_{n=1}^{2\tilde{L}}  \!\frac{1}{(\mathbf{J}_{\boldsymbol{\xi}^{S}_n,\boldsymbol{\xi}^{S}_n})^2} \big(\!(\bar{\mathbf{A}}^n_{S})^{\!*}  (\boldsymbol{\varphi}^{\text{\Rmnum{2}}}_c)^{\!k} \\+ (\bar{\mathbf{b}}^n_{S})^{\!*} \big) 
\!+\! \sum_{n=1}^{2O} \!  \frac{1}{(\mathbf{J}_{\boldsymbol{\xi}^{O}_n,\boldsymbol{\xi}^{O}_n})^2} \big((\bar{\mathbf{A}}^n_{c,O})^{\!*}  (\boldsymbol{\varphi}^{\text{\Rmnum{2}}}_c)^{\!k} \!+\! (\bar{\mathbf{b}}^n_{c,O})^{\!*} \big) \!+\!
\sum_{n=1}^{2}\!\frac{1}{(\mathbf{J}_{\boldsymbol{\xi}^{U}_n,\boldsymbol{\xi}^{U}_n})^2}  \big((\bar{\mathbf{A}}^n_{U})^* (\boldsymbol{\varphi}^{\text{\Rmnum{2}}}_c)^k + (\bar{\mathbf{b}}^n_{U})^* \big)    \big]  $.

\par After obtaining the Riemannian gradient, we transplant the conjugate gradient method from the Euclidean space to the Riemannian manifold. Specifically, the search direction $\mathbf{d}^{k}$ at point $\boldsymbol{\varphi}^k$ is given by 
%\begin{equation}
	$\mathbf{d}^{k} =  -G_{\mathcal{M}} f(\boldsymbol{\varphi}^k ) + \varrho_R T_{\boldsymbol{\varphi}^{k}} (\mathbf{d}^{k-1})$,  
%\end{equation}
where $\mathbf{d}^{0} =  -G_{\mathcal{M}} f(\boldsymbol{\varphi}^0 )$, $\varrho_R$ is the iteration parameter selected according to the Fletcher-Reeves principle \cite{absil2009optimization} and $T_{\boldsymbol{\varphi}^{k}} (\mathbf{d}^{k-1}) = \mathbf{d}^{k-1} - \Re\{ \mathbf{d}^{k-1} \circ (\boldsymbol{\varphi}^{k})^*  \} \circ \boldsymbol{\varphi}^{k}$ maps the previous search direction $\mathbf{d}^{k-1}$ to the current tangent space  $\mathcal{T}_{\boldsymbol{\varphi}^k}\mathcal{M}$. Subsequently, $\boldsymbol{\varphi}^k$ is updated by
\begin{equation}
	\label{update for phi_k}
	\dot{\boldsymbol{\varphi}}^k = \boldsymbol{\varphi}^k + \varrho_A \mathbf{d}^{k},
\end{equation}
where $\varrho_A$ is the step size determined by the Armijo rule \cite{bertsekas1997nonlinear}. As $\dot{\boldsymbol{\varphi}}^k$ generally does not belong to  $\mathcal{M}$, it is essential to apply the retraction operation to project $\dot{\boldsymbol{\varphi}}^k$ onto $\mathcal{M}$, i.e., 
$\boldsymbol{\varphi}^{k+1} = \dot{\boldsymbol{\varphi}}^k \circ \frac{1}{|\dot{\boldsymbol{\varphi}}^k|}$.  The proposed RCG-based algorithm for solving problem \eqref{approximate problem formulation} is summarized in Algorithm \ref{Algorithm3, RCG-based Algorithm}, which is guaranteed to converge to a stationary point \cite{absil2009optimization}.

\section{Simulation results}
\label{sec:simulation}
In this section, we provide numerical results to validate the effectiveness of the proposed SI-JSCE scheme. In our simulations, a 2D coordinate system is considered and we  place the self-sensing IRS  with $N_p=192$ reflecting elements and $N_s = 160$ sensors at  $[22.5\, \text{m}, 0.4\, \text{m}]$. The BS is equipped with $M = 160$ antennas and placed at  $[-22.5\, \text{m}, 0.4\, \text{m}]$. We consider a $40\, \text{m} \times 40\, \text{m}$ area for  $\mathcal{R}$ and a $15\, \text{m} \times 15\, \text{m}$ area for  $\mathcal{R}_u$, respectively. The corresponding numbers of grids are set to $Q = 64$ and $P = 9$, respectively. Within  $\mathcal{R}$, there are $K_B = 3$ block targets and $L_B = 4$ block scatterers. Each block target or scatterer occupies two adjacent grids in the vertical direction. Consequently, the total numbers of  individual targets and scatterers are $K = 2K_B = 6$ and $L = 2L_B = 8$, respectively. Besides, we define the overlapping ratio as $\gamma_{o} = \frac{O}{K+L-O}$, and  the number of overlapping individual targets and scatterers is set to $O = 4$, resulting in an overlapping ratio of $\gamma_{o} = 0.4$. Without loss of generality, the other system parameters are set as follows unless otherwise specified: $P_c = P_u= P_T$, carrier frequency  28 GHz, $\sigma^2 = -100$ dBm, $T_1 = T_2 = T_3 = T_4 = 2$. The scanning-based IRS reflection coefficients employed in phase \Rmnum{1} are generated according to \cite[Section \Rmnum{3}-C]{xiao2016codebook}, where we set the codebook layers $k_r$ and $k_c$ for the  design of $\boldsymbol{\Phi}^{\text{\Rmnum{1}}}_r$ and $\boldsymbol{\Phi}^{\text{\Rmnum{1}}}_c$ to  $\log_2(T_1)$ and $\log_2(T_2)$, respectively. Moreover, the beam coverage is limited to $\pi/2$. The large-scale path losses of the communication channels are modeled as in \cite{chen2023jsdmIRS}. For comparison, we consider the following benchmark algorithms:
\begin{itemize}
	\item A two-phase orthogonal matching pursuit (TP-OMP) algorithm with optimized IRS reflection coefficients: this algorithm is similar to the proposed AS-TVBI algorithm, except that the estimation algorithm in the E step is replaced by the OMP algorithm.
	\item A TP-SBL algorithm which replaces the OMP algorithm in TP-OMP by the sparse Bayesian learning (SBL) algorithm.
%	A two-phase sparse Bayesian learning (SBL) algorithm with optimized IRS reflection coefficients (namely the TP-SBL algorithm): this algorithm is similar to the TP-OMP algorithm, but with the OMP algorithm replaced by the SBL algorithm, which also fails to exploit the structured sparsity of the sparse SAC channels.
	\item A single-phase AS-TVBI algorithm (SP-TVBI) with scanning-based IRS reflection coefficients: this algorithm is obtained by removing phase \Rmnum{2} in the proposed AS-TVBI algorithm, but allocating  more pilots in phase \Rmnum{1}, i.e., $T_1 = T_2 = 4$.
	\item A genie-aided AS-TVBI algorithm, where the true values of the target/scatterer/user positions and SAC channel coefficients are available  for IRS reflection design.
\end{itemize}
	\vspace{-0.0cm}
\begin{figure}[htbp]
	\setlength{\abovecaptionskip}{-0.1cm}
	\centering
	\includegraphics[width=0.3\textwidth]{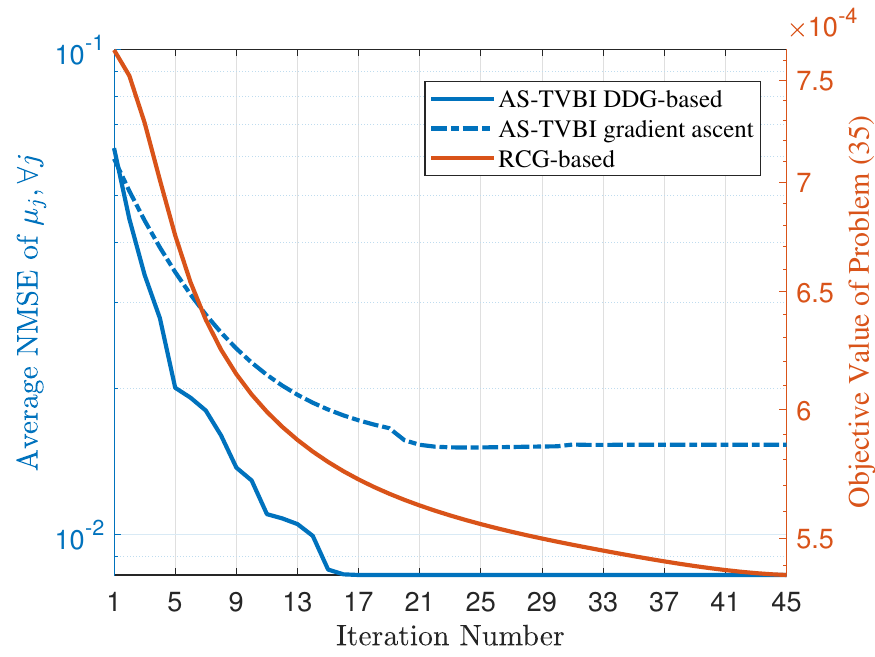}
	\caption{Convergence behavior of the proposed AS-TVBI and RCG-based algorithms.}
	\label{pic:convergence}	
	\vspace{-0.5cm}
\end{figure}

\par Prior to performance comparison, we first  examine the convergence behavior of the proposed AS-TVBI and RCG-based  algorithms in Fig. \ref{pic:convergence}. As can be seen, the proposed AS-TVBI algorithm using the DDG-based method achieves convergence in about 15 iterations, indicating that the obtained approximate marginal posteriors $\psi(\mathbf{w}), \forall \mathbf{w} \in \Omega$ are sufficiently accurate and have minimal impact on the convergence process. Moreover, we can also observe that using the DDG-based method can achieve better performance than the gradient ascent method in our AS-TVBI algorithm, which verifies the superiority of the DDG-based method.  Besides, the proposed RCG-based algorithm is able to converge in about 40 iterations.

\subsubsection{Impact of the Transmit Power, $P_T$} 
\begin{figure}[htbp]
	\setlength{\abovecaptionskip}{-0.1cm}
	\centering
	\subfigure{
		\centering
		\includegraphics[width=0.2\textwidth]{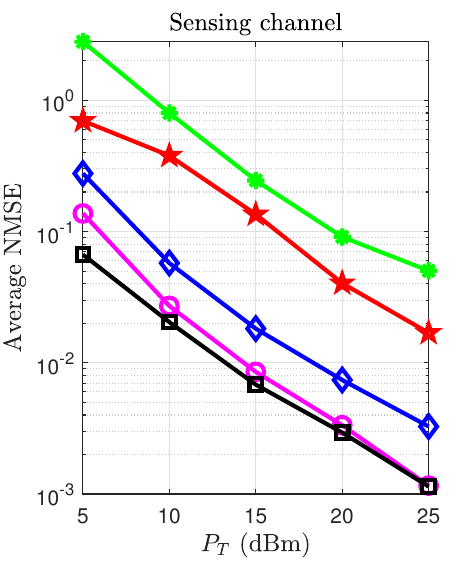}}
	\subfigure{
		\centering
		\includegraphics[width=0.2\textwidth]{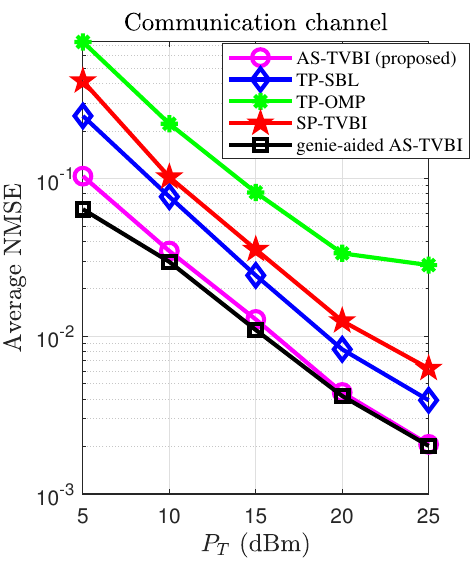}}

	\caption{Average NMSE of SAC channel estimation versus the transmit power, $P_T$.}
	\label{pic:NMSE_versus_PT}	
	\vspace{-0.6cm}
\end{figure}

In Fig. \ref{pic:NMSE_versus_PT}, we investigate the average normalized mean squared error (NMSE) performance of SAC channel estimation achieved by the considered algorithms versus the transmit power $P_T$. It is observed that the proposed two-phase AS-TVBI algorithm with optimized IRS reflection coefficients outperforms the TP-OMP and TP-SBL algorithms for both  channels. This is mainly because the proposed algorithm adopts the turbo framework, where  the POS sparsity and 2D block sparsity of the SAC channels can be effectively utilized  through the skillfully designed support vectors $\mathbf{s}$ and $\mathbf{s}_{U}$.
Moreover, we observe that the NMSE performance of the proposed algorithm  deteriorates when only one-phase estimation is considered. The reason is that the simple scanning-based IRS reflection design fails to  provide sufficient beamforming gain given  limited sensing/CE pilots. 
Furthermore, we notice that the NMSE performance achieved by  the genie-aided AS-TVBI algorithm is only slightly better than the proposed algorithm, and the performance gap diminishes as the transmit power $P_T$ increases. This observation validates the effectiveness of the two-phase estimation approach in the proposed SI-JSCE scheme, where the estimation in phase \Rmnum{2} is based on the coarse results obtained in phase \Rmnum{1}.

\par
\begin{figure}[htbp]
		\setlength{\abovecaptionskip}{-0.1cm}
		\centering
		{\includegraphics[width=0.3\textwidth]{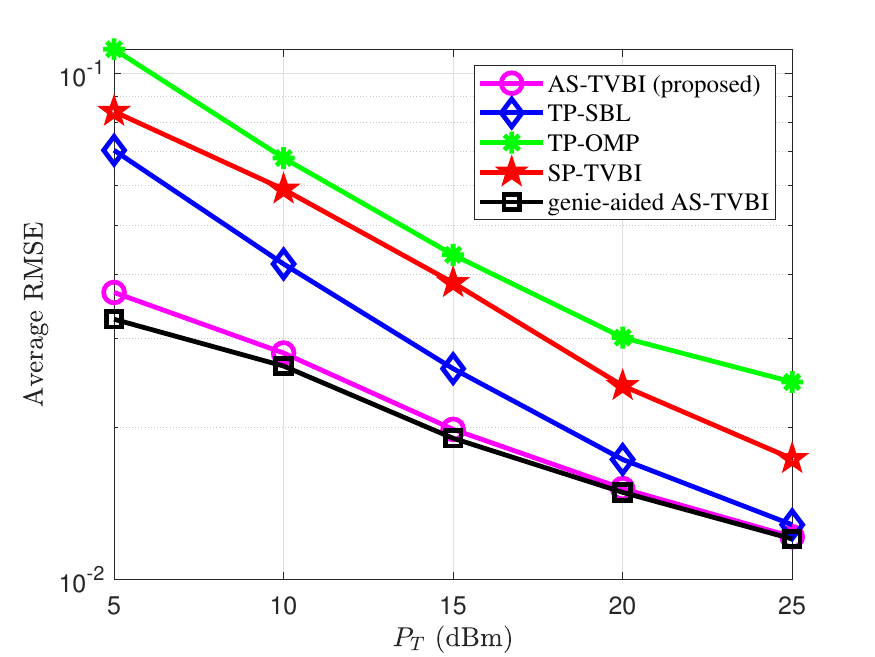}}
		\caption{Average RMSE of target/scatterer/user location sensing versus the transmit power, $P_T$.}
		\label{pic:RMSE_versus_PT}
		\vspace{-0.3cm}
\end{figure}

 Next, in Fig. \ref{pic:RMSE_versus_PT}, we plot the average root mean squared error (RMSE) performance  of target/scatterer/user location  sensing versus the transmit power $P_T$. As expected, the proposed AS-TVBI algorithm exhibits superior location sensing performance. This is mainly due to its ability to provide more accurate approximate marginal posteriors $\psi(\mathbf{w}), \forall \mathbf{w} \in \Omega$ to the M step compared to the other algorithms, enabling a  more precise estimation of the position offset $\{ \Delta \mathbf{r}, \Delta \mathbf{z}\}$. The gap between the performance of TP-SBL and AS-TVBI expeditiously increases when $P_T$ decreases, because SBL cannot provide an accurate estimation in the E step when $P_T$ is small due to its failure in capturing the structured sparsity inherent in the SAC channels. This observation also reveals the efficiency of AS-TVBI in the small $P_T$ regime.

%\begin{figure}[tbp]
%	\setlength{\abovecaptionskip}{-0.4cm}
%	\centering
%	\includegraphics[width=0.5\textwidth]{NMSE_versus_PT.eps}
%	\caption{Average NMSE of SAC channels estimation versus the transmit power, $P_T$.}
%	\label{pic:NMSE_versus_PT}	
%	\vspace{-0.5cm}
%\end{figure}

\vspace{-0.0cm}
\begin{figure}[htbp]
	\setlength{\abovecaptionskip}{-0.1cm}
	\centering
	\subfigure{
		\centering
		\includegraphics[width=0.2\textwidth]{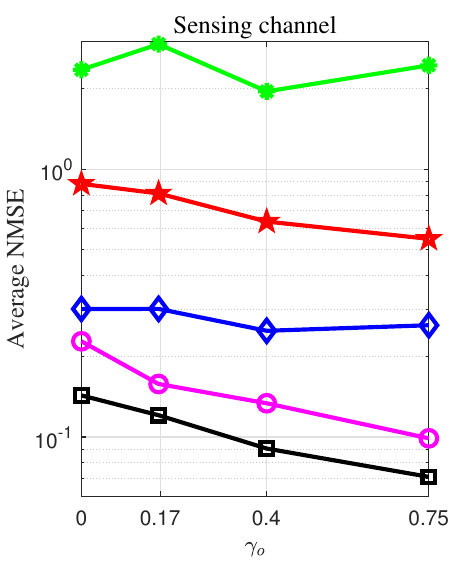}}
	\subfigure{
		\centering
		\includegraphics[width=0.2\textwidth]{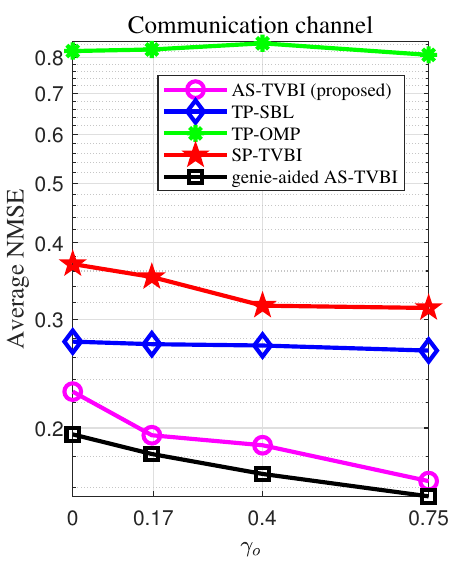}}
	
	\caption{Average NMSE of SAC channel estimation versus the overlapping ratio, $\gamma_{o}$.}
	\label{pic:NMSE_versus_ovlratio}	
	\vspace{-0.5cm}
\end{figure}

\subsubsection{Impact of the Overlapping Ratio, $\gamma_o$}
\par In Fig. \ref{pic:NMSE_versus_ovlratio}, we investigate the average NMSE performance of SAC channel estimation versus the overlapping ratio $\gamma_o$, under $P_T=5$ dBm. 
%It is noteworthy that in our simulations, to vary the overlapping ratio, we first generate the scatterer positions and then select a portion of them to also serve as the target positions.  
One can observe from Fig. \ref{pic:NMSE_versus_ovlratio} that the NMSE performance of the SP-TVBI, AS-TVBI and genie-aided AS-TVBI algorithms improves with the increasing of $\gamma_o$, while this trend does not hold for the other benchmark algorithms. This is because the aforementioned three algorithms can effectively exploit the POS sparsity of the SAC channels, and a larger overlapping ratio results in stronger mutual enhancement. Besides, we can see that  the NMSE performance of  sensing channel estimation achieved by the TP-OMP and TP-SBL algorithms fluctuates as $\gamma_{o}$ changes. This behavior arises because  the target positions are not the same for different values of $\gamma_o$, and  the large-scale channel parameters, such as the large-scale path loss and the RCS, are dynamically changing with varying $\gamma_o$.

%\begin{figure}[tbp]
%	\setlength{\abovecaptionskip}{-0.4cm}
%	\centering
%	\includegraphics[width=0.5\textwidth]{NMSE_versus_ovlratio.eps}
%	\caption{Average NMSE of SAC channels estimation versus the overlapping ratio, $\gamma_{o}$.}
%	\label{pic:NMSE_versus_ovlratio}	
%	\vspace{-0.55cm}
%\end{figure}

\subsubsection{Impact of the Number of Reflecting Elements, $N_p$}
\begin{figure}[htbp]
		\setlength{\abovecaptionskip}{-0.1cm}
		\centering
		{\includegraphics[width=0.3\textwidth]{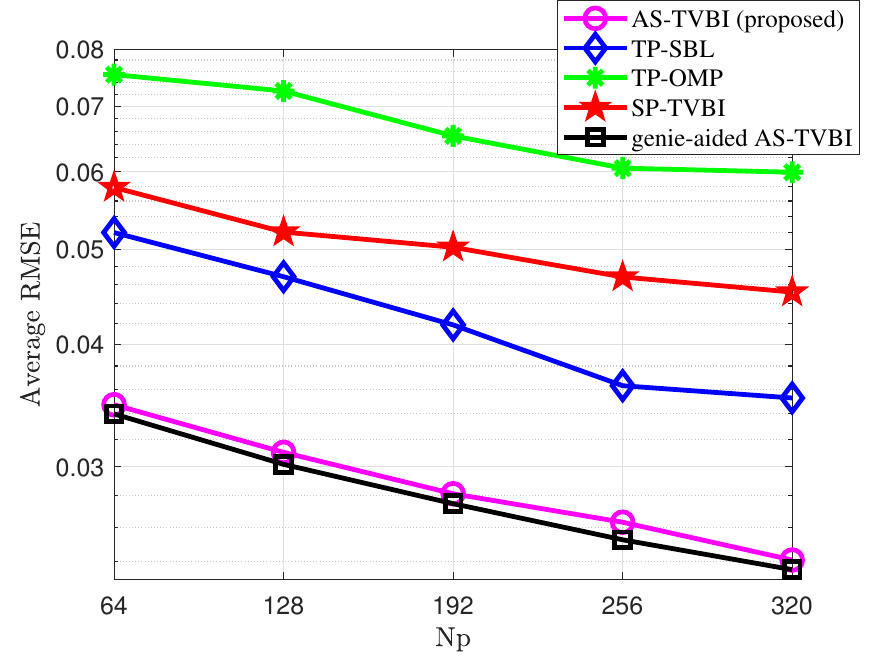}}
		\caption{Average RMSE of target/scatterer/user location sensing versus the number of reflecting elements, $N_p$.}
		\label{pic:RMSE_versus_Np}
		\vspace{-0.4cm}
\end{figure}

 Finally, we plot in Fig. \ref{pic:RMSE_versus_Np} the average RMSE performance of target/scatterer/user location sensing versus the number of reflecting elements $N_p$, under  $P_T = 10$ dBm. It can be observed that  the RMSE performance of all the considered algorithms improves as $N_p$ increases. The reason behind this phenomenon is twofold. First, in phase \Rmnum{1}, a larger number of reflecting elements enables higher beamforming gain for all the sensing/CE pilots, leading to  more precise estimation results for phase \Rmnum{2}. Second, in phase \Rmnum{2}, with an increased number of reflecting elements, more high-gain IRS beams can be leveraged to improve the received SNR for both sensing and communication  using the proposed RCG-based algorithm.
% \vspace{-0.2cm}
% \begin{figure}[htbp]
% 	\centering
% 	\begin{minipage}{.24\textwidth}
% 		\setlength{\abovecaptionskip}{-0.1cm}
% 		\centering
% 		{\includegraphics[width=0.93\textwidth]{RMSE_versus_PT.eps}}
% 		\caption{Average RMSE of target/scatterer/user location sensing versus the transmit power, $P_T$.}
% 		\label{pic:RMSE_versus_PT}
% 	\end{minipage}
% 	\begin{minipage}{.24\textwidth}
% 		\setlength{\abovecaptionskip}{-0.1cm}
% 		\centering
% 		{\includegraphics[width=0.9\textwidth]{RMSE_versus_Np.eps}}
% 		\caption{Average RMSE of target/scatterer/user location sensing versus the number of reflecting elements, $N_p$.}
% 		\label{pic:RMSE_versus_Np}
% 	\end{minipage}
% 	\vspace{-0.6cm}
% \end{figure}
\vspace{-0.3cm}
 
\vspace{-0.0cm}
\section{Conclusions}
\label{sec:conclusion}
In this paper, we studied the joint location sensing and CE problem in  a self-sensing IRS aided mmWave ISAC system. To address this problem, we first proposed a two-phase SI-JSCE scheme, where refined sensing/CE results can be obtained in phase \Rmnum{2} using the coarse  estimation results from phase \Rmnum{1}. Then, a novel AS-TVBI algorithm was designed to simultaneously obtain the approximate marginal posteriors of the SAC channels and the ML estimates of the position offsets. Since the proposed algorithm is able to exploit  the POS sparsity and 2D block sparsity inherent in the SAC channels, the location sensing and CE performance can be further enhanced. In order to obtain the optimized IRS reflection coefficients for phase \Rmnum{2}, we  formulated a CRB minimization problem which can be efficiently solved  using the proposed  RCG-based algorithm.   Simulation results validated  the effectiveness of the proposed transmission scheme and algorithms.

\vspace{-0.3cm}

\begin{appendix}
	\subsection{Derivation of the Update Equations for  $\psi(\mathbf{x})$, $\psi(\bm{\rho})$   and $\psi(\mathbf{s})$}	
	\label{Appendix: Derivation of the VBI update expression}	
	According to \eqref{stationary solution for KLD min}, $\psi(\mathbf{x}_{\text{ITS}}) = \mathcal{CN}(\mathbf{x}_{\text{ITS}}; \boldsymbol{\mu}_{\text{ITS}},\boldsymbol{\Sigma}_{\text{ITS}})$ in \eqref{update for x} can be derived as
%	\begin{equation}
%	\begin{aligned}
	\begin{align}
		&\ln \! \psi(\mathbf{x}_{\text{ITS}}) \!\!\propto\!\! \langle \ln p(\mathbf{y}_{\!s,r}|\mathbf{x}_{\text{ITS}},\mathbf{x}_{\text{CTS}};\Delta\mathbf{r},\Delta\mathbf{z})  \rangle_{\psi(\mathbf{x}_{\text{CTS}})} \!+\! \langle \ln p(\mathbf{x}_{\text{ITS}}| \nonumber \\ &\boldsymbol{\rho}_{\text{ITS}})  \rangle_{\psi(\boldsymbol{\rho}_{\text{ITS}})}  \!\propto \! \!-\!\frac{1}{\sigma^2} \!\big\langle\! \|\! \mathbf{y}_{\!s,r} \!\!-\!\! \mathbf{F}_{\text{\!ITS}}(\!\Delta\mathbf{r}\!)\mathbf{x}_{\text{ITS}} \!\!-\!  \mathbf{F}_{\!\text{CTS}}(\!\Delta\mathbf{r}\!)\mathbf{x}_{\text{CTS}}\! \|^{\!2}  \big\rangle_{\!\psi(\!\mathbf{x}_{\text{CTS}})}  \nonumber \\
		& \!\!-\! \!
		\mathbf{x}_{\text{ITS}}^{\!H} \text{diag}^{\!-\!1}\!\big(\!\langle\boldsymbol{\rho}_{\text{ITS}}\rangle_{\!\psi(\!\boldsymbol{\rho}_{\!\text{ITS}})}\!\big)\!\mathbf{x}_{\text{ITS}} \! \!\propto\!\! -(\mathbf{x}_{\text{ITS}} \!\!-\!\! \boldsymbol{\mu}_{\text{ITS}})^{\!H}\! \boldsymbol{\Sigma}_{\text{ITS}}^{\!-\!\!1}(\mathbf{x}_{\!\text{ITS}} \!\!-\!\! \boldsymbol{\mu}_{\text{ITS}}).
	\end{align}
%	\end{aligned}
%	\end{equation}
	Following a similar derivation, we can obtain the remaining update equations for $\psi(\mathbf{x}_j)$. 
	$\psi(\boldsymbol{\rho}_j) = \prod_{q \in \mathcal{Q}}   \Gamma(\rho_{j,q};\tilde{a}_{j,q},\tilde{b}_{j,q})$, $\forall j \in \mathcal{J}_{1}$ in \eqref{update for rho} can be derived as
	\begin{equation}
	\begin{aligned}
		&\ln \psi(\boldsymbol{\rho}_j) \propto \langle \ln p(\mathbf{x}_j|\boldsymbol{\rho}_j)  \rangle_{\psi(\mathbf{x}_j)} + \langle \ln p(\boldsymbol{\rho}_j|\mathbf{s}_T)  \rangle_{\psi(\mathbf{s}_T)} \\
		& \propto \sum_{q \in \mathcal{Q} } \!
		\big[\tilde{\pi}_{T,q}a_{j,q} \!+\! (1\!-\!\tilde{\pi}_{T,q})\bar{a}_{j,q}\big]\ln \rho_{j,q}\!-\! \big[\tilde{\pi}_{T,q}b_{j,q}\! \\ 
		&+  \!(1\!-\!\tilde{\pi}_{T,q})\bar{b}_{j,q} \!+\! \langle|x_{j,q}|^2\rangle_{\psi(x_{j,q})} \big] \rho_{j,q}, 
	\end{aligned}
	\end{equation}
	where $\langle|x_{j,q}|^2\rangle_{\psi(x_{j,q})} = |\mu_{j,q}|^2 + \Sigma_{j,q}$. Similarly, we can obtain the other update equations for $\psi(\boldsymbol{\rho}_j)$, $\forall j \in \mathcal{J}_2 \cup \mathcal{J}_3$. 
	$\psi(\mathbf{s}_T) = \prod_{q\in \mathcal{Q}} (\tilde{\pi}_{T,q})^{\frac{1+s_{T,q}}{2}}(1-\tilde{\pi}_{T,q})^{\frac{1-s_{T,q}}{2}}$ in \eqref{update for s} can be derived as
	\vspace{-0.1cm}
	\begin{align}
		&\ln \psi(\mathbf{s}_T)  \propto \sum_{j \in \mathcal{J}_1 }  \langle \ln p(\boldsymbol{\rho}_j | \mathbf{s}_T)  \rangle_{\psi(\boldsymbol{\rho}_j)} + \ln \hat{p}(\mathbf{s}_T)  \propto \sum_{j \in \mathcal{J}_1 }  \sum_{q \in \mathcal{Q} }  \nonumber  \\
		&
		\frac{1 \!\!+\!\! s_{T,q}}{2}\Big[\!\ln \frac{b_{j,q}^{a_{j,q}}}{\Gamma(a_{\!j,q})} \!+\! (a_{\!j,q} \!-\! 1) \langle \ln \rho_{j,q}  \rangle \!-\! b_{\!j,q}\langle \rho_{j,q}  \rangle \!+\! \ln \pi_{T,q}  \Big] \!+ \nonumber  \\
		&  	\frac{1 \!\!-\!\! s_{T,q}}{2}\Big[\!\ln \frac{\bar{b}_{\!j,q}^{\bar{a}_{\!j,q}}}{\Gamma(\bar{a}_{\!j\!,q})} \!+\! (\bar{a}_{\!j,q} \!-\! 1) \langle \ln \rho_{\!j,q}  \rangle \!-\! \bar{b}_{\!j,q}\langle \rho_{\!j,q}  \rangle \!+\! \ln (1\!\!-\!\pi_{\!T\!,q})  \Big] \nonumber  \\
		& \propto \ln \prod_{q \in \mathcal{Q}} (\tilde{\pi}_{T,q})^{\frac{1+s_{T,q}}{2}}(1-\tilde{\pi}_{T,q})^{\frac{1-s_{T,q}}{2}}.
	\end{align}
	The update equations for $\psi(\mathbf{s}_{NL})$ and $\psi(\mathbf{s}_{L})$ can be derived in a similar manner.
	
	\vspace{-0.3cm}
	\subsection{Message Update Equations for Module B}	
	\label{Appendix: message passing over the support factor graph}	
	\subsubsection{Message passing over the path $g_{T,q}^B \to s_{T,q} \to f_{T,q} \to s_{U,q}$}
	The factor node $g_{T,q}^B$ represents the output message from Module A, which is given by $g_{T,q}^B = \psi(s_{T,q})/g_{T,q}^A(s_{T,q}) \propto \pi_{T,q}^B \delta(s_{T,q}\!-\!1) \!+\! (1\!-\!\pi_{T,q}^B) \delta(s_{T,q}\!+\!1)$, where
	$\pi_{T,q}^B = \frac{\tilde{\pi}_{T,q}(1 - \gamma_{T,q})}{\tilde{\pi}_{T,q}(1 - \gamma_{T,q}) + (1-\tilde{\pi}_{T,q}) \gamma_{T,q}}$ with $\gamma_{T,q}$ being the probability of $s_{T,q} = 1$ in the output message $g_{T,q}^A(s_{T,q})$ from Module B (detailed expression of $\gamma_{T,q}$ will be specified later). The messages from $g_{T,q}^B$ to $s_{T,q}$ (denoted by $\nu_{g_{T,q}^B \to s_{T,q}}(s_{T,q})$)  and from $s_{T,q}$ to $f_{T,q}$ (denoted by $\nu_{ s_{T,q} \to f_{T,q} }(s_{T,q})$) are the same as $g_{T,q}^B$. Then the message from $f_{T,q}$ to $s_{U,q}$ is given by 
	$\nu_{  f_{T,q} \to s_{U,q} }(s_{U,q}) \propto \sum_{s_{T,q}} \nu_{ s_{T,q} \to f_{T,q} }(s_{T,q}) \times f_{T,q}(s_{T,q}, s_{U,q}) \propto \pi_{T,q}^{in} \delta(s_{U,q}\!-\!1) \!+\! (1\!-\!\pi_{T,q}^{in}) \delta(s_{U,q}\!+\!1)$, where $\pi_{T,q}^{in} = \frac{\pi_{T,q}^B p_T + (1-\pi_{T,q}^B)(1-p_T)}{\pi_{T,q}^B p_T + (1-\pi_{T,q}^B)(1-p_T) + (1 - \pi_{T,q}^B)}$.
	\subsubsection{Message passing over the path $g_{N\!L,q}^B \!\!\to\!\! s_{N\!L,q} \!\!\to\!\! f_{N\!L,q} \!\!\to\!\! s_{U,q}$} Following a similar derivation, we can obtain $g_{N\!L,q}^B  \propto \pi_{N\!L,q}^B \delta(s_{N\!L,q}\!-\!1) \!+\! (1\!-\!\pi_{N\!L,q}^B) \delta(s_{N\!L,q}\!+\!1)$ and $\nu_{  f_{N\!L,q} \to s_{U,q} }(s_{U,q}) \propto  \pi_{N\!L,q}^{in} \delta(s_{U,q}\!-\!1) \!+\! (1\!-\!\pi_{N\!L,q}^{in}) \delta(s_{U,q}\!+\!1)$, where	$\pi_{N\!L,q}^B = \frac{\tilde{\pi}_{N\!L,q}(1 - \gamma_{N\!L,q})}{\tilde{\pi}_{N\!L,q}(1 - \gamma_{N\!L,q}) + (1-\tilde{\pi}_{N\!L,q}) \gamma_{N\!L,q}}$ and $\pi_{N\!L,q}^{in} = \frac{\pi_{N\!L,q}^B p_{N\!L} + (1-\pi_{N\!L,q}^B)(1-p_{N\!L})}{\pi_{N\!L,q}^B p_{N\!L} + (1-\pi_{N\!L,q}^B)(1-p_{N\!L}) + (1 - \pi_{N\!L,q}^B)}$ with $\gamma_{N\!L,q}$ defined similar to $\gamma_{T,q}$.
	\subsubsection{Message passing over the MRF of $\mathbf{s}_U$} To facilitate the subsequent calculation, we define a fusion message of $\nu_{  f_{T,q} \to s_{U,q} }(s_{U,q})$ and $\nu_{  f_{N\!L,q} \to s_{U,q} }(s_{U,q})$ as $\nu_{  f_{q} \to s_{U,q} }(s_{U,q}) \propto \nu_{  f_{T,q} \to s_{U,q} }(s_{U,q}) \times \nu_{  f_{N\!L,q} \to s_{U,q} }(s_{U,q}) \propto \pi_{q}^{in} \delta(s_{U,q}\!-\!1) \!+\! (1\!-\!\pi_{q}^{in}) \delta(s_{U,q}\!+\!1)$, where $ \pi_{q}^{in} = \frac{\pi_{T,q}^{in} \pi_{N\!L,q}^{in} }{\pi_{T,q}^{in} \pi_{N\!L,q}^{in} + (1 - \pi_{T,q}^{in}) (1 - \pi_{N\!L,q}^{in} )}$. Considering the factor graph of the MRF shown in Fig. \ref{pic:facgph_MRF}, we define $s_{q_l} = s_{U,q-Qy}$, $s_{q_r} = s_{U,q+Qy}$, $s_{q_t} = s_{U,q-1}$ and $s_{q_b} = s_{U,q+1}$ to denote the neighboring variable nodes of $s_{U,q}$ for clarity. Then, the input message of $s_{U,q}$ from the left is given by $\nu_{q}^l \!\propto\! \sum_{s_{q_l}} \nu_{f_{q_l}\to s_{q_l}}(\!s_{q_l}) \!\prod_{k \in \{l,t,b\}}\nu_{q_l}^k \omega(s_{q_l})\varpi(s_{U,q}, \\ s_{q_l}) \propto \lambda_{q}^l \delta(s_{U,q}-1) + (1-\lambda_{q}^l) \delta(s_{U,q}+1)$, where 
	\begin{equation}
		\lambda_{q}^l  \!\!=\!\! \frac{\pi_{\!q_l}^{\!in} \!\prod_{k \!\in\! \{l\!,t\!,b\}}\lambda_{\!q_l}^{\!k} e^{\!-\alpha \!+\! \beta} \!\!+\! \! (1\!-\!\pi_{\!q_l}^{\!in}) \! \prod_{k \!\in\! \{l\!,t\!,b\}}(1\!-\!\lambda_{\!q_l}^{\!k}) e^{\alpha \!-\! \beta}  }{ (e^{\!\beta} \!\!+\!\! e^{\! -\!\! \beta})(\pi_{\!q_l}^{\!in} e^{\!-\!\alpha} \!\prod_{k \!\in\! \{l\!,t\!,b\}}\!\lambda_{\!q_l}^{\!k}  \!\!+\!\! (1\!\!-\!\!\pi_{\!q_l}^{\!in}) e^{\!\alpha}\! \prod_{k \!\in\! \{l\!,t\!,b\}}(1\!\!-\!\!\lambda_{\!q_l}^{\!k}) ) }.
	\end{equation}
	The input messages of $s_{U,q}$ from the top, right and bottom, denoted by $\nu_{q}^t$, $\nu_{q}^r$ and $\nu_{q}^b$, can be obtained using a similar expression as $\nu_{q}^l$.
	
	\subsubsection{Message passing over the path $s_{U,q} \to  f_{T,q} \to s_{T,q}$}
	\label{subsucsec: MP over sUq-fTq-sTq}
	The message from $s_{U,q}$ to $f_{T,q}$ can be calculated as $\nu_{s_{U,q} \to f_{T,q}}(s_{U,q}) \propto \prod_{k \in \{l,r,t,b\}}\nu_{q}^k \omega(s_{U,q}) \nu_{f_{NL,q}\!\to\! s_{U,q}}(s_{U,q}) \!\propto\!\!\!\!\! \\ \pi_{T,q}^{out} \delta(s_{U,q} - 1) + (1-\pi_{T,q}^{out}) \delta(s_{U,q} + 1)$, where
	\begin{equation}
		\!\!\!\!\!\!\pi_{\!T\!,q}^{\!out} \!\!=\!\! \frac{\pi_{NL,q}^{in} e^{-\alpha} \prod_{k \in \{l,r,t,b\}}\lambda_{q}^k }{\pi_{\!N\!\!L\!,q}^{\!in} e^{\!-\!\alpha} \!\!\prod_{\!k \!\in\! \{l\!,r\!,t,b\}}\!\lambda_{\!q}^{\!k}  \!\!+\!\! (1\!\!-\!\!\pi_{\!N\!\!L\!,q}^{in}) e^{\!\alpha}\! \!\prod_{\!k \!\in\! \{l\!,r\!,t\!,b\}}(1\!\!-\!\!\lambda_{\!q}^{\!k}) }. \!\!
	\end{equation}
	The message from $f_{T,q}$ to $s_{T,q}$ is given by $\nu_{f_{T,q} \to s_{T,q}}(s_{T,q}) \propto \sum_{s_{U,q}} \nu_{s_{U,q}\to f_{T,q}}(s_{U,q}) \times f_{T,q}(s_{T,q}, s_{U,q}) \propto \gamma_{T,q} \delta(s_{T,q}-1) + (1-\gamma_{T,q})\delta(s_{T,q}+1)$, where $\gamma_{T,q} = \pi_{T,q}^{out}p_T$. 
	The messages corresponding to the path $s_{U,q} \to  f_{N\!L,q} \to s_{N\!L,q}$ have similar expressions. 
	\vspace{-0.4cm}
	
	\subsection{Derivation of the Gradients $\mathbf{g}_{\text{BS}}^r$, $\mathbf{g}_{\text{IRS}}^r$, $\mathbf{g}_{\text{BS}}^z$ and $\mathbf{g}_{\text{IRS}}^z$}	
	\label{Appendix: gradient Mstep}
	Let $\Delta \mathbf{r}_q = [\Delta r_q^x,\Delta r_q^y]$, $\forall q \in \mathcal{Q}$, where $\Delta r_q^x$ and $\Delta r_q^y$ denote the x-axis and y-axis position offsets of $\mathbf{r}_q = [r_q^x, r_q^y]$, respectively. Then the partial derivative $\frac{\partial Q(\!\Delta \mathbf{r}\!, \Delta \mathbf{z};\Delta \mathbf{r}^{\!n}, \Delta \mathbf{z}^{n}) }{ \partial \Delta r_q^x} $ can be expressed as $ \frac{\partial Q(\!\Delta \mathbf{r}\!, \Delta \mathbf{z};\Delta \mathbf{r}^{\!n}, \Delta \mathbf{z}^{n}) }{ \partial \Delta r_q^x} = \mathbf{g}_{\text{BS},q}^{r,x} + \mathbf{g}_{\text{IRS},q}^{r,x} $. Therein, $\mathbf{g}_{\text{IRS},q}^{r,x}$ is given by
	$\mathbf{g}_{\text{IRS},q}^{r,x} = 2\Re\bigg\{ \sum_{t=1}^{T_s} \Big[ \tilde{\boldsymbol{\varphi}}_{r}^H\!(t) \Big( \mathbf{a}_{IS}^{\prime}(\theta_{I,q}^r)\! \sum_{k = 1}^Q (\mu_{\text{ITS},q}^* \mu_{\text{ITS},k} + [\boldsymbol{\Sigma}_{\text{ITS}}^*]_{q,k})\mathbf{a}_S(\theta_{I,k}^r)\mathbf{a}_I^H(\theta_{I,k}^r) +  \mathbf{a}_I^\prime(\theta_{I,q}^r) 
	\mathbf{a}_B^H(\theta_{B,q}^r)  \sum_{k = 1}^Q (\mu_{\text{ITB},q}^*\\ \mu_{\text{ITB},k} + [\boldsymbol{\Sigma}_{\text{ITB}}^*]_{q,k})\mathbf{a}_B\!(\theta_{B,k}^r)\mathbf{a}_I^H(\theta_{I,k}^r)     \Big) \tilde{\boldsymbol{\varphi}}_{r}(t) \!+\! 
	\tilde{\boldsymbol{\varphi}}_{r}^H(t)\! 
	\Big(\! \mathbf{a}_{IS}^{\prime}(\theta_{\!I\!,q}^r)   \\
	\Big( \sum_{k\!=\!1}^Q \!\mu_{\text{ITS},q}^*  
	 \mu_{\text{CTS},k} \mathbf{a}_S(\theta_{I\!,k}^r) \!- \! \mu_{\text{ITS},q}^*\mathbf{y}_{s,r}(t)\Big) \!\!+\! \mathbf{a}_I^\prime(\theta_{\!I,q}^r)\mathbf{a}_B^H(\theta_{\!B,q}^r) \\   
	\Big( \sum_{k=1}^Q \!\mu_{\text{ITB},q}^* \mu_{\text{CTB},k} \mathbf{a}_B(\theta_{B\!,k}^r) \!-\! \mu_{\text{ITB},q}^*\mathbf{y}_{B,r}(t)\Big) \!+\!  
	 \sum_{j=1}^Q  \mu_{\text{ITS},j}^* \\
	\mu_{\text{CTS},q} \mathbf{a}_I(\theta_{I\!,j}^r)  \mathbf{a}_S^H(\theta_{\!I,j}^r) \mathbf{a}_S^\prime(\theta_{\!I,q}^r)\! \Big) 
	\!+ \! (\mathbf{a}_S^\prime(\theta_{I,q}^r))^H    \Big( \sum_{k\!=\!1}^Q \big( \mu_{\text{CTS},q}^* \\
	 \mu_{\text{CTS},k} + [\boldsymbol{\Sigma}_{\text{CTS}}^*]_{q,k} \big) 
	\mathbf{a}_S(\theta_{I,k}^r) -   \mu_{\text{CTS},q}^*\mathbf{y}_{s,r}(t) \Big) \Big] C_{I,q}^x +  \sum_{t=1}^{T_c} \\  \Big[   (\mathbf{a}_I^\prime(\theta_{I,q}^r))^H  \sum_{k\!=\!1}^Q\mu_{\text{INL},q}^* \mu_{\text{BNL},k} \mathbf{R}^H\!(t)\mathbf{a}_B(\theta_{B,k}^r) \!+\! (\mathbf{a}_S^\prime\!(\theta_{\!I,q}^r))^H \\ 
	  \Big( \sum_{k\!=\!1}^P \mu_{\text{INL},q}^* \mu_{\text{IL},k} \mathbf{a}_S(\theta_{\!I,k}^z) \!\!-\!\! \mu_{\text{INL},q}^*\mathbf{y}_{s,c}(t) \Big) \! \Big] C_{I,q}^x \!+\! (\mathbf{a}_I^\prime(\theta_{I,q}^r))^H \\
 	\Big[ \sum_{k=1}^Q \big( \mu_{\text{INL},q}^*  \mu_{\text{INL},k} [\boldsymbol{\Sigma}_{\text{INL}}^*]_{q,k}  \big) \mathbf{R}^{H}\mathbf{R} 
	\mathbf{a}_I(\theta_{I,k}^r) \!+\! \mathbf{R}^{H} \bar{\mathbf{A}}_B^z  \mu_{\text{INL},q}^* \\
	\boldsymbol{\mu}_{\text{BL}} + \mathbf{R}^{H}\mathbf{R} \mathbf{A}_I^z  \mu_{\text{INL},q}^* \boldsymbol{\mu}_{\text{IL}} - \mu_{\text{INL},q}^*\mathbf{R}^{H}\!\mathbf{y}_{B,c} \Big] C_{I,q}^x \bigg\}$,
	where $\theta_{I,q}^r$ and $\theta_{I,p}^z$ denote $\theta_{I}(\mathbf{r}_q + \Delta \mathbf{r}_q)$ and $\theta_{I}(\mathbf{z}_p + \Delta \mathbf{z}_p)$, respectively,  $\mathbf{a}_{IS}^{\prime}(\theta_{I,q}^r) \!=\! \mathbf{a}_I^\prime(\theta_{I,q}^r) \mathbf{a}_S^H(\theta_{I,q}^r)   + \mathbf{a}_I(\theta_{I,q}^r) (\mathbf{a}_S^\prime(\theta_{I,q}^r))^H$ with $\mathbf{a}_I^\prime(\theta_{I,q}^r)$ and $\mathbf{a}_S^\prime(\theta_{I,q}^r)$ being the partial derivatives $\frac{\partial \mathbf{a}_I(\theta_{I,q}^r)}{\partial \theta_{I,q}^r}$ and $\frac{\partial \mathbf{a}_S(\theta_{I,q}^r)}{\partial \theta_{I,q}^r}$, $\mathbf{R} = [\mathbf{R}(1);\cdots;\mathbf{R}(T_c)]$ with $\mathbf{R}(t) = \boldsymbol{\varphi}_c^T(t) \odot \mathbf{H}_{IB}$, $\bar{\mathbf{A}}_B^z = \mathbf{1}_{T_c} \otimes \mathbf{A}_{B}(\Delta \mathbf{z})$, $\mathbf{A}_I^z = \mathbf{A}_I(\Delta\mathbf{z})$, $C_{I,q}^x  = \frac{\partial \theta_{I,q}^r}{\partial \Delta r_q^x} = -\frac{r_q^y + \Delta r_q^y - p_I^y}{\| \mathbf{r}_q + \Delta\mathbf{r}_q - \mathbf{p}_I\|^2}$, $T_s \!=\! T_1$ and $T_c \!=\! T_2$ for phase \Rmnum{1}, $T_s \!=\! T_1 \!+\! T_3$ and $T_c \!=\! T_2 \!+\! T_4$ for phase \Rmnum{2}, $\tilde{\boldsymbol{\varphi}}_{r}(t) = \tilde{\boldsymbol{\varphi}}^{\text{\Rmnum{1}}}_{r}(t)$ when $t \leq T_1$; otherwise, $\tilde{\boldsymbol{\varphi}}_{r}(t) = \tilde{\boldsymbol{\varphi}}^{\text{\Rmnum{2}}}_{r}(t-T_1)$, $\boldsymbol{\varphi}_{c}(t) = \boldsymbol{\varphi}^{\text{\Rmnum{1}}}_{c}(t)$ when $t \leq T_2$; otherwise, $\boldsymbol{\varphi}_{c}(t) = \boldsymbol{\varphi}^{\text{\Rmnum{2}}}_{c}(t-T_2)$. Furthermore, $\mathbf{g}_{\text{BS},q}^{r,x}$ is given by
	$\mathbf{g}_{\text{BS},q}^{r,x} \!=\! 2\Re\bigg\{\! \sum_{t\!=\!1}^{T_s} \Big[ \tilde{\boldsymbol{\varphi}}_{r}^H\!(t) \Big( \mathbf{a}_I(\theta_{I,q}^r)(\mathbf{a}_B^\prime(\theta_{B,q}^r))^H  \sum_{k \!=\! 1}^Q (\mu_{\text{ITB},q}^* \mu_{\text{ITB},k} \!+\! [\boldsymbol{\Sigma}_{\text{ITB}}^*]_{q,k})\mathbf{a}_B\!(\theta_{B,k}^r)\mathbf{a}_I^H(\theta_{I,k}^r)   \!\Big)\! \tilde{\boldsymbol{\varphi}}_{r}(t)
	 \!\!+ \!\! \tilde{\boldsymbol{\varphi}}_{r}^{\!H}\!(t)\!
	\Big(\! \mathbf{a}_I(\!\theta_{\!I,q}^{r})  (\mathbf{a}_{\!B}^\prime(\!\theta_{B,q}^r))^{\!H} \\ 
	\Big( \sum_{k=1}^Q \!\mu_{\text{ITB},q}^* \mu_{\text{CTB},k} \mathbf{a}_B(\theta_{B\!,k}^r) \!-\! \mu_{\text{ITB},q}^*\mathbf{y}_{B,r}(t)\Big)  \!\!+\! \sum_{j=1}^Q \mu_{\text{ITB},j}^* \\  
	\mu_{\text{CTB},q} \mathbf{a}_I(\!\theta_{I\!,j}^r)   
	\mathbf{a}_B^H(\!\theta_{B\!,j}^r) \mathbf{a}_B^\prime(\theta_{B,q}^r) \Big) 
	\!\!+\!\!  (\mathbf{a}_B^\prime(\!\theta_{B,q}^r))^H \Big(\! \sum_{k\!=\!1}^Q \!\!
	\big( \mu_{\\\text{CTB},q}^* \\
	\mu_{\text{CTB},k} \!\!+\!\! [\boldsymbol{\Sigma}_{\text{CTB}}^*]_{q,k} \big)  \mathbf{a}_B(\theta_{B,k}^r) \!-\!  \mu_{\text{CTB},q}^*\mathbf{y}_{B,r}(t) \Big) \Big] C_{B,q}^x 
	\!+ \!   \sum_{t\!=\!1}^{\!T_c}    (\mathbf{a}_{\!B}^\prime(\!\theta_{B,q}^r))^H \Big( \!\sum_{k\!=\!1}^Q\! \big(\mu_{\text{BNL},q}^* \mu_{\text{BNL},k} \!+\! [\boldsymbol{\Sigma}_{\text{BNL}}^*]_{q,k} \! \big) 
	\mathbf{a}_{\!B}(\theta_{B,k}^r) + \\
	\sum_{k\!=\!1}^Q \! \mu_{\!\text{BNL},q}^* \mu_{\!\text{INL},k} \mathbf{R}(t) \mathbf{a}_I(\theta_{I,k}^r)
	\!+\! \sum_{k=1}^P  \mu_{\text{BNL},q}^*  \mu_{\text{BL},k}  \mathbf{a}_B(\theta_{B,k}^z) \\
	+ \sum_{k=1}^P  \mu_{\text{BNL},q}^* \mu_{\text{IL},k} \mathbf{R}(t) \mathbf{a}_I(\theta_{I,k}^z)  - \mu_{\text{BNL},q}^* \mathbf{y}_{B,c}(t)   \Big) C_{B,q}^x \bigg\}$,
	where $C_{B,q}^x  = \frac{\partial \theta_{B,q}^r}{\partial \Delta r_q^x} = -\frac{r_q^y + \Delta r_q^y - p_B^y}{\| \mathbf{r}_q + \Delta\mathbf{r}_q - \mathbf{p}_B\|^2}$. Similarly, we have $ \frac{\partial Q(\!\Delta \mathbf{r}\!, \Delta \mathbf{z};\Delta \mathbf{r}^{\!n}, \Delta \mathbf{z}^{n}) }{ \partial \Delta r_q^y} = \mathbf{g}_{\text{BS},q}^{r,y} + \mathbf{g}_{\text{IRS},q}^{r,y} $, where   $\mathbf{g}_{\text{IRS},q}^{r,y}$ and $\mathbf{g}_{\text{BS},q}^{r,y}$ can be obtained by replacing $C_{I,q}^x$ and $C_{B,q}^x$ by $C_{I,q}^y  = \frac{r_q^x + \Delta r_q^x - p_I^x}{\| \mathbf{r}_q + \Delta\mathbf{r}_q - \mathbf{p}_I\|^2}$ and $C_{B,q}^y  =  \frac{r_q^x + \Delta r_q^x - p_B^x}{\| \mathbf{r}_q + \Delta\mathbf{r}_q - \mathbf{p}_B\|^2}$ in the expressions of  $\mathbf{g}_{\text{IRS},q}^{r,x}$ and $\mathbf{g}_{\text{BS},q}^{r,x}$, respectively. The expressions for $\mathbf{g}_{\text{IRS},p}^{z} = [\mathbf{g}_{\text{IRS},p}^{z,x},\mathbf{g}_{\text{IRS},p}^{z,y}]$ and $\mathbf{g}_{\text{BS},p}^{z} = [\mathbf{g}_{\text{BS},p}^{z,x},\mathbf{g}_{\text{BS},p}^{z,y}]$ are similar to $\mathbf{g}_{\text{IRS},q}^{r} =[\mathbf{g}_{\text{IRS},q}^{r,x},\mathbf{g}_{\text{IRS},q}^{r,y}]$ and $\mathbf{g}_{\text{BS},q}^{r} = [\mathbf{g}_{\text{BS},q}^{r,x},\mathbf{g}_{\text{BS},q}^{r,y}]$.  Due to space limitation, the details are omitted here.
	\vspace{-0.1cm}
	
	\subsection{Derivation of the Submatrices in \eqref{FIM}}
	\label{Appendix: submatrices in FIM}
	First, we focus on the calculation of $\mathbf{J}_{\!\boldsymbol{\xi}^{\dot{T}}\!,\boldsymbol{\xi}^{\dot{T}}}$. Let $\boldsymbol{\mu}^{\text{\Rmnum{1}}}_{s,r} \!\!=\!  (\mathbf{I}_{\!N_s} \!\otimes (\boldsymbol{\Phi}^{\text{\Rmnum{1}}}_r)^T) \text{vec}(\tilde{\mathbf{H}}_{CTS}^T) + (\mathbf{I}_{\!N_s} \!\otimes \mathbf{1}_{\!T_1})\mathbf{h}_{CTS}$, $\boldsymbol{\mu}^{\text{\Rmnum{1}}}_{B,r} = (\mathbf{I}_{M} \otimes (\boldsymbol{\Phi}^{\text{\Rmnum{1}}}_r)^T)\text{vec}(\tilde{\mathbf{H}}_{CTB}^{\!T}) + (\mathbf{I}_{M} \otimes \mathbf{1}_{\!T_1})\mathbf{h}_{CTB}$, $\boldsymbol{\mu}^{\text{\Rmnum{2}}}_{s,r} =  (\mathbf{I}_{N_s} \otimes (\boldsymbol{\Phi}^{\text{\Rmnum{2}}}_r)^T) \text{vec}(\tilde{\mathbf{H}}_{CTS}^T) + (\mathbf{I}_{N_s} \otimes \mathbf{1}_{T_3})\mathbf{h}_{CTS}$ and $\boldsymbol{\mu}^{\text{\Rmnum{2}}}_{B,r} = (\mathbf{I}_{M} \otimes (\boldsymbol{\Phi}^{\text{\Rmnum{2}}}_r)^T)\text{vec}(\tilde{\mathbf{H}}_{CTB}^T) + (\mathbf{I}_{M} \otimes \mathbf{1}_{T_3})\mathbf{h}_{CTB}$ with $\tilde{\mathbf{H}}_{CTS} = \mathbf{H}_{ITS}\text{diag}(\mathbf{h}_{CI})$ and $\tilde{\mathbf{H}}_{CTB} = \mathbf{H}_{ITB}\text{diag}(\mathbf{h}_{CI})$. Then, $\mathbf{J}_{\boldsymbol{\xi}^{\dot{T}},\boldsymbol{\xi}^{\dot{T}}}$ can be expressed as
	\begin{align}
	\mathbf{J}_{\boldsymbol{\xi}^{\dot{T}},\boldsymbol{\xi}^{\dot{T}}} =& \frac{2}{\sigma^2}\Re\bigg\{\frac{ \partial(\boldsymbol{\mu}^{\text{\Rmnum{1}}}_{s,r})^H  \partial\boldsymbol{\mu}^{\text{\Rmnum{1}}}_{s,r}}{ \partial\boldsymbol{\xi}^{\dot{T}} \partial (\boldsymbol{\xi}^{\dot{T}})^T}  + \frac{ \partial(\boldsymbol{\mu}^{\text{\Rmnum{1}}}_{B,r})^H \partial\boldsymbol{\mu}^{\text{\Rmnum{1}}}_{B,r}}{ \partial\boldsymbol{\xi}^{\dot{T}}\partial (\boldsymbol{\xi}^{\dot{T}})^T } \nonumber \\
	 &+ \frac{ \partial(\boldsymbol{\mu}^{\text{\Rmnum{2}}}_{s,r})^H \partial\boldsymbol{\mu}^{\text{\Rmnum{2}}}_{s,r}}{ \partial\boldsymbol{\xi}^{\dot{T}} \partial (\boldsymbol{\xi}^{\dot{T}})^T} + \frac{ \partial(\boldsymbol{\mu}^{\text{\Rmnum{2}}}_{B,r})^H \partial\boldsymbol{\mu}^{\text{\Rmnum{2}}}_{B,r}}{ \partial\boldsymbol{\xi}^{\dot{T}} \partial (\boldsymbol{\xi}^{\dot{T}})^T} \bigg\}, 
	\end{align}
	where $\frac{ \partial(\boldsymbol{\mu}^{\text{\Rmnum{1}}}_{s,r})^H  \partial\boldsymbol{\mu}^{\text{\Rmnum{1}}}_{s,r}}{ \partial\boldsymbol{\xi}^{\dot{T}} \partial (\boldsymbol{\xi}^{\dot{T}})^T}$ $\!=\!$
%		\begin{equation}
	$\sum_{n = 1}^{N_s} \!\sum_{t = 1}^{T_1} \! \nabla_{\! \small{\boldsymbol{\xi}^{\dot{T}}}} \bar{\mathbf{h}}_{CTS,n} \bar{\boldsymbol{\varphi}}^{\text{\Rmnum{1}}}_r(t)  (\bar{\boldsymbol{\varphi}}^{\text{\Rmnum{1}}}_r(t))^{\!H}\! \\ (\nabla_{\! \small{\boldsymbol{\xi}^{\dot{T}}}} \bar{\mathbf{h}}_{CTS,n})^{\!H}$
%		\end{equation}
	with $\bar{\boldsymbol{\varphi}}^{\text{\Rmnum{1}}}_r(t) \!=\! [(\boldsymbol{\varphi}^{\text{\Rmnum{1}}}_r(t))^{\!*};1]$ and $\nabla_{\! \small{\boldsymbol{\xi}^{\dot{T}}}} \bar{\mathbf{h}}_{CTS,n} = \\ \frac{\partial}{\partial \boldsymbol{\xi}^{\!\dot{T}}}\!\big[ [\tilde{\mathbf{H}}_{CTS}]_{\!n,:}^*, \mathbf{h}_{CTS,n}^* \! \big]$; $\frac{ \partial(\boldsymbol{\mu}^{\!\text{\Rmnum{1}}}_{\!B\!,r}\!)^{\!H} \!\partial\boldsymbol{\mu}^{\!\text{\Rmnum{1}}}_{\!B\!,r}}{ \partial\boldsymbol{\xi}^{\dot{T}}\! \partial (\boldsymbol{\xi}^{\dot{T}})^T} \!\!=\!\! \sum_{n \!=\! 1}^{M}  \!
	 \sum_{t \!=\! 1}^{T_1} \!\!  
	 \nabla_{ \small{\boldsymbol{\xi}^{\dot{T}}}} \bar{\mathbf{h}}_{C\!T\!B,n} \\ \bar{\boldsymbol{\varphi}}^{\text{\Rmnum{1}}}_r(t) (\bar{\boldsymbol{\varphi}}^{\text{\Rmnum{1}}}_r(t))^{H} 
	(\nabla_{\!\! \small{\boldsymbol{\xi}^{\dot{T}}}} \bar{\mathbf{h}}_{CTB,n})^{\!H}$
	with $\nabla_{\!\! \small{\boldsymbol{\xi}^{\dot{T}}}} \bar{\mathbf{h}}_{CTB,n} \!\!=\!\! \frac{\partial}{\partial \boldsymbol{\xi}^{\dot{T}}} \! \big[ [\tilde{\mathbf{H}}_{CTB}]_{n,:}^*, \mathbf{h}_{CTB,n}^*  \big]$; $\frac{ \partial(\boldsymbol{\mu}^{\text{\Rmnum{2}}}_{s,r}\!)^{\!H}  \partial\boldsymbol{\mu}^{\text{\Rmnum{2}}}_{s,r}}{ \partial\boldsymbol{\xi}^{\dot{T}} \partial (\boldsymbol{\xi}^{\dot{T}})^T} $ and $ \frac{ \partial(\boldsymbol{\mu}^{\text{\Rmnum{2}}}_{B,r}\!)^{\!H}  \partial\boldsymbol{\mu}^{\text{\Rmnum{2}}}_{B,r}}{ \partial\boldsymbol{\xi}^{\dot{T}} \partial (\boldsymbol{\xi}^{\dot{T}})^T}  $ exhibit similar expressions to $\frac{\partial (\boldsymbol{\mu}^{\text{\Rmnum{1}}}_{s,r}\!)^{\!H}\partial\boldsymbol{\mu}^{\text{\Rmnum{1}}}_{s,r}}{\partial \boldsymbol{\xi}^{\dot{T}} \partial (\boldsymbol{\xi}^{\dot{T}})^T }$ and $\frac{\partial (\boldsymbol{\mu}^{\text{\Rmnum{1}}}_{B,r}\!)^{\!H} \partial\boldsymbol{\mu}^{\text{\Rmnum{1}}}_{B,r}}{\partial \boldsymbol{\xi}^{\dot{T}} \partial (\boldsymbol{\xi}^{\dot{T}})^T }$ but with $\bar{\boldsymbol{\varphi}}^{\text{\Rmnum{1}}}_r(t)$ and $T_1$ replaced by $\bar{\boldsymbol{\varphi}}^{\text{\Rmnum{2}}}_r(t)$ and $T_3$, respectively.
		
	\par Next, let  $\boldsymbol{\mu}^{\text{\Rmnum{1}}}_{s,c} =   (\mathbf{I}_{N_s} \otimes \mathbf{1}_{T_2})\mathbf{h}_{SU}$, $\boldsymbol{\mu}^{\text{\Rmnum{1}}}_{B,c} = (\mathbf{I}_{M} \otimes (\boldsymbol{\Phi}^{\text{\Rmnum{1}}}_c)^T)\text{vec}(\tilde{\mathbf{H}}_{BU}^T) + (\mathbf{I}_{M} \otimes \mathbf{1}_{T_2})\mathbf{h}_{BU}$, $\boldsymbol{\mu}^{\text{\Rmnum{2}}}_{s,c} =   (\mathbf{I}_{N_s} \otimes \mathbf{1}_{T_4})\mathbf{h}_{SU}$ and $\boldsymbol{\mu}^{\text{\Rmnum{2}}}_{B,c} = (\mathbf{I}_{M} \otimes (\boldsymbol{\Phi}^{\text{\Rmnum{2}}}_c)^T)\text{vec}(\tilde{\mathbf{H}}_{BU}^T) + (\mathbf{I}_{M} \otimes \mathbf{1}_{T_4})\mathbf{h}_{BU}$ with $\tilde{\mathbf{H}}_{BU} = \mathbf{H}_{IB}\text{diag}(\mathbf{h}_{IU})$. Then, $\mathbf{J}_{\boldsymbol{\xi}^{S},\boldsymbol{\xi}^{S}}$ can be expressed as
	\begin{align}
	\mathbf{J}_{\boldsymbol{\xi}^{S},\boldsymbol{\xi}^{S}} =& \frac{2}{\sigma^2}\Re\bigg\{\frac{ \partial(\boldsymbol{\mu}^{\text{\Rmnum{1}}}_{s,c})^H \partial\boldsymbol{\mu}^{\text{\Rmnum{1}}}_{s,c}}{ \partial\boldsymbol{\xi}^{S} \partial (\boldsymbol{\xi}^{S})^T} +\frac{ \partial(\boldsymbol{\mu}^{\text{\Rmnum{1}}}_{B,c})^H  \partial\boldsymbol{\mu}^{\text{\Rmnum{1}}}_{B,c}}{ \partial\boldsymbol{\xi}^{S} \partial (\boldsymbol{\xi}^{S})^T} \nonumber \\
	&+ \frac{ \partial(\boldsymbol{\mu}^{\text{\Rmnum{2}}}_{s,c})^H \partial\boldsymbol{\mu}^{\text{\Rmnum{2}}}_{s,c}}{ \partial\boldsymbol{\xi}^{S} \partial (\boldsymbol{\xi}^{S})^T} + \frac{ \partial(\boldsymbol{\mu}^{\text{\Rmnum{2}}}_{B,c})^H \partial\boldsymbol{\mu}^{\text{\Rmnum{2}}}_{B,c}}{ \partial\boldsymbol{\xi}^{S}\partial (\boldsymbol{\xi}^{S})^T }\bigg\}, 
	\end{align}
	where $\frac{ \partial(\boldsymbol{\mu}^{\text{\Rmnum{1}}}_{s,c})^H \partial\boldsymbol{\mu}^{\text{\Rmnum{1}}}_{s,c}}{ \partial\boldsymbol{\xi}^{S}\partial (\boldsymbol{\xi}^{S})^T }$ $\!=\!$
	$T_2 \frac{\partial \mathbf{h}_{SU}^H \partial \mathbf{h}_{SU}}{\partial\boldsymbol{\xi}^{S} \partial (\boldsymbol{\xi}^{S})^T }$, $\frac{ \partial(\boldsymbol{\mu}^{\text{\Rmnum{1}}}_{B,c}\!)^{\!H} \!\partial\boldsymbol{\mu}^{\text{\Rmnum{1}}}_{B,c}}{ \partial\boldsymbol{\xi}^{S}\! \partial (\boldsymbol{\xi}^{S})^T} $ $\!=\!\!$
	$\sum_{n = 1}^{M} \!\sum_{t = 1}^{T_2}\! \! \nabla_{\!\! \small{\boldsymbol{\xi}^{S}}} \bar{\mathbf{h}}_{BU,n} \bar{\boldsymbol{\varphi}}^{\text{\Rmnum{1}}}_c(t) (\!\bar{\boldsymbol{\varphi}}^{\text{\Rmnum{1}}}_c(t)\!)^{\!H}  (\nabla_{\!\! \small{\boldsymbol{\xi}^{S}}} \bar{\mathbf{h}}_{BU,n})^{\!H}$
	with  $\bar{\boldsymbol{\varphi}}^{\text{\Rmnum{1}}}_c(t) = [(\boldsymbol{\varphi}^{\text{\Rmnum{1}}}_c(t))^*;1]$  and $\nabla_{\!\! \small{\boldsymbol{\xi}^{S}}} \bar{\mathbf{h}}_{BU,n} \!\!=\!\! \frac{\partial}{\partial \boldsymbol{\xi}^{S}} \! \big[ [\tilde{\mathbf{H}}_{BU}]_{n,:}^*,  \mathbf{h}_{BU,n}^*  \big]$.
	$\frac{ \partial(\boldsymbol{\mu}^{\text{\Rmnum{2}}}_{s,c})^H \partial\boldsymbol{\mu}^{\text{\Rmnum{2}}}_{s,c}}{ \partial\boldsymbol{\xi}^{S} \partial (\boldsymbol{\xi}^{S})^T}$ and $ \frac{ \partial(\boldsymbol{\mu}^{\text{\Rmnum{2}}}_{B,c})^H \partial\boldsymbol{\mu}^{\text{\Rmnum{2}}}_{B,c}}{ \partial\boldsymbol{\xi}^{S}\partial (\boldsymbol{\xi}^{S})^T }$ can be obtained similarly by replacing  $T_2$ and $\bar{\boldsymbol{\varphi}}^{\text{\Rmnum{1}}}_c(t)$  by  $T_4$ and  $\bar{\boldsymbol{\varphi}}^{\text{\Rmnum{2}}}_c(t) $.
		
	\par Finally, by following a similar derivation, we can readily obtain the expressions for $\mathbf{J}_{\boldsymbol{\xi}^{\dot{T}},\boldsymbol{\xi}^{O}}$,  $\mathbf{J}_{\boldsymbol{\xi}^{O},\boldsymbol{\xi}^{S}}$, $\mathbf{J}_{\boldsymbol{\xi}^{O},\boldsymbol{\xi}^{O}}$ and $\mathbf{J}_{\boldsymbol{\xi}^{U},\boldsymbol{\xi}^{U}}$. The details are omitted here for brevity.
\vspace{-0.0cm}
	
\end{appendix}
%%%%%%%%%%%%%%%%%%%%%%%%%%%%%% Bibliography %%%%%%%%%%%%%%%%%%%%%%%%%%%%%%%%%%%%
\bibliographystyle{IEEEtran}
%\bibliography{IEEEabrv,bib}
\bibliography{bib_ISAC_IRS}
\end{document}